\documentclass[a4paper, 12pt]{article}
\usepackage[dvips]{graphicx}
\usepackage{natbib}
\usepackage{lscape}
\usepackage{color}
\usepackage{ulem}
\begin{document}
\begin{center}
  \thispagestyle{empty}
  \textbf{\Large{Statistical Analysis of Global Connectivity and Activity Distributions in Cellular Networks}}\\[5ex]
  \textbf{Adri\'an L\'opez Garc\'ia de Lomana$^{1}$, Qasim K. Beg$^{2,3}$, G. de Fabritiis$^{1}$ and Jordi Vill\`a-Freixa}$^{1,*}$\\[10ex]
\end{center}
$^{1}$Computational Biochemistry and Biophysics Laboratory, Research Unit on Biomedical Informatics, IMIM / Universitat Pompeu Fabra, c/ Dr. Aiguader 88, 08003, Barcelona, Spain.\\ 
$^{2}$Department of Pathology, 3550 Terrace St., University of Pittsburgh, Pittsburgh, PA 15261, USA.\\
$^{3}$Current Address: Department of Biomedical Engineering, 44 Cummington St., Boston University, Boston, MA 02215, USA.\\
$^{*}$ Corresponding author, jordi.villa@upf.edu.
\section*{Abstract}
Various molecular interaction networks have been claimed to follow power-law
decay for their global connectivity distribution. It has been proposed that
there may be underlying generative models that explain this heavy-tailed
behavior by self-reinforcement processes such as classical or hierarchical
scale-free network models. Here we analyze a comprehensive data set of
protein-protein and transcriptional regulatory interaction networks in yeast,
an \textit{E. coli} metabolic network, and gene activity profiles for
different metabolic states in both organisms. We show
that in all cases the networks have a heavy-tailed distribution, but most of
them present significant differences from a power-law model according to a stringent
statistical test. Those few data sets that have a statistically significant
fit with a power-law model follow other distributions equally well. Thus,
while our analysis supports that both global connectivity interaction networks
and activity distributions are heavy-tailed, they are not generally described
by any specific distribution model, leaving space for further inferences on
generative models.\\[2ex]
\textbf{Key words:} cellular networks, fat-tailed distributions, maximum likelihood estimation, hypothesis test.
\newpage
\section{Introduction}
Over the last few years it has been extensively argued that the
connectivity distributions of various molecular interaction networks follow
power-law distributions that are best approximated by classical \citep{barabasi99}
or hierarchical \citep{ravasz02} scale-free network models (reviewed in Refs.
\citep{albert05, barabasi04}). Although this does not imply that power-law
distributions do in fact describe the observed degree distributions, it has been
used as motivation for developing generative models that yield power laws.
Moreover, power-law models also appear to characterize the distribution of
activity levels. For example, both calculated metabolic flux values
\citep{almaas04}, and measured gene expression values \citep{ueda04} seem to
follow such a distribution, which is in agreement with theoretical predictions for
load distribution on scale-free networks \citep{almaas04, goh01}.

Using a statistical framework similar to the one employed in this paper, a recent
work \citep{edwards07} revisited several enhanced datasets of foraging patterns in
wild animals and concluded that the distributions previously classified as
power-law were better described by a Gamma distribution. The previous spurious
results were attributed in part to the limited magnitude of the dataset (also
typical of biological datasets) and to the graphical method used to assess the
character of the distributions. 
It has been suggested \citep{khanin06, stumpf07, tanaka05a} that the charactheristic observation that biological networks follow a power-law distribution may have been reached due to the same methodological shortcomings. 
Specifically, it has been argued that analysing relatively small cellular networks, having
only a few hundred to a few thousands data elements using the commonly employed
frequency-degree or intensity plots, does not have sufficient power to differentiate
among various network models having heavy-tailed distributions, and that the use
of rank-degree (intensity) plots proves superior for this purpose \citep{khanin06,
stumpf07, tanaka05a}. Furthermore, a rigorous quantification of goodness-of-fit
is also required to establish relatedness to various network models, and the
quality of the underlying data set (i.e., the quality of network reconstruction)
is critical for proper analysis.
\subsection{The Models and the MLE Analytical Framework}
Assessment of the best model explaining a given data distribution has typically been done using simple linear regression methods. These methods are suitable
for normal distribution functions, but not for highly skewed distribution functions. Essentially, the
problem arises from the fact that skewed distributions are characterized by the scale of the tail, which forms most of the support for the distribution but
that is barely populated (i.e. contains less than 10\% of the data points).
Because of this, simple least square fits of the probability density distribution
computed via histogram methods are very poor estimators of the distribution
parameters (see \citep{tanaka05a} for a review of possible problems) due to
the noisy poor sampling of the tails.

Therefore, it has been argued that density plots should not be used as a base for the fitting of these types of data, as several more reliable methods are available. 
A simple and better strategy is to use rank-plots as commonly used in engineering and economics \citep{tanaka05a}. 
Logarithmic binning \citep{albert99} has also been used as a more robust alternative, but it has been reported that this procedure fails to retrieve the value of the exponent as the slope of the graph for power-law distributions \citep{tanaka05a,goldstein04,clauset07}. 
Finally, one could apply a logarithmic transformation to the data and fit the corresponding distribution function \citep{de_fabritiis03} thus avoiding the problem of skewed data from the outset. 
In this case the appropriate transformation of the probability distribution has to be performed (for instance a normal distribution for log-transformed, log-normally distributed data), a procedure which could be cumbersome for some distribution functions.

Below we focus on the discrimination of different models (see Section \ref{fdf}) for several types of molecular interaction and expression data sets by using probability-based techniques to perform comparative analyses. 
In order to have a good mathematical representation of the probability distribution we use cumulative distribution functions (cdf) that are directly related to rank-plots \citep{tanaka05a}. 
In addition, instead of graphical-based estimation methods we utilize maximum likelihood estimation (MLE) (see Section \ref{methods}), which is not dependent on the graphical representation and allows us to perform a statistical test of the proposed models \citep{goldstein04,stumpf05,hoogenboom06,clauset07}.

Based on these considerations, here we reexamine both the global topological connectivity distributions and absolute gene and protein expression value distributions in cellular networks, with a focus on the most completely reconstructed, highest quality data sets. 
We have tested the empirical distributions against several plausible models with heavy-tailed distribution.  
Note also that generative models in the context of biological networks have been proposed for power-law \citep{barabasi99,qian01,rzhetsky01,bhan02,pastor-satorras03} and broad-scale \citep{amaral00} distributions but not for the rest of the tested distributions. 
We examine the extent to which several types of distributions, some with plausible underlying generative models, can provide a similar probabilistic framework for the experimental data being analyzed. 
We show that for the analyzed molecular interaction distributions, only the high-throughput protein-protein interaction data set and the out-degree distribution of the transcriptional regulatory network significantly fitted the power-law model. 
In the case of the distributions of global gene and protein expression values, only two mid-log cultures of \textit{E. coli} showed a statistically significant fit.
At any rate, all experimental data sets having significant fits for the
power-law distribution, showed equally good fits for other distributions which
leads to inconclusive results for the support of a single specific
distribution. 
We discuss the implications of this finding for the uniqueness of generative models solely from topology and activity data distributions.
\section{Materials and Methods} \label{methods}
For model classification, we have basically used the methods developed and implemented in \citep{clauset07}.
However, we provide here sufficient details about the mathematical foundations to make the analytical procedure clear.
\subsection{Probability Based Estimation Method} \label{estimationMethod}
The likelihood of a given probability distribution $p_{\mathbf{v}}(x_i)$ depending on parameters $\mathbf{v}$ and describing a given dataset of independent data $(x_1, ..., x_N)$, is defined as $l(\mathbf{v}|x_1, ..., x_N)=\displaystyle\prod_{i}^{N}p_{\mathbf{v}}(x_i)$.
Our datasets are built of the (discrete) number of edges in interaction networks or the (continuous) amount of expression of the nodes in activity networks. 
Therefore, from a group of edges or nodes $(x_1, ..., x_N)$, we obtain the probability associated with each value $(p_{\mathbf{v}}(x_1), ..., p_{\mathbf{v}}(x_N))$. 

Here we use different model distributions with their corresponding set of
parameters $\mathbf{v}$. In each case a maximum likelihood estimation (MLE) based on
the log-likelihood function, $L(\mathbf{v}|x) = \mathrm{log}\;l(\mathbf{v}|x)$,
is used to compute the set of parameters $\mathbf{v}_{opt}$ which maximizes $L$.
A special case represents the parameter $x_{min}$ which is the lower bound of the
power-law distribution (see Section \textit{SI} 2). For its calculation we
followed the alternative procedures developed and implemented at
\citep{clauset07}. Once the vector $\mathbf{v}_{\mathrm{opt}}$ has been obtained,
we move on to find if there are differences between the empirical data and the
model. An intuitive approach would be to perform a Kolmogorov-Smirnov (KS)
test using the empirical distribution set and the estimated model distribution
\citep{goldstein04}. However the two distributions are not independent, which is
one of the required premises of the KS test, because the model was estimated from
the same data with which we want to perform the test. Instead, a  Monte Carlo
protocol  can be used to avoid such direct comparison. As described in
\citep{clauset07}, the KS statistic $D$ is calculated for many Monte Carlo
generated data sets from the estimated distribution. Our $p$-value will be simply
the fraction of times that $D$ for the empirical data is larger than $D$ for the
generated data sets \citep{clauset07}. We define the null hypothesis $H_0$ as, no
statistical difference between the data and the model. In our case, the
confidence value of $p=0.1$ will in fact be more appropriate than $p=0.05$ for
the statistical test, following \citep{clauset07,mayo06}. The rationale is that we
are looking for differences in only one tail of the distribution, we look for
$p$-values that are larger than that associated with the experimental fit.

\subsection{Distribution Functions Estimation} \label{fdf}
Both continuous and discrete model distributions are used depending on the nature
of the data analyzed in each case. As described in Section
\ref{fittedDistributionFunction}, we have tested seven models of probability
distribution against the studied data. In turn, five model distributions were
tested against the continuous data sets: power-law, log-normal, exponential,
Weibull and broad-scale. It is worth noting that only some of the distributions
tested (e.g., power-law and broad-scale) have been put forward as distributions
agreeing with plausible generative models in the biological context
\citep{barabasi99,amaral00}. The intention of this work is to expand the
application of the methods to other \textit{a priori} suitable distributions to
give a more general scope to the study.
\subsubsection{Distribution Functions} \label{fittedDistributionFunction}
\textit{Power-law model:} 
A random variable $X$ is said to follow a power-law distribution with index
$\alpha$ when the scaling law of $P[X>x]$ is power-law, such as $P[X>x] = 1-P[X \leq
x] \approx cx^{-\alpha}$ for x$\rightarrow\infty$. We used the Pareto
distribution for the continuous data: probability density function (pdf):
$p(x)=\frac{\alpha-1}{x_{min}}\left(\frac{x}{x_{min}}\right)^{-\alpha}$,
cumulative distribution function (cdf): $P[X \leq
x]=\left(\frac{x}{x_{min}}\right)^{-\alpha+1}$ and for the discrete case we used
Zipf's law with the following terms, pdf: $p(k)=\frac{k^{-\alpha}}{\zeta
\;(\alpha,k_{min})}$, cdf: $P[K \leq k]=\frac{\zeta \;(\alpha,k)}{\zeta
\;(\alpha,k_{min})}$. The Hurwitz zeta function is defined as $\zeta \;(\alpha,k)
= \displaystyle\sum_{n=0}^{\infty}(n+k)^{-\alpha}$. The reader is referred to
\citep{goldstein04} and \citep{clauset07} for further explanations. Specifically, for the
determination of $x_{min}$ (or $k_{min}$), we have used the methods described in
\citep{clauset07}.

\textit{Log-normal model:} 
A random variable $X$ is log-normally distributed if $Y=log(X)$ follows a normal distribution. 
We have used the following distributions for the continuous case: pdf: $p(x) = \frac{1}{x\sigma \sqrt{2\pi}}e^{-\frac{(ln(x)-\mu)^2}{2\sigma^2}}$, cdf: $P[X\leq x]=\frac{1}{2}+\frac{1}{2}\mathrm{erf}(\frac{ln(x)-\mu}{\sigma \sqrt{2}})$. 
Numerical approximations of \citep{clauset07} have been used for the discrete models.

\textit{Poisson model:}
We have used the discrete Poisson distribution to model the discrete data sets using the following probability mass function (pmf): $f(k,\lambda)=\frac{\lambda^ke^{-\lambda}}{k!}$.

\textit{Yule model:} 
Again this is a discrete distribution that we have only used for discrete data sets. 
We used the following forms, pdf: $p(k)=\rho\mathbf{B}(k,\rho+1)$ and cdf: $P[K \leq k]=1-k\mathbf{B}(k,\rho+1)$, where $\rho>0$ and $\mathbf{B}$ is the Beta function.

\textit{Exponential model:} 
Exponential models have no biological relevance in the framework of our study.
However, we have included it in some cases for methodological comparison as a representative model of a non fat-tailed distribution. 
For a random variable $X$ distributed exponentially we have used the following forms: pdf: $p(x)=\lambda e^{-\lambda x}$, cdf: $P[X\leq x]=1-e^{-\lambda x}$ for the continuous data sets; we have taken the pmf: $f(k,\lambda)=(1-e^{-\lambda})e^{\lambda k_{min}}\;e^{\lambda k}$ as the discrete model.

\textit{Weibull model:} 
Also referred to as stretched exponential, we have used pdf: $p(x)=(k/\lambda)(x/\lambda)^{k-1}e^{-(x/\lambda)^k}$ and cdf: $P[X\leq x]=1-e^{-(x/\lambda)^k}$ for the continuous data sets. For the discrete data we have used the code available from \citep{clauset07} that uses the Nakagawa-Osaki \citep{nakagawa75} method for the discretization of the Weibull distribution. pmf: $f(k;q,\beta)=q^{k^{\beta}}$. 

\textit{Broad-scale model:} 
Broad-scale distribution functions, also referred to here as power-law with cut-off
or power-law plus exponential, identify a class of distributions that are
characterized by a scaling law $p(x) \approx ce^{-\lambda x}x^{-\alpha}$ for $x
\rightarrow \infty$. Note that the broad-scale is very similar to the Gamma
distribution. Both contain a shape and a scale parameter. The Gamma distribution in addition
contains a normalizing constant. For the Gamma distribution, the pdf
would be $p(x) = \frac{\beta^{\alpha}}{\Gamma(x)}e^{-\beta x}x^{\alpha-1}$, being
$\Gamma(x)=(\alpha-1)!$ for $\alpha > 0$. 
The broad-scale distribution is a type of nested function, which implies a specific statistical treatment in Section \ref{distributionComparison}.
In the current case the broad-scale consists of a power-law behavior and after a given threshold it has an exponential
decay. Numerical routines available from \citep{clauset07} have been used to
calculate the corresponding pdf and cdf.

\subsection{Distribution Comparison} \label{distributionComparison}
To compare the feasibility of the different models with respect to the power-law
distribution, we applied a likelihood ratio test, which compares the fits of
two given competing distributions. Note that we are not comparing a model against
the empirical distribution but two different distributions with each other.
Strictly speaking in this case we compare different models with each other to
indirectly assess how well the distributions explain the data. We have again
applied the methods developed in \citep{clauset07} and \citep{vuong89} to evaluate
the normalized log-likelihood ratio (NLLR). If the NLLR has a positive value the
power-law model is supported, whereas the alternative distribution offers a
better fit if NLLR has a negative value. In order to determine significant
positive or negative values, we calculated the associated $p$-value. Given
the different experimental design with respect to the one described in Section
\ref{estimationMethod}, we take here as confidence value the most commonly used
$p=0.05$ \citep{mayo06}. We search for differences on both sides of the distribution and therefore we use
$p=0.05$ as level of significance.

The broad-scale distribution requires a special comment, as it is a nested
function containing both power-law and exponential terms. In fact, if we
compare the power-law distribution fit with the broad-scale distribution fit, the
NLLR test will always be zero or negative, favoring the broad-scale, as it
reproduces the power-law behavior plus an additional term. To solve this problem,
we used a correction of the $p$-value calculated for the log-likelihood ratio
(LLR) as described in \citep{vuong89} and \citep{clauset07}.
\subsection{Data Sets Used}
Molecular interaction maps were obtained from different sources: \textit{S.
cerevisiae} protein-protein interaction networks \citep{reguly06}, \textit{S.
cerevisiae} transcriptional regulatory network \citep{macisaac06}, \textit{E.
coli} metabolic networks \citep{feist07}. Yeast expression values were taken from
\citep{ghaemmaghami03}. Global mRNA expression data for \textit{E. coli} cells and
protein expression data for \textit{S. cerevisiae} cells were obtained from
\citep{beg07}.
\section{Results}
\subsection{Global Topological Organization of Molecular Interaction Networks}
The aim of this paper is to test by means of a probabilistic approach the ability
of heavy-tailed distribution functions to explain the experimental data
distributions for given datasets. Due to the nature of the experimental data
being considered here, we have differentiated between global interaction networks
and global activity networks. Global interaction measurements refer to the number
of interactions of a given molecule. Thus, measurements are counted as natural numbers
and the studied models will be discrete. On the other hand, global activity
interactions are defined in terms of cellular expression, measurements are
expressed using positive real numbers and the applied models will be continuous.
\subsubsection{Global Interaction Networks}
To assess the degree distribution of an undirected homotypic molecular
interaction network we used data obtained from baker's yeast, \textit{S.
cerevisiae}, in \citep{reguly06} that, to date, is the most comprehensive
information for a species-specific protein-protein interaction (PPI) network. We
analyzed two data sets built using two different methodologies. The first data
set was created from the combination of five different studies based on
experimental high-throughput techniques, which we refer to as HTP data
\citep{reguly06}. The second data set was curated manually from the literature (LC
data set), and is assumed to be more accurate than the former \citep{reguly06}.
After the removal of redundant and self-interactions \citep{reguly06}, we obtained
a data set of 11,571 interactions between 4,474 proteins for the HTP data and
8,165 interactions between 2,689 proteins for the LC data. Table
\ref{tableDiscrete} shows that after fitting the parameters using an MLE, only the
HTP data set provides a statistically sufficient goodness-of-fit, based on the KS
test (see \textit{Materials and Methods}). Whether or not the high rate of
false-discovery interactions from yeast-two hybrid experiments in the HTP
data set leads to the acceptance of the null hypothesis for the power-law model is an open issue for
discussion \citep{huang07}. In fact, for the same network, if the data is
retrieved manually from the literature (LC data set), the power-law model is
rejected. Furthermore, for the case of HTP data, exponential and Poisson
distributions behave significantly worse than the power-law model while for the
LC data set, log-normal, Yule, Weibull and broad-scale are better models than the
power-law. Fig. 1{\bf a} shows the degree distribution of the LC network with
its corresponding best fitted power-law model. An equivalent plot for the HTP
data set is shown in Fig. \textit{SI} 1{\bf a}.

We also analyzed the degree distribution of a directed heterotypic molecular interaction network, the transcriptional regulatory (TR) network of \textit{S. cerevisiae} \citep{macisaac06}, the edges of which represent interactions between transcription factors (TF) and gene regulatory regions, distinguishing out-degree from in-degree interactions. 
In total the data includes 99 TFs and 1,851 TF-regulated genes connected through 3,394 links. 
Table \ref{tableDiscrete} clearly indicates that the degree distribution of out-degree interactions (TF$\rightarrow$gene) provide a statistically significant fit for the power-law model.
However, all other tested distributions except the Poisson fit the empirical data equally well which prevents us from establishing unequivocally the power-law distribution as the statistical model to describe the data. 
This result should be taken with caution as the sample size ($N=99$) is really on the limit to be considered too small \citep{hart99}.
For the case of in-degree interactions (number of TF affecting a given gene) the range of connectivities is too small to provide conclusive results, although it appears that all other tested models apart from the Poisson model offer significantly better fits than the power-law.
A graphical representation of the out-degree distribution with their best-fitted power-law model is shown in Fig. 1{\bf b}.
Fig. \textit{SI} 1{\bf c} shows the corresponding plot for the in-degree distribution.

Finally, we assessed the recent high quality reconstruction of the \textit{E. coli} metabolic network \citep{feist07}, in which substrates are connected to each other through edges that represent the actual metabolic reactions \citep{jeong00} with a total of 2,381 reactions, of which 304 are exchange, 553 reversible and 1,524 irreversible reactions. 
We have studied separately the in-degree distribution with 1,657 nodes and 3,050 links and the out-degree distribution with 1,656 metabolites and 2,788 links. 
Both networks behave very similarly: both display significant differences with the power-law model.
While the log-normal model fits the data equally as well as the power-law model, all other tested models behave significantly worse.
The empirical distributions with their corresponding best-fitted power-law models are represented graphically in Fig. \textit{SI} 1{\bf e},{\bf f}.
\subsubsection{Clustering Coefficients}
To gain further insight into the topological organization of molecular interaction networks, we have analyzed the $C(k)$ \textit{vs.} $k$ relationships for the HTP and LC reconstructed PPI networks of yeast \citep{reguly06}, the TR regulatory network of yeast \citep{macisaac06}, and the metabolic network of \textit{E. coli} \citep{feist07}. 
Fig. \textit{SI} 2 shows how the dependence of the clustering coefficient in the three types of molecular interaction networks is not uniform. 
While the \textit{E. coli} metabolic network displays a distribution close to $C(k)=k^{-1}$ (Fig. \textit{SI} 2{\bf d}), as previously described \citep{ravasz02}, the distribution observed in the other two networks is significantly less organized. 
\subsection{Global Activity Level Distributions in Molecular Interaction Networks}
Next we assessed the distribution of activity levels achieved on the underlying network topology. 
First, we examined single time point snapshots for mRNA and protein expression in \textit{E. coli} and in \textit{S. cerevisiae}, respectively. 

We used global gene expression data from \textit{E. coli} cells grown in mid-log phase batch cultures with single carbon source media, using acetate, galactose, glucose, glycerol and maltose as individual carbon sources \citep{beg07}.
Signals for 3,977 probes were examined for the expression level of the corresponding genes based on their hybridization intensity (Table \ref{tableMidlog}). 
Activity distribution of acetate growing cells significantly fits the power-law model, in fact representing the best fit with the data in all cases analyzed in this study.
Interestingly though the broad-scale model gives us a significantly much better fit.
Maltose is another carbon source where gene expression distribution of \textit{E. coli} shows significant fit with the power-law model. 
However, three other distributions, log-normal, Weibull and broad-scale provide significantly better fits than the power-law model.
Of the two cases, in which the power-law model displays no significant differences from the empirical data but other models are also plausible, the classification of power-law character remains uncertain. 
None of the other mid-log cultures provide a significant fit with the power-law distribution.
For glucose, galactose and glycerol, log-normal, Weibull and broad-scale are significantly better models than the power-law, while exponential is significanlty worse model.
Worth noting is the graphical comparison of Fig. 1 {\bf c} and 1{\bf d}.
The acetate distribution displays no statistical difference with the power-law model while the galactose distribution does differ from it.
This important difference in the nature of the empirical data is not revealed from the inspection of the graphical representations but uniquely from the statistical test.

We also examined the transcriptome state of \textit{E. coli} cells grown in chemostat cultures at different dilution rates, representing various steady-state growth rates at different culture densities \citep{vazquez08}.
We find that the general pattern for gene expression distribution is largely invariant in all different growth media and at different growth rates, but for the steady-state cultures none of the empirical data sets fit the power-law model significantly (Table \textit{SI} 1 and Fig. \textit{SI} 4).
Curiously the broad-scale model outperforms for all steady-state cases.
Also note that the log-normal and Weibull distributions display superior fits for dilutions 0.25 and 0.4.
In all the cases, the exponential distribution fits the empirical data significantly worse than the power-law distribution. 

Protein expression values in mid-log phase of \textit{S. cerevisiae} cell cultures from 3,868 different strains expressing a single GFP or TAP tagged protein in a rich growth media \citep{ghaemmaghami03} displayed a slightly different expression mode (Fig. \textit{SI} 5), suggesting difference in mRNA and protein expression value distributions under these growth conditions. 
Yet again the empirical dataset shows significant differences with respect to the power-law model (Table \textit{SI} 2).
Comparing models, the power-law distribution fits significantly better than the exponential, although similarly to the log-normal and Weibull distributions. 
However the broad-scale model offers a significantly better fit that the power-law model.

Finally, to assess the transcriptome state of \textit{E. coli} cells under the
most complex growth conditions, we have examined the dynamical microarray profile
of \textit{E. coli} cells grown in batch culture in a mixed-substrate medium
\citep{beg07}. At all sampled time points, the distribution of expression of the
transcriptome was differing significantly from the power-law distribution (Table
\textit{SI} 3). In this mRNA expression data set the log-normal, Weibull and
broad-scale models show better statistical fit with the experimental data at all
time points, while the exponential distribution is also better for some of the
cases. A graphical representation of the best fitting power-law distribution
against the empirical data is shown in Fig. \textit{SI} 6.

\section{Discussion}

The structure and activity of intracellular molecular interaction networks
represents a basic tenet of life. Thus it is of great importance to correctly
identify their true structural and functional properties at the global,
intermediate and local levels.
Heavy-tailed probability distributions have been used in the literature to
explain the topological features of complex biological networks. In this work we
have analyzed to what extent, among others, three well known heavy-tailed
distributions (power-law, broad-scale and log-normal) can be uniquely used to
represent the observed experimental data for protein interaction, transcription
regulation, metabolic pathways and expression experiments. Our results
demonstrate that the large-scale topology of the molecular interaction networks
and the global mRNA and protein expression distributions examined here do not
 strictly  follow power-law distributions. Moreover, none of the three
 heavy-tailed models tested had a universal agreement with the empirical data even when using
the highest quality data sets available. Distributions are evidently heavy-tailed
and for this type of data MLE analyses prove superior to graphical methods for
assessing different tested distributions. However, we could not assess any of the
studied models with statistical reliability. 
Our results could be explained by several non-exclusive arguments. First, our
analyses could be affected by a biased acquisition of the empirical data, a
systematic error that would affect, for instance, proteins which have been
studied more extensively because they have a high biomedical interest. Also, such
simple mathematical frameworks as considered here may not represent entirely the
biological mechanisms at work, but also be the convolution of the physicochemical
constraints of the cell \citep{beg07}. Finally, population-level evolutionary
processes, such as activation of a foraging program upon extracellular
substrate exhaustion, clearly affect the function of cellular networks
\citep{beg07} and are likely to influence the nature of the underlying molecular interactions as well.

Additional problems can have their origin in data acquisition.
In some cases, the experimental data might be incomplete or noisy, specially for the HTP data which derives partially from yeast two-hybrid experiments.
There has been several efforts \citep{huang07,scholtes08} to infer statistically the actual PPI network from such experiments and consequently our results of the analysis refer to the data, not the actual network.
Indeed, the analysis of the reconstructed network from literature, in a manually curated way, considered to be more correct, yields different results.
The inference of the actual network is out of the scope of the current work.

We should also note that the topology of cellular networks can be viewed at
various levels of complexity. For example, in the metabolic representation
considered here each metabolite that participates in an interconversion reaction
is considered equal \citep{jeong00}. However, in alternative representations
molecules that only act as donor or acceptor (e.g., ATP or ADP) or that do not
contribute a carbon or nitrogen atom to a metabolic reaction can be excluded
\citep{arita04}, or they can be pre-classified according to their perceived
biochemical role \citep{tanaka05b}. Similarly, in TR networks, genes and their
protein products are usually defined as common nodes with regulatory links among
them being mediated by the binding of TFs to the promoter regions of the genes
\citep{shen-orr02}. At other times, however, genes and their protein products are
considered as separate nodes \citep{yeger-lotem04}. While certain properties are
unaffected by alternative network representations, others are affected
\citep{arita04,tanaka05b} potentially complicating the interpretation of
subsequent analytical results.

Finally, our analyses also reveal highly similar, but dynamically regulated
global mRNA and protein expression profiles, in which expression values are
significantly variable. Thus, while the global function of cellular networks is
expected to be greatly influenced by their underlying topology
\citep{almaas04,ueda04,goh01}, mRNA or protein expression data reflect the dynamic
physicochemical state of the cell, likely necessitating an explicit dynamical
model for establishing a relationship between topology and expression.
\section*{Acknowledgments}
We acknowledge A. Clauset and C. R. Shalizi for making publicly available the implementation of the methods.
We thank R.V. Sol\'e, R. Albert, P. Delicado, P. Ru\'e, M. Dies and S. Laurie
for comments at different stages of the work. We also thank the constructive
comments from two anonymous reviewers. A.L.G. de L. acknowledges Generalitat de
Catalunya for FI and BE grants. G.D.F. acknowledges the Ram\'on y Cajal granting scheme. This
research was supported in part by EC funded FP6 STREP projects BioBridge (contract number
FP6-037909), QosCosGrid (contract number FP6-033883), and VPH NoE (contract
number FP7-223920).
\section*{Disclosure Statement}
No competing financial interests exist.

\newpage
%\bibliography{alopezLibrary}
%\bibliographystyle{plainnat}
%%%%%%%%%%%%%%%%%%%%%%%%%%%%%%%%%%%%%%%%%%%%%%%%% 
\begin{figure}[!htpb]
   \begin{center}
     \begin{tabular}{cc}
       \multicolumn{1}{l}{\mbox{\bf a.}} & \multicolumn{1}{l}{\mbox{\bf b.}} \\
       \rotatebox{270}{\includegraphics[scale=.25]{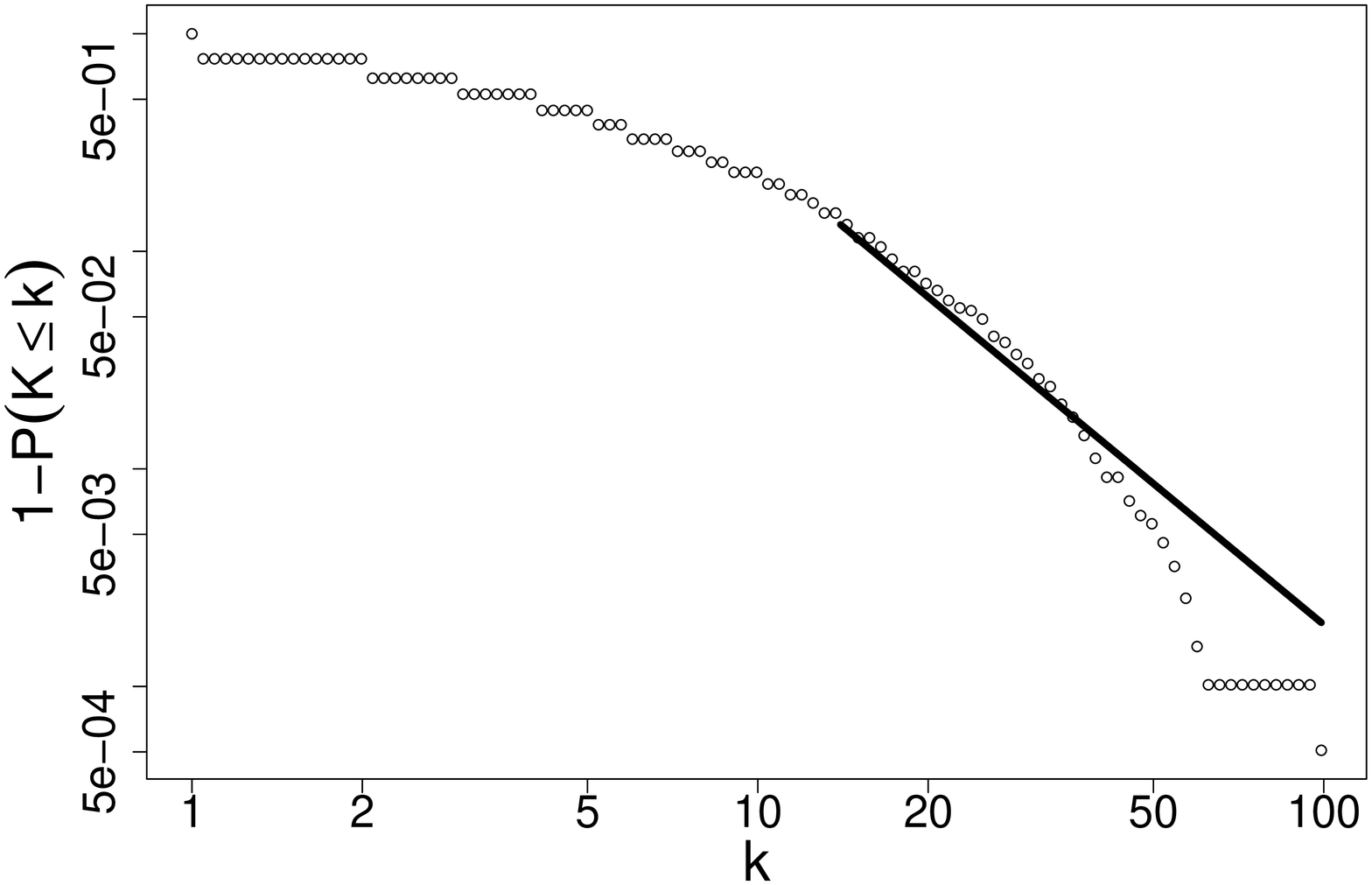}} & \rotatebox{270}{\includegraphics[scale=.25]{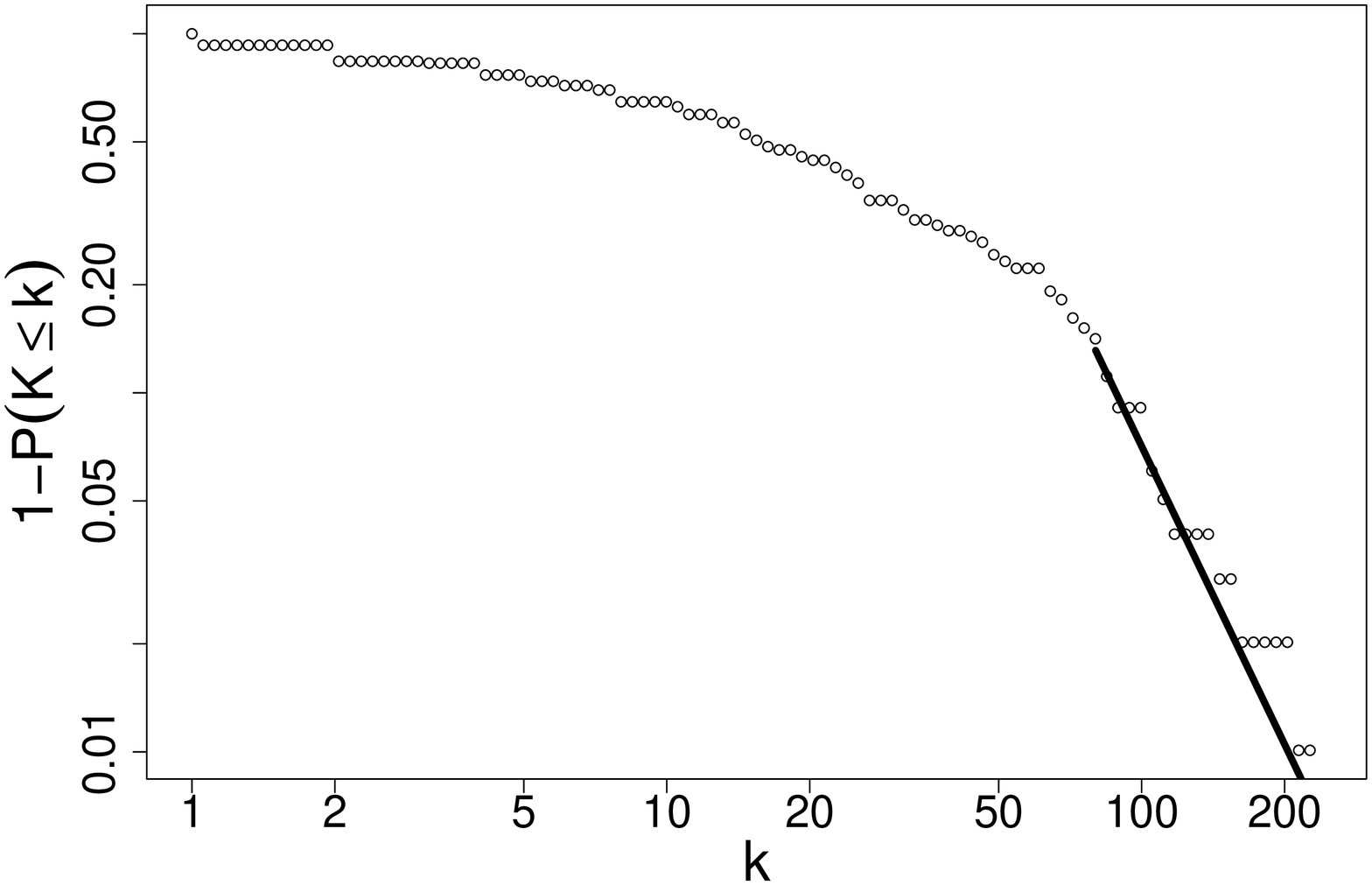}}\\
       \multicolumn{1}{l}{\mbox{\bf c.}} & \multicolumn{1}{l}{\mbox{\bf d.}} \\
       \rotatebox{270}{\includegraphics[scale=.25]{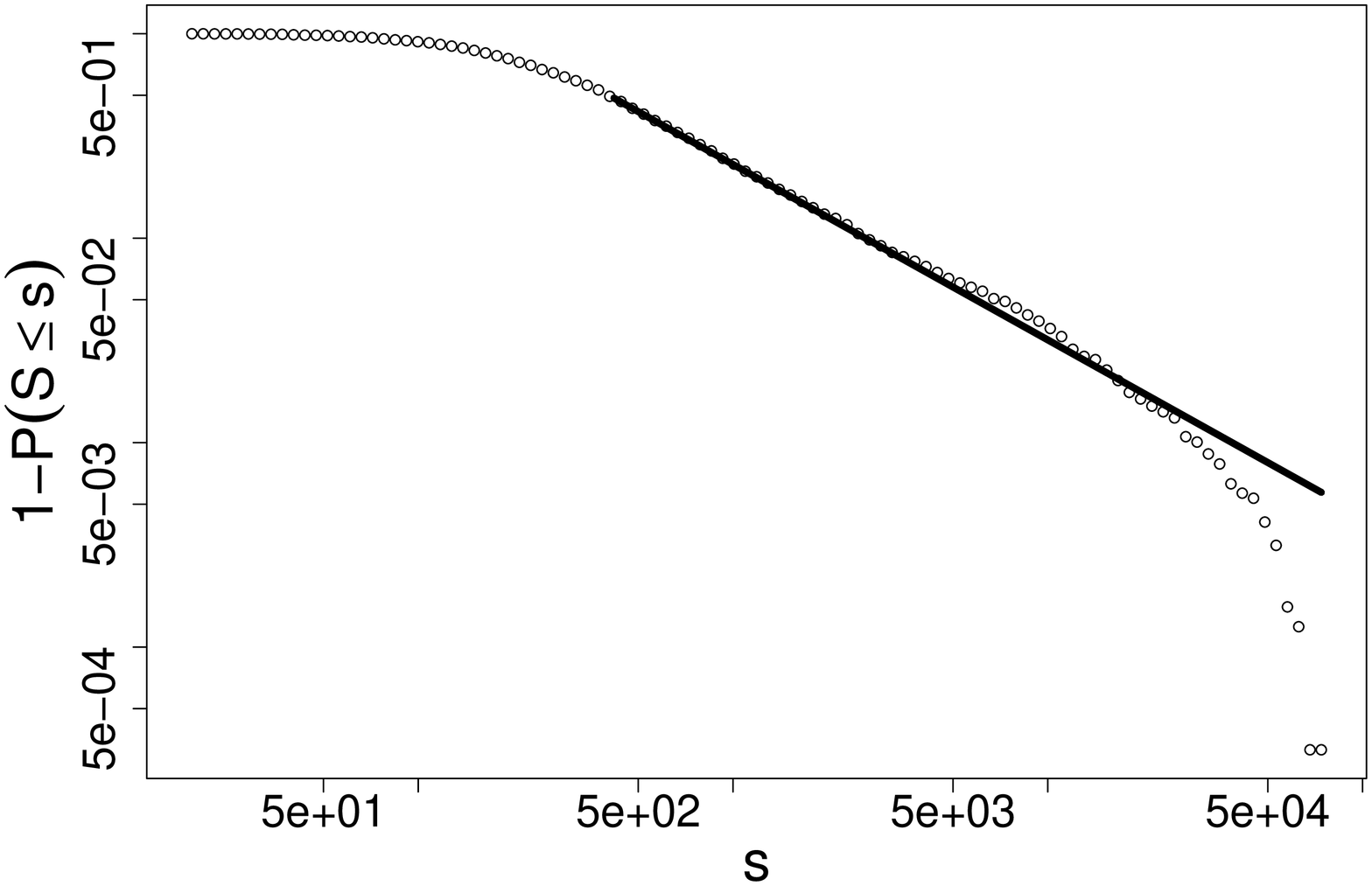}} & \rotatebox{270}{\includegraphics[scale=.25]{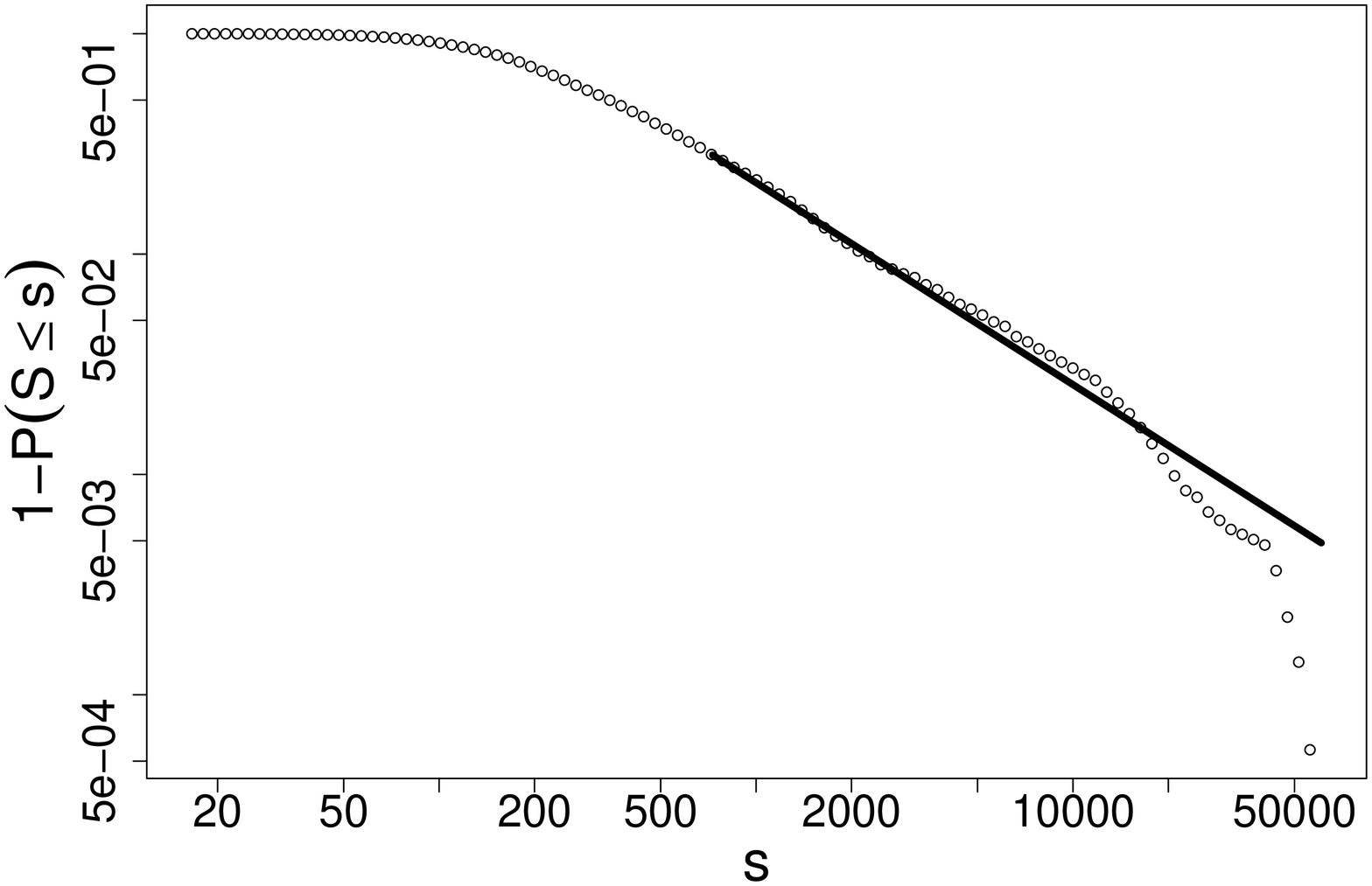}}\\
     \end{tabular}
     \caption{\footnotesize 
       The connectivity degree distribution of ({\bf a.}) LC derived \textit{S. cerevisiae} protein-protein interaction; ({\bf b.}) out-degree distribution of the transcriptional regulatory network of \textit{S. cerevisiae} and ({\bf c., d.}) intensity distribution of mRNA expression values in \textit{E. coli} cells from mid-log (OD ~0.2) cultures with ({\bf c.}) acetate and ({\bf d.}) galactose as single carbon source.
       Points represent the empirical distribution, lines represent the most likely power-law models.
       \label{distributions}}
   \end{center}
 \end{figure}
%%%%%%%%%%%%%%%%%%%%%%%%%%%%%%%%%%%%%%%%%%%%% 
%%%%%%%%%%%%%%%%%%%%%%%%%%%%%%%%%%%%%%%%%%%%% 
 \begin{landscape}
   \begin{table*}
     \begin{center}
       \begin{tabular}{ c | c | c  c | c  c | c  c | c  c | c  c | c  c | c}
         & Power-Law & \multicolumn{2}{|c|}{Log-Normal} & \multicolumn{2}{|c|}{Poisson} & \multicolumn{2}{|c|}{Yule} & \multicolumn{2}{|c|}{Exponential} & \multicolumn{2}{|c|}{Weibull} & \multicolumn{2}{|c|}{PL + Exp.} & \\
         data set & $p$ & NLLR & $p$ & NLLR & $p$ & NLLR & $p$ & NLLR & $p$ &  NLLR & $p$ & LLR & $p$ & diagnosis\\ \hline
         HTP & \textbf{0.76} & -0.67 & 0.50 & 3.62 & \textbf{0.00} & -1.00 & 0.32 & 2.71 & \textbf{0.01} & -0.25 & 0.80 & -0.74 & 0.22 & ambiguous\\
         LC & 0.01 & -2.14 & \textbf{0.03} & 4.30 & \textbf{0.00} & -4.29 & \textbf{0.00} & -0.52 & 0.60 & -2.16 & \textbf{0.03} & -6.86 & \textbf{0.00} & reject\\
         TR In & 0.00 & -4.49 & \textbf{0.00} & 2.08 & \textbf{0.04} & -7.98 & \textbf{0.00} & -2.86 & \textbf{0.00} & -4.64 & \textbf{0.00} & -26.40 & \textbf{0.00} & reject\\
         TR Out & \textbf{0.75} & -0.23 & 0.82 & 2.22 & \textbf{0.03} & -0.24 & 0.81 & 0.25 & 0.80 & -0.23 & 0.82 & -0.10 & 0.65 & ambiguous\\
         Met In & 0.04 & -1.95 & 0.05 & 2.24 & \textbf{0.02} & 2.94 & \textbf{0.00} & 4.00 & \textbf{0.00} & 5.66 & \textbf{0.00} & 0.00 & 1.00 & reject\\
         Met Out & 0.00 & -1.11 & 0.27 & 2.62 & \textbf{0.01} & 4.59 & \textbf{0.00} & 4.68 & \textbf{0.00} & 3.03 & \textbf{0.00} & 0.00 & 1.00 & reject\\
       \end{tabular}
     \end{center}
     \caption{\footnotesize
       Summary of probabilistic analyses for representative examples of the discrete data sets. 
       The Supplementary Information includes the further results discussed in the text.
       The second column shows the $p$-value associated with the differences between the data and the power-law model.
       The next six columns correspond to the log-likelihood ratio tests comparing the power-law model with other plausible models.
       Positive values support the power-law model.
       The normalized log-likelihood ratio (NLLR) is used for non-nested functions while the raw log-likelihood ratio (LLR) is used for the power-law with exponential cutoff model.
       $p$-values here are associated with the differences between the two models.
       Significant $p$-values are denoted in bold.
       \label{tableDiscrete}}
   \end{table*}
 \end{landscape}
%%%%%%%%%%%%%%%%%%%%%%%%%%%%%%%%%%%%%%%%%%%%% 
%%%%%%%%%%%%%%%%%%%%%%%%%%%%%%%%%%%%%%%%%%%%% 
\begin{table}[!htbp]
  \begin{center}
    {\begin{tabular}{ c | c | c  c | c  c | c  c | c  c | c}
        & Power-Law & \multicolumn{2}{|c|}{Log-Normal} & \multicolumn{2}{|c|}{Exponential} & \multicolumn{2}{|c|}{Weibull} & \multicolumn{2}{|c|}{PL + Exp.} & \\
        data set & $p$ & NLLR & $p$ & NLLR & $p$ &  NLLR & $p$ & LLR & $p$ & diagnosis\\ \hline
        Acetate & \textbf{0.28} & -1.86 & 0.06 & 14.49 & \textbf{0.00} & -1.89 & 0.06 & -10.61 & \textbf{0.00} & ambiguous\\ 
        Galactose & 0.00 & -2.19 & \textbf{0.02} & 10.84 & \textbf{0.00} & -2.26 & \textbf{0.02} & -10.91 & \textbf{0.00} & reject\\ 
        Glucose & 0.05 & -2.40 & \textbf{0.01} & 12.16 & \textbf{0.00} & -2.48 & \textbf{0.01} & -13.31 & \textbf{0.00} & reject\\ 
        Glycerol & 0.04 & -2.68 & \textbf{0.01} & 11.38 & \textbf{0.00} & -2.77 & \textbf{0.00} & -14.58 & \textbf{0.00} & reject\\ 
        Maltose & \textbf{0.21} & -2.26 & \textbf{0.02} & 12.90 & \textbf{0.00} & -2.32 & \textbf{0.02} & -12.33 & \textbf{0.00} & ambiguous\\
      \end{tabular}}
  \end{center}  
  \caption{\footnotesize
    Summary of probabilistic analyses for representative examples of the continuous data sets. 
    The Supplementary Information includes the other results discussed in the text.
    Interpretation of the table as for Table \ref{tableDiscrete}.
    \label{tableMidlog}
  }
\end{table}
\end{document}

\begin{center}
  \thispagestyle{empty}
  \underline{\textbf{\Large{Supplementary Information}}}\\[5ex]
  \textbf{\Large{Global Connectivity and Activity Distributions in Cellular Networks}}\\[5ex]
  \textbf{Adri\'an L\'opez Garc\'ia de Lomana$^{\rm 1}$, Qasim K. Beg$^{\rm 2,\#}$, G. de Fabritiis$^{\rm 1}$ and Jordi Vill\`a-Freixa$^{\rm 1, *}$}\\[10ex]
  $^{\rm 1}$Computational Biochemistry and Biophysics Laboratory, Research Unit on Biomedical Informatics, IMIM / Universitat Pompeu Fabra, c/ Dr. Aiguader 88, 08003, Barcelona, Spain\\[2ex]$^{\rm 2}$Department of Pathology, 3550 Terrace St., University of Pittsburgh, Pittsburgh, PA 15261, USA\\[2ex]
  $^{\rm \#}$Current Address: Department of Biomedical Engineering, 44 Cummington St., Boston University, Boston, MA 02215, USA\\[2ex]
$^{\rm *}$To whom correspondence should be addressed: J. V.-F., jordi.villa@upf.edu\\[10ex]
\end{center}
\newpage
\section{Clustering Coefficient Analysis}
\begin{figure}
  \begin{center}
    \begin{tabular}{cc}
      \multicolumn{1}{l}{\mbox{\bf a.}} & \multicolumn{1}{l}{\mbox{\bf b.}} \\
      \includegraphics[scale=.3]{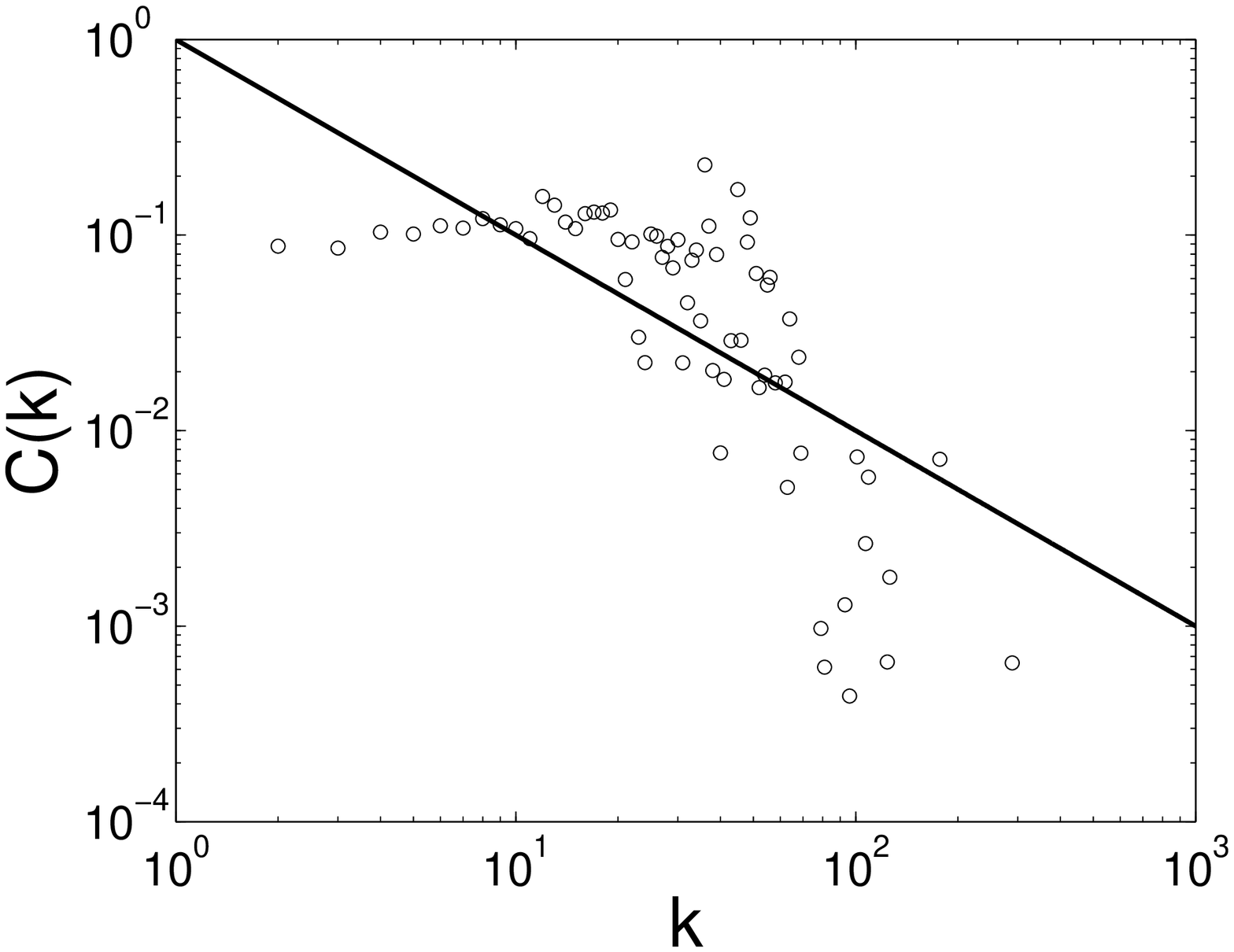}&\includegraphics[scale=.3]{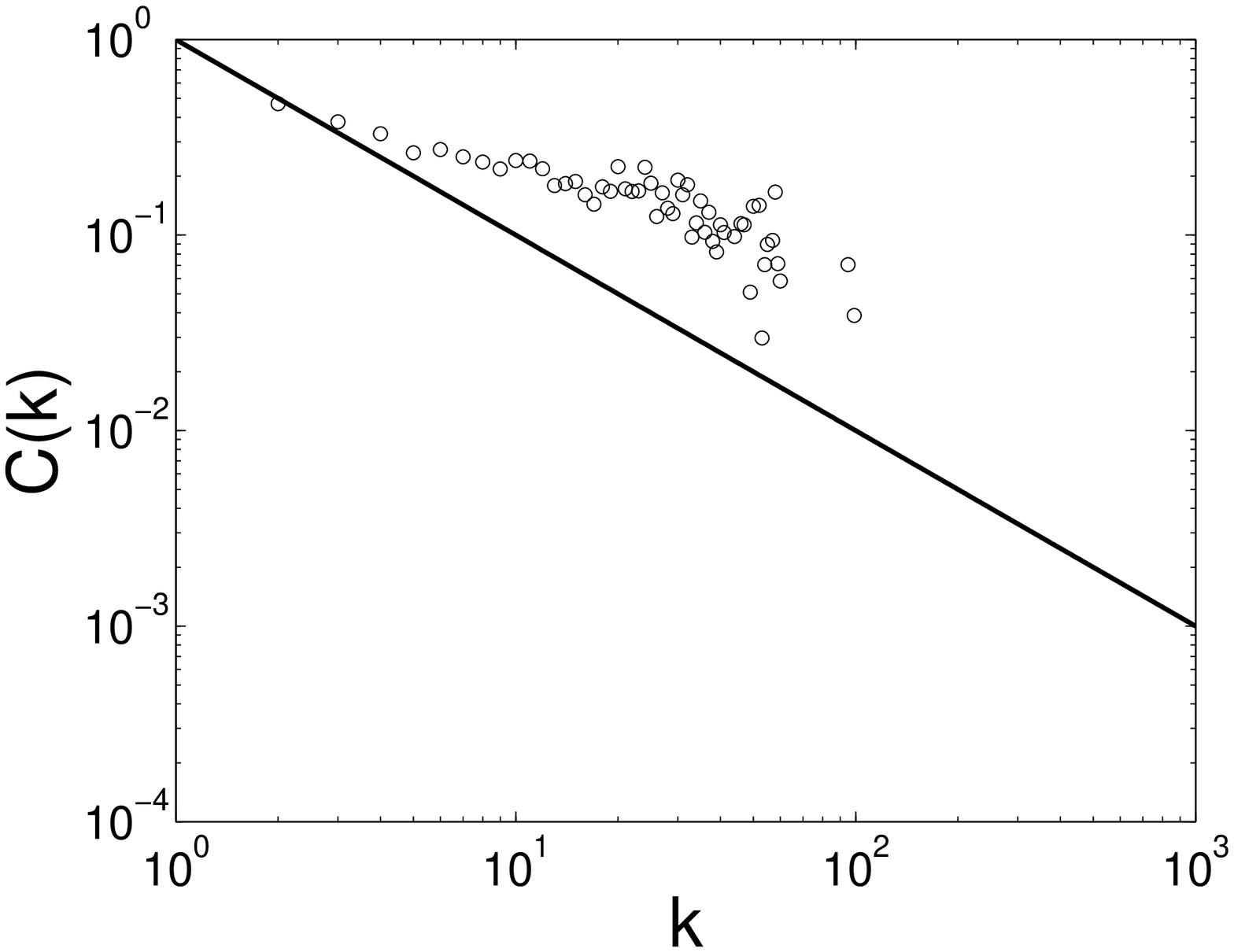}\\
      \multicolumn{1}{l}{\mbox{\bf c.}} & \multicolumn{1}{l}{\mbox{\bf d.}} \\
      \includegraphics[scale=.3]{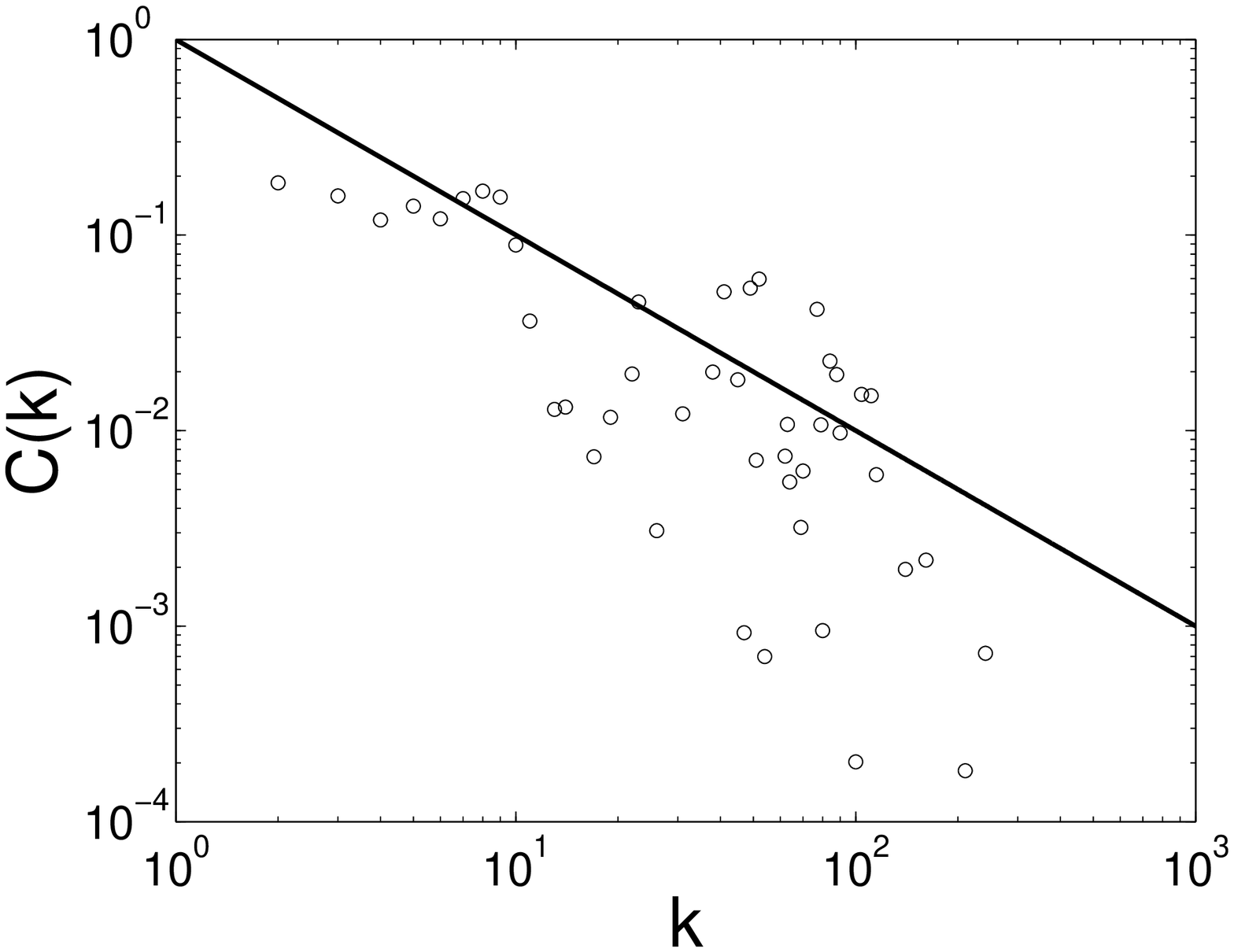}&\includegraphics[scale=.3]{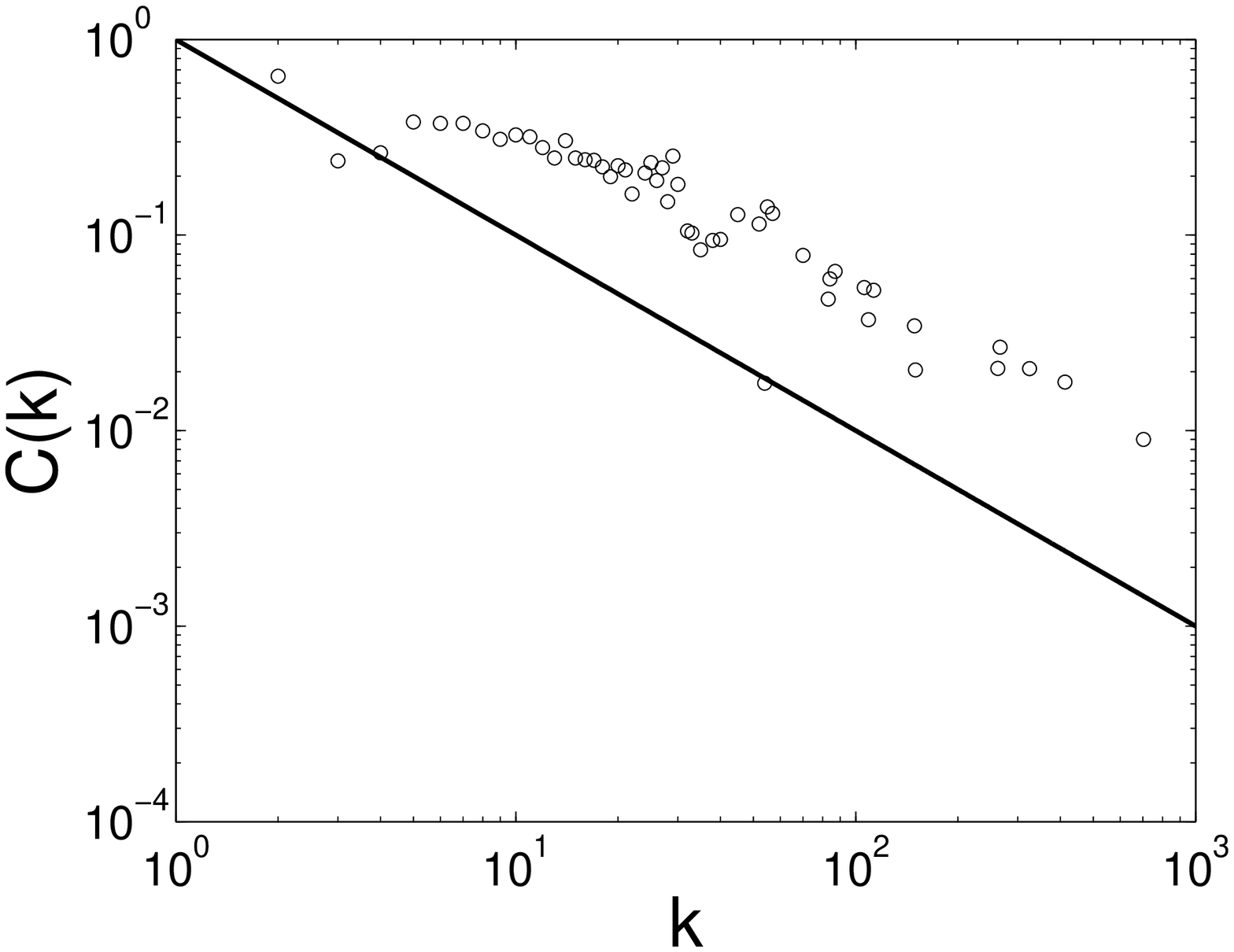}\\
    \end{tabular}
    \caption{\footnotesize 
      {\bf a., b.} Clustering coefficients ($C(k)$) for the degree distribution of nodes in the \textit{S. cerevisiae} protein-protein interaction network built from HTP and LC data respectively; {\bf c.} \textit{S. cerevisiae} transcriptional regulatory network, and {\bf d.} the \textit{E. coli} metabolic network. 
      Circles represent empirical distributions, while lines represent $C(k)=k^{-1}$. 
      \label{clustering}}
  \end{center}
\end{figure}

The clustering coefficient is defined as $C(i)=\frac{2e_i}{k_i(k_i-1)}$, $i$ denoting a given node, $k_i$ the number of links of $i$, and $e_i$ the number of connections among the links. 
The clustering coefficient ranges from 0 (low connected network) to 1 (highly connected network). 
In relation to the connectivity distribution, $k$, scale-free and modular networks have been reported to display constant $C(k)$ distributions \citep{ravasz02}, while hierarchical networks display $C(k)=k^{-1}$ distributions \citep{ravasz02}. 
In our analyses, we do not find a conclusive agreement between any of the models and the experimental data (Figure \ref{clustering}).
\section{Analysis Protocol}
In order to give a broad idea of the methods we used, we explain here the mathematical protocol we followed, which is the same as the one described in \cite{clauset07}.
The protocol was newly developed by them and we used their own implementation in \texttt{MATLAB} \citep{clauset07}. 
We have applied it to the biological data sets of interest.

The main idea of the protocol is to answer two very well defined questions:
\it
\begin{itemize}
\item Given the power-law model, which is the most probable parameter combination for a given data set?
\item Once we have an inferred model, is there any significative difference between the model and the data set?
\end{itemize}
\rm
For the first question we use maximum likelihood methods to find the most probable parameters from a given model in order to explain the data set.
In the case of the power-law model, we have two parameters, $\alpha$, which is the scaling parameter and $x_{min}$ which is the lowest $x$ from which the distribution behaves as power-law.
The parameter $x_{min}$ is inferred in a special way as it represents one of the bounds of the distribution but does not affect the shape of it. 
The method used, described and implemented by \cite{clauset07}, consists in a very practical approach:
$x_{min}$ is chosen so the difference between the inferred model and the data set is minimized.
The rationale behind it is that if $x_{min}$ is underestimated, the scaling parameter will be wrongly inferred while if $x_{min}$ is overestimated the reduced data size will provoke random fluctuations on the estimation.

Once we have in hands a data set and its corresponding inferred model, we want to know if there are differences between the data and the model in order to know whether or not the data set follows the power-law distribution.
In this context of non-normality an appropriate and common test used to compare a reference distribution and an empirical distribution is the Kolmogorov-Smirnov (KS) test.
However one of the premises of the KS is that the two distributions need to be independent.
In our case the model is inferred from the same data set we want to compare it with and the premise of independency is not hold as correlations are introduced.
Therefore Monte Carlo methods are used to generate synthetic data from the inferred model.
A $D$ statistic of the KS test is calculated for each synthetic data set.
At this point we have a distribution of $D$ values.
The position of the $D$ statistic calculated from the empirical data set and the inferred model, within the distribution of $D$ values calculated from the Monte Carlo generated data sets, gives us the $p$-value associated to the empirical $D$.
Specifically the way of calculating the $p$-value is simply counting the number of times the empirical $D$ is larger than the synthetic $D$.
As we are looking for deviations in one side (larger than) the confidence value that should be used is $p=0.1$.
We have elaborated a simplistic graphical flow of the protocol to facilitate the comprehension of the methods used (Figure \ref{figureProtocol}).

Finally, we have an additional question:
\it
\begin{itemize}
\item Is there any other model that fits better the data than the power-law distribution?
\end{itemize}
\rm
Once we know to which extent is the power-law model able to explain the empirical data distribution, we wonder if there is any other model that could equally well describe the experimental data.
For that purpose, the normalized-log-likelihood ratio test is used.
It is a simple test that compare the log-likelihood values of two competing distributions with respect to the empirical data. 
For the case of nested functions, which is the case of the broad-scale, the $p$-value should be corrected.

All code and data sets necessary to reproduce the results are provided upon request as a tar file.
\newpage
%\bibliography{library}
%\bibliographystyle{plainnat}

\newpage
\begin{figure}
  \begin{center}
    \begin{tabular}{cc}
      \multicolumn{1}{l}{\mbox{\bf a.}} & \multicolumn{1}{l}{\mbox{\bf b.}} \\
      \rotatebox{270}{\includegraphics[scale=.25]{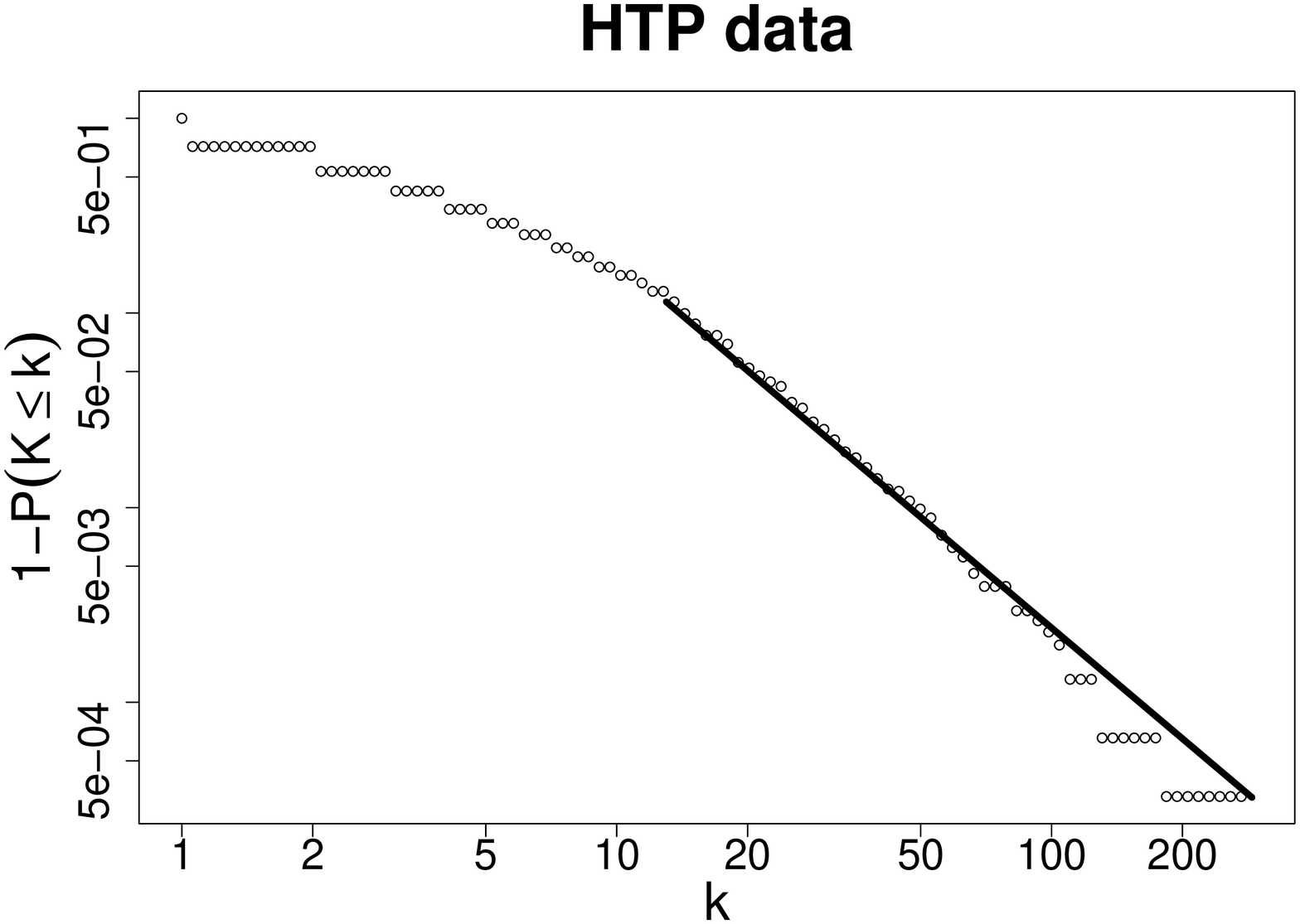}} & \rotatebox{270}{\includegraphics[scale=.25]{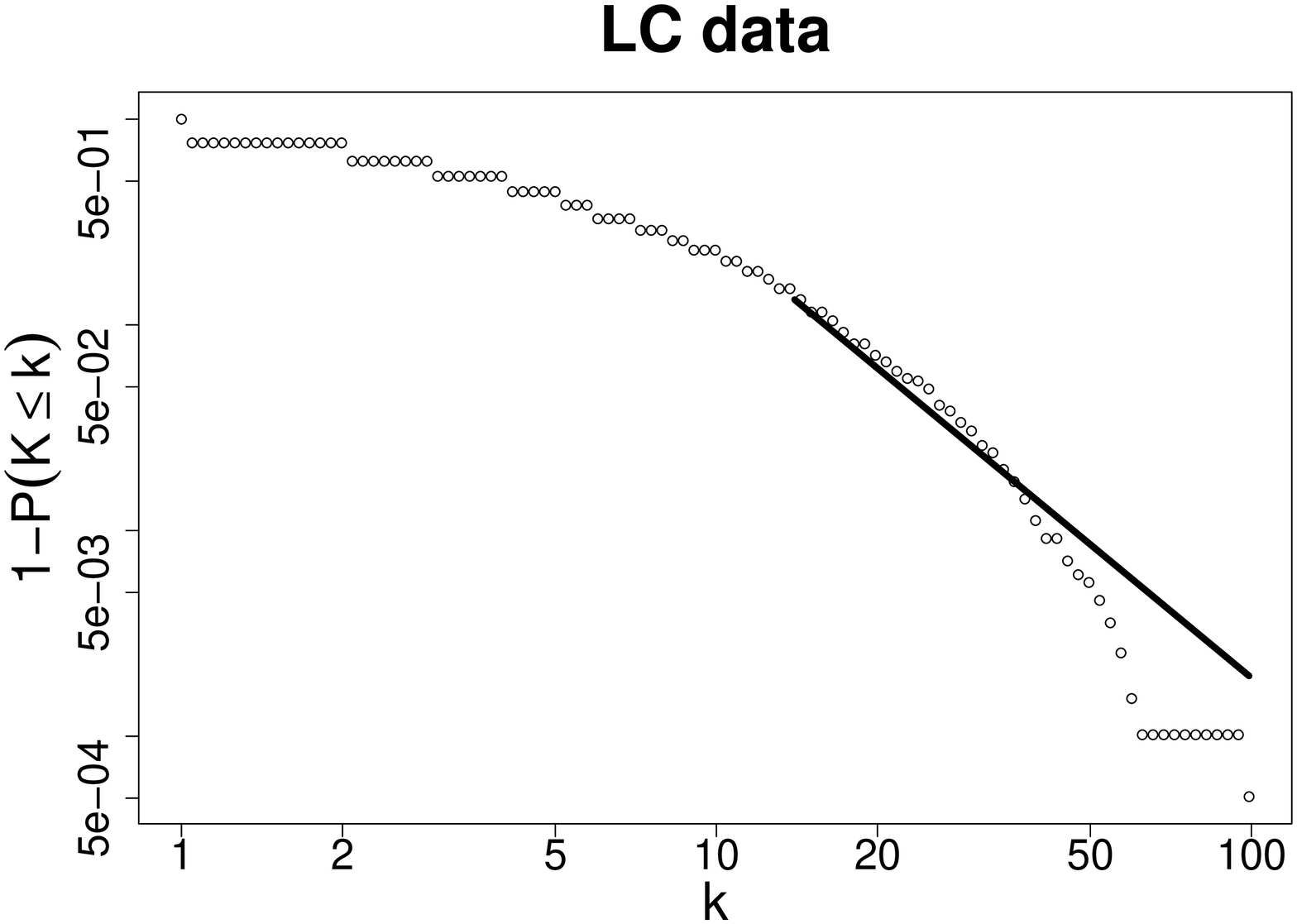}}\\
      \multicolumn{1}{l}{\mbox{\bf c.}} & \multicolumn{1}{l}{\mbox{\bf d.}} \\
      \rotatebox{270}{\includegraphics[scale=.25]{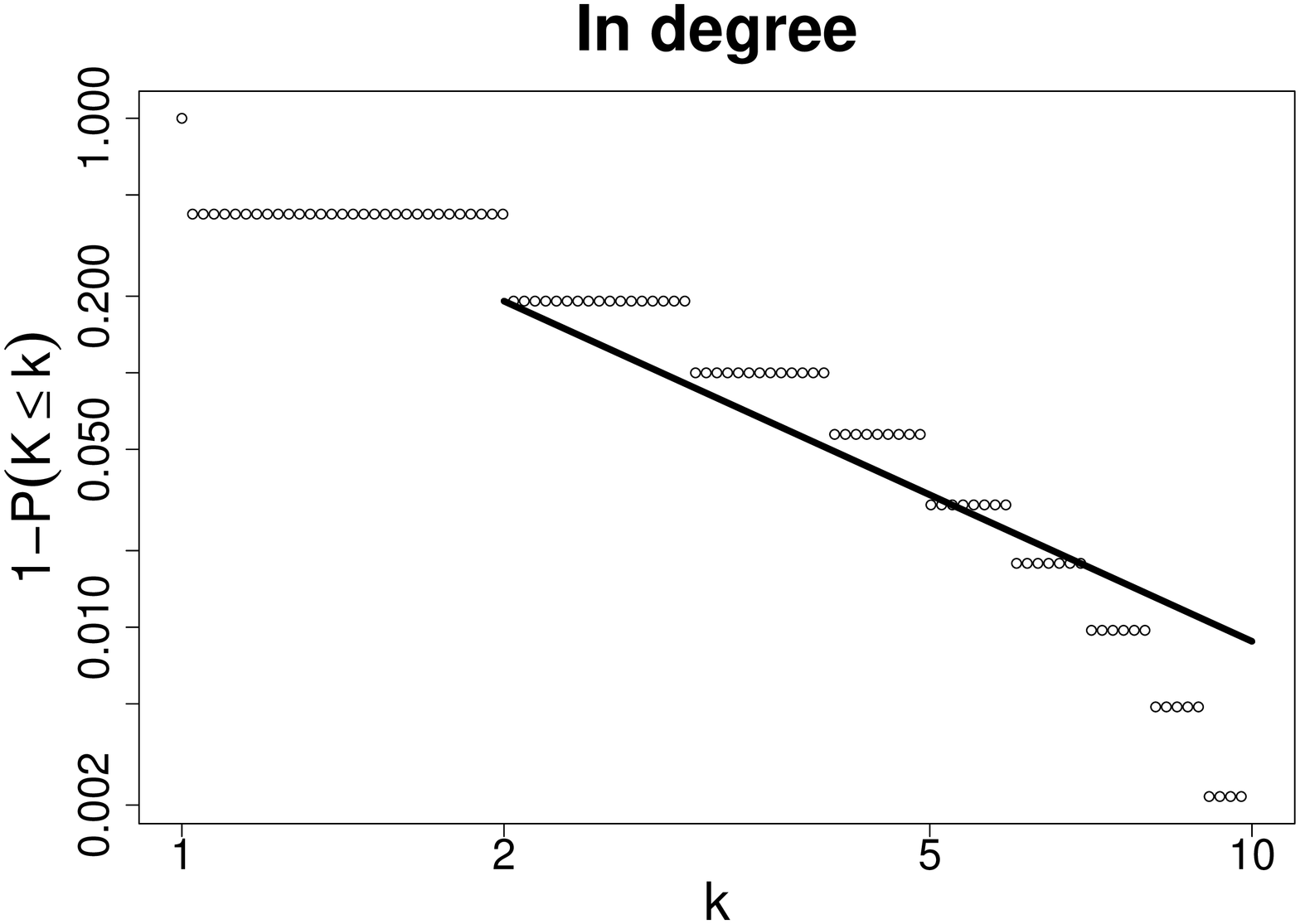}} & \rotatebox{270}{\includegraphics[scale=.25]{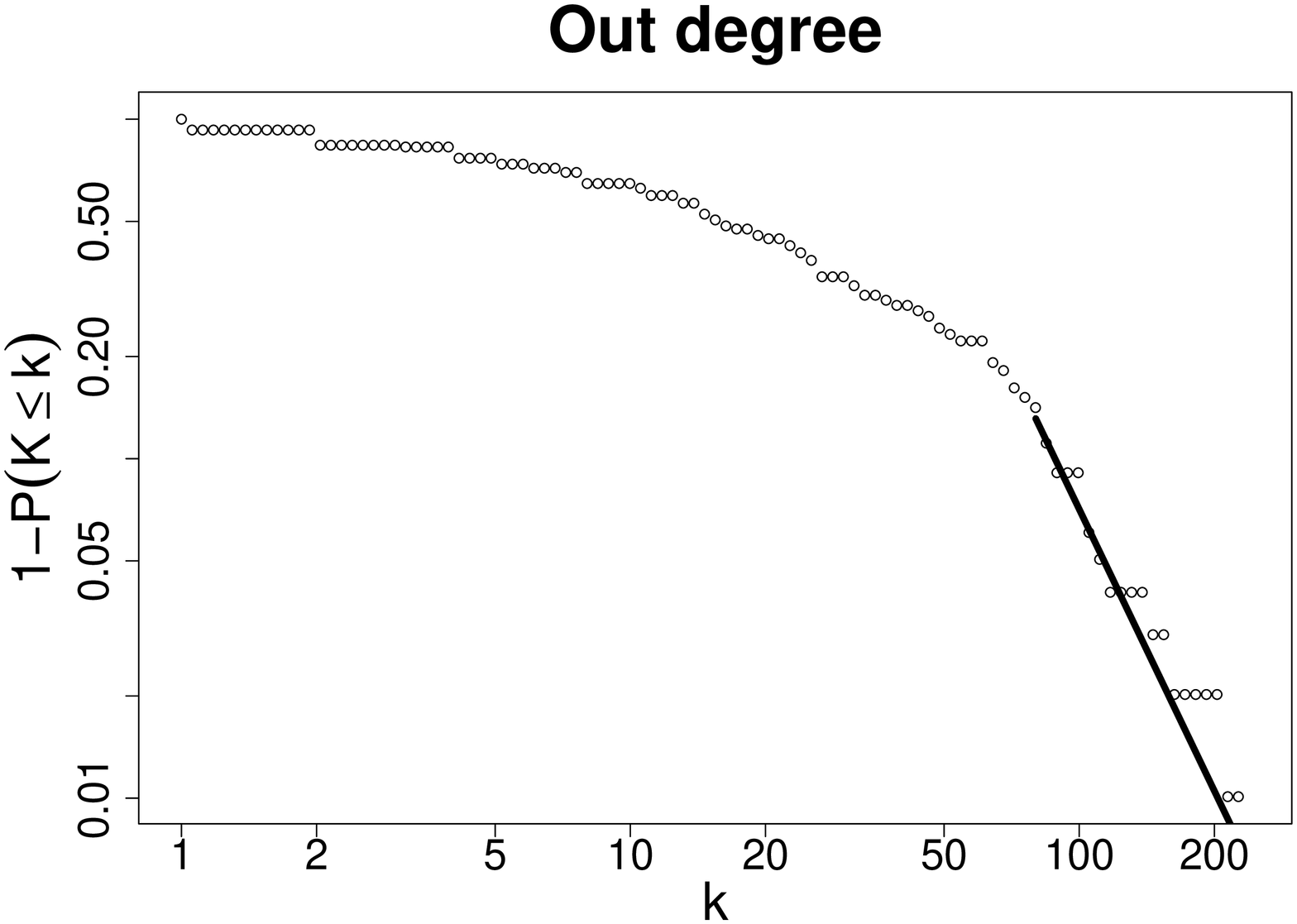}}\\
      \multicolumn{1}{l}{\mbox{\bf e.}} & \multicolumn{1}{l}{\mbox{\bf f.}}\\
      \rotatebox{270}{\includegraphics[scale=.25]{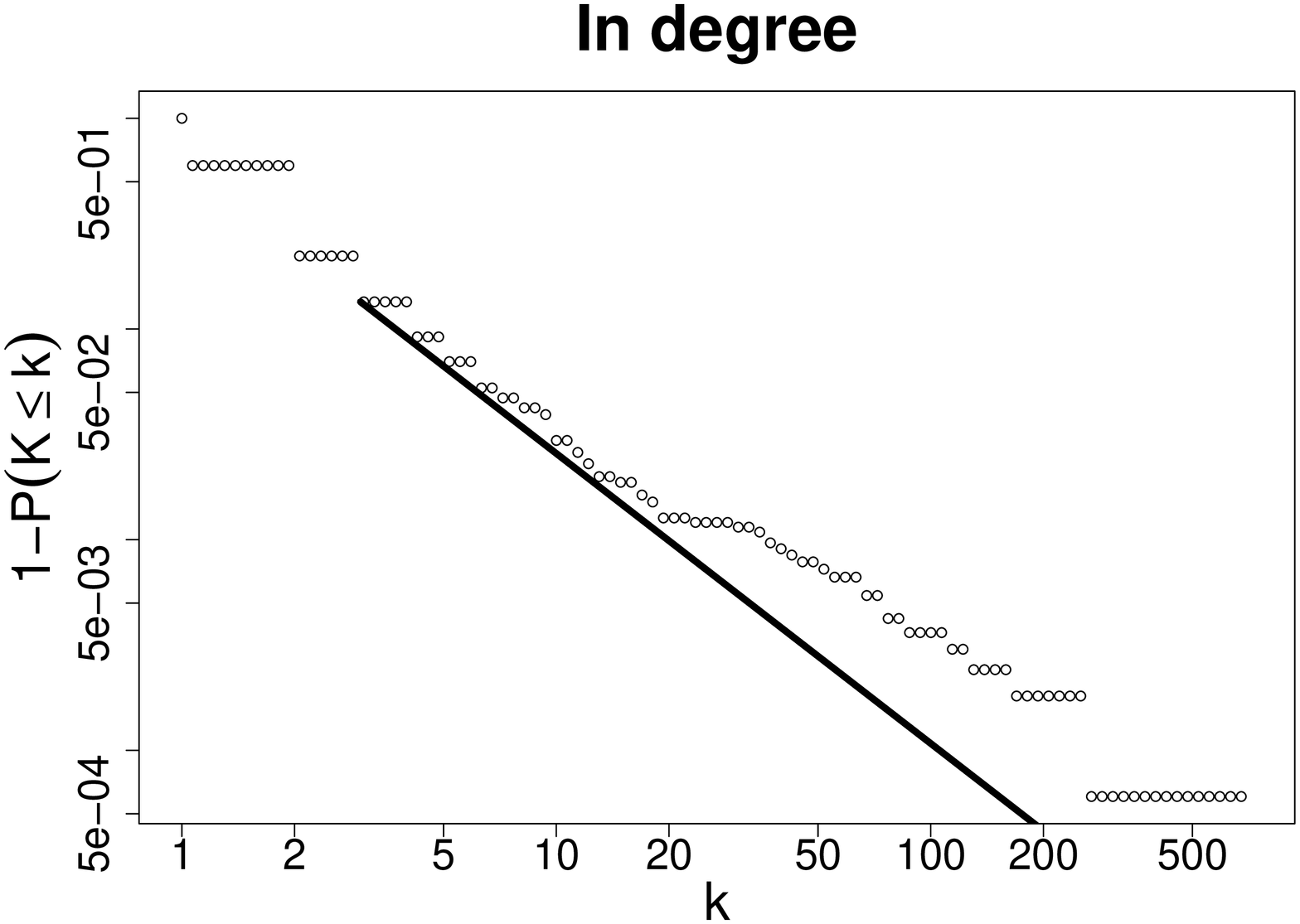}} & \rotatebox{270}{\includegraphics[scale=.25]{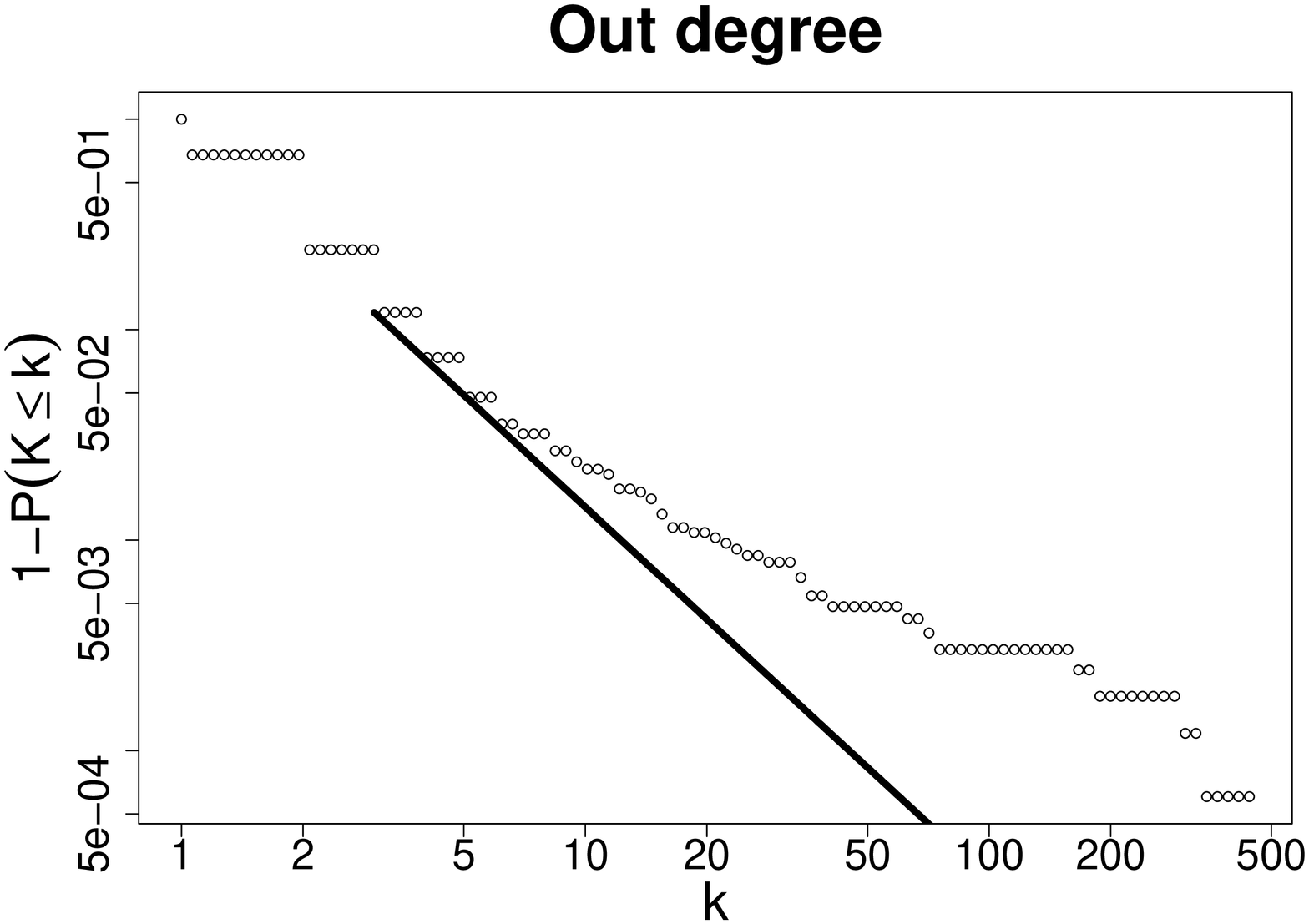}}\\
    \end{tabular}
    \caption{\footnotesize 
      {\bf a., b.} The degree distribution of connectivity of \textit{S. cerevisiae} protein-protein interaction network; {\bf c., d.} transcriptional regulatory network and {\bf e., f.} the \textit{E. coli} metabolic network.
      Points represent the empirical distribution, lines represent the most likely models.
      \label{distributions}}
  \end{center}
\end{figure}
%%%%%%%%%%%%%%%%%%%%%%%%%%%%%%%%%%%%%%%%%%%%%%%%%%%
%%%%%%%%%%%%%%%%%%%%%%%%%%%%%%%%%%%%%%%%%%%%%%%%%%%
\begin{figure}
  \begin{center}
    \begin{tabular}{cc}
      \multicolumn{1}{l}{\mbox{\bf a.}} & \multicolumn{1}{l}{\mbox{\bf b.}} \\
      \rotatebox{270}{\includegraphics[scale=.25]{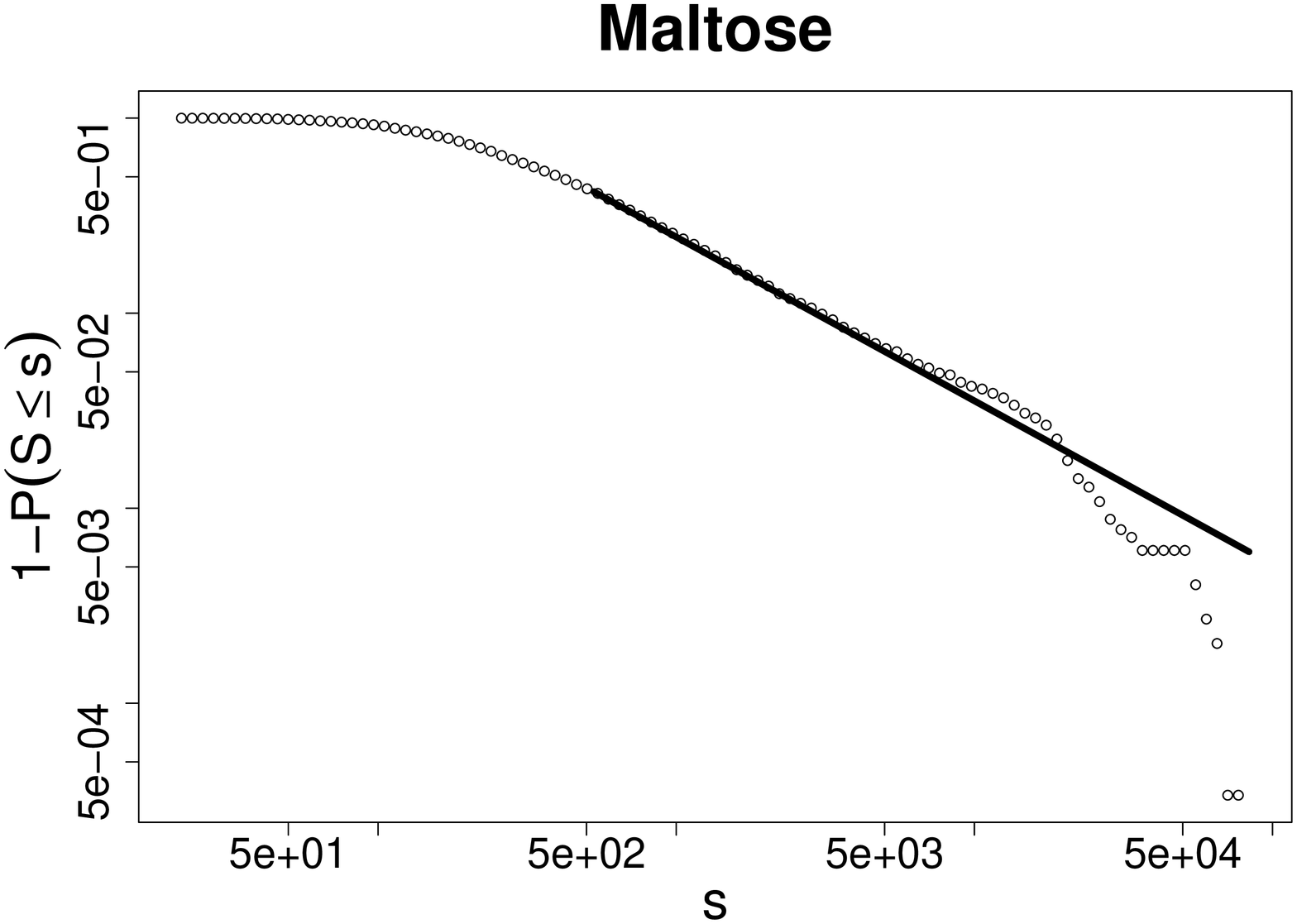}} & \rotatebox{270}{\includegraphics[scale=.25]{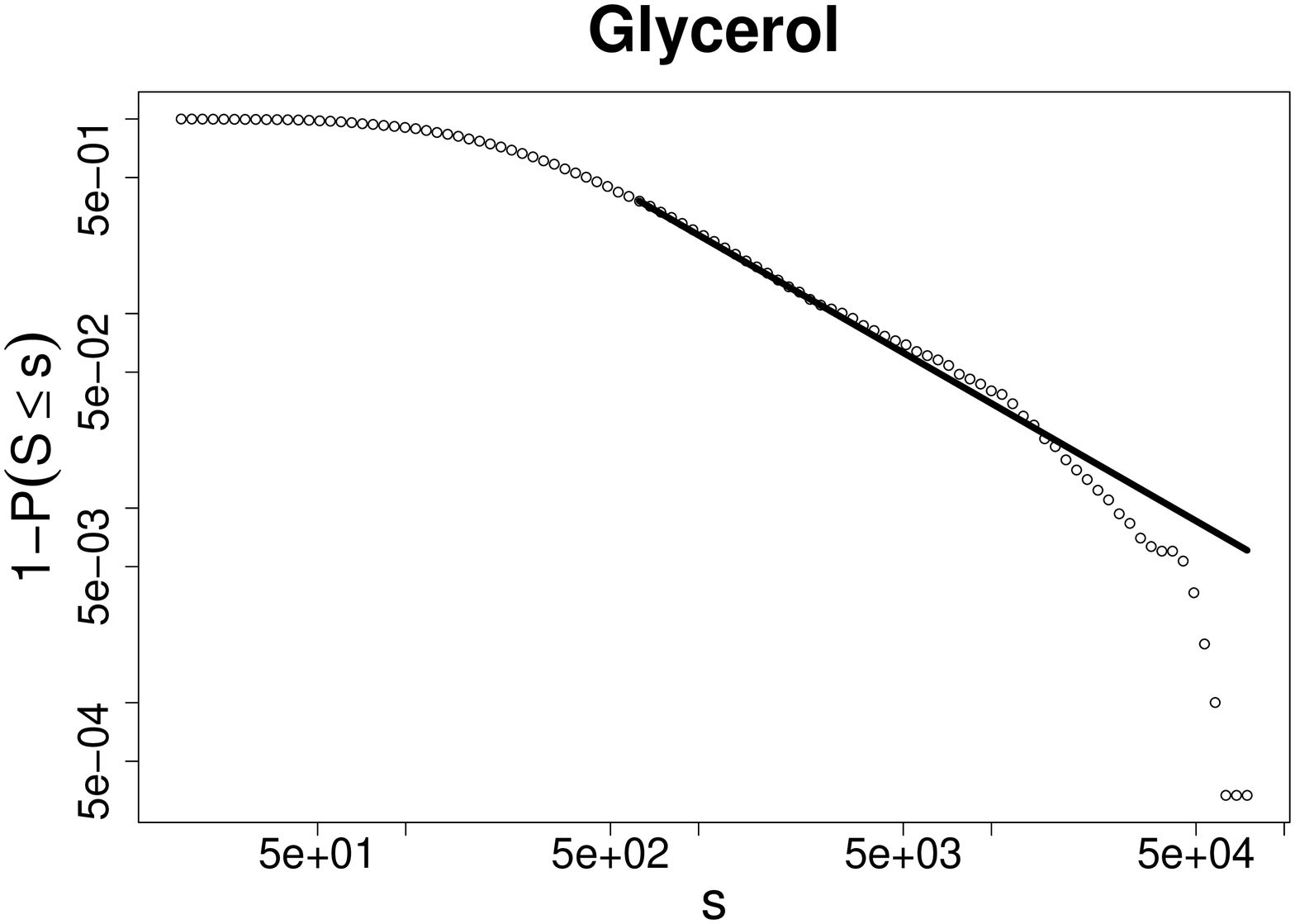}}\\
      \multicolumn{1}{l}{\mbox{\bf c.}} & \multicolumn{1}{l}{\mbox{\bf d.}} \\
      \rotatebox{270}{\includegraphics[scale=.25]{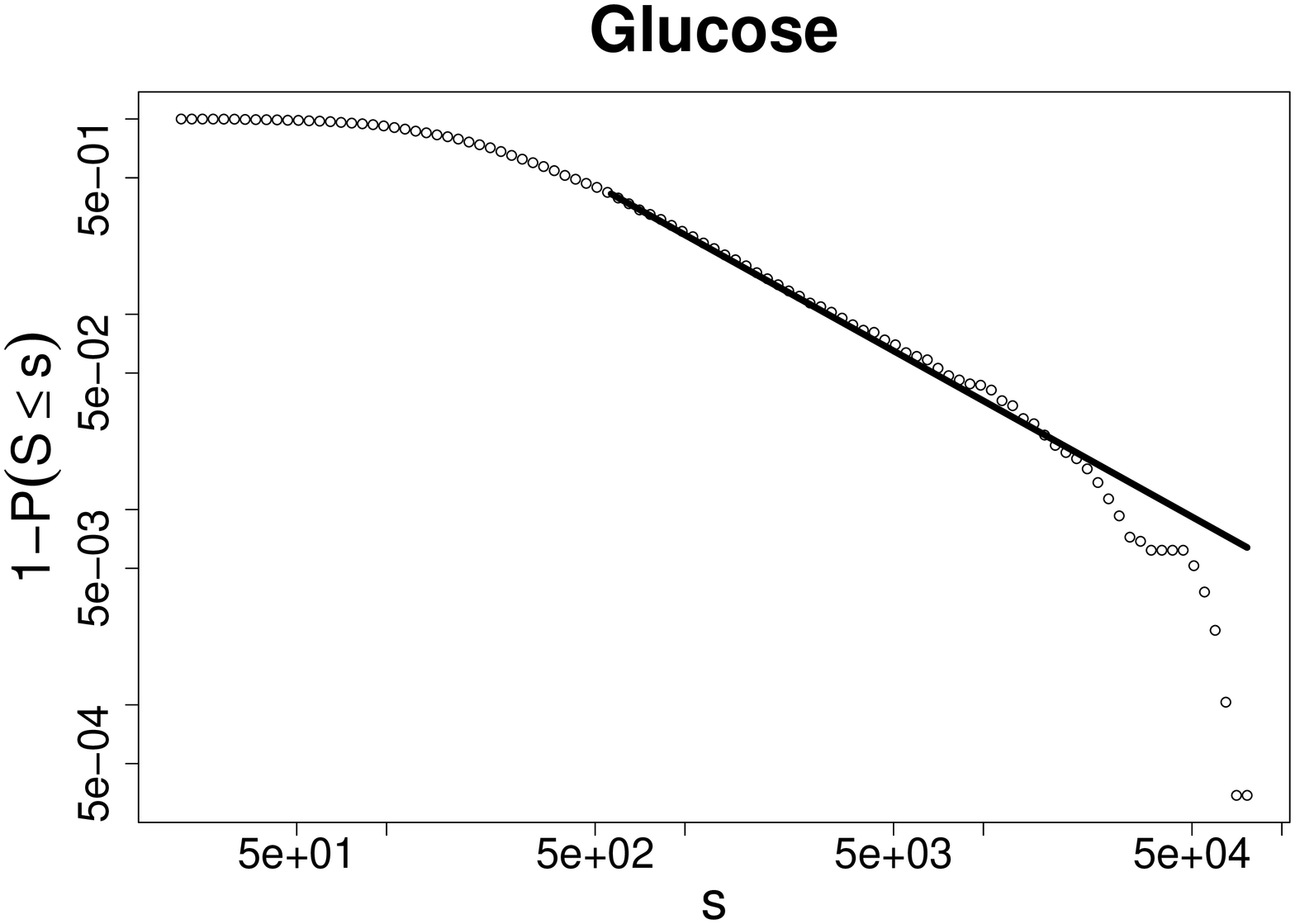}} & \rotatebox{270}{\includegraphics[scale=.25]{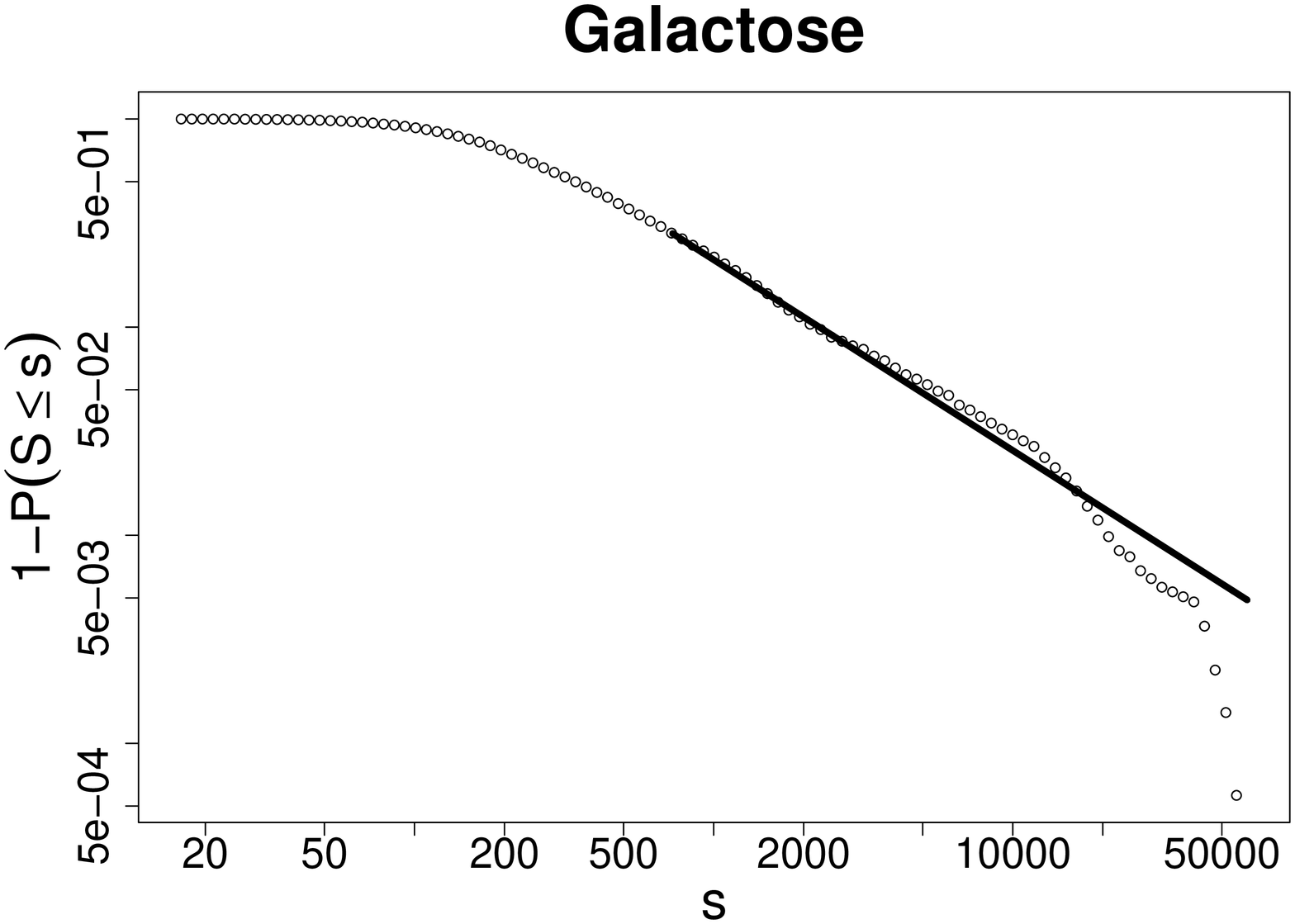}}\\
      \multicolumn{1}{l}{\mbox{\bf e.}}&\\
      \rotatebox{270}{\includegraphics[scale=.25]{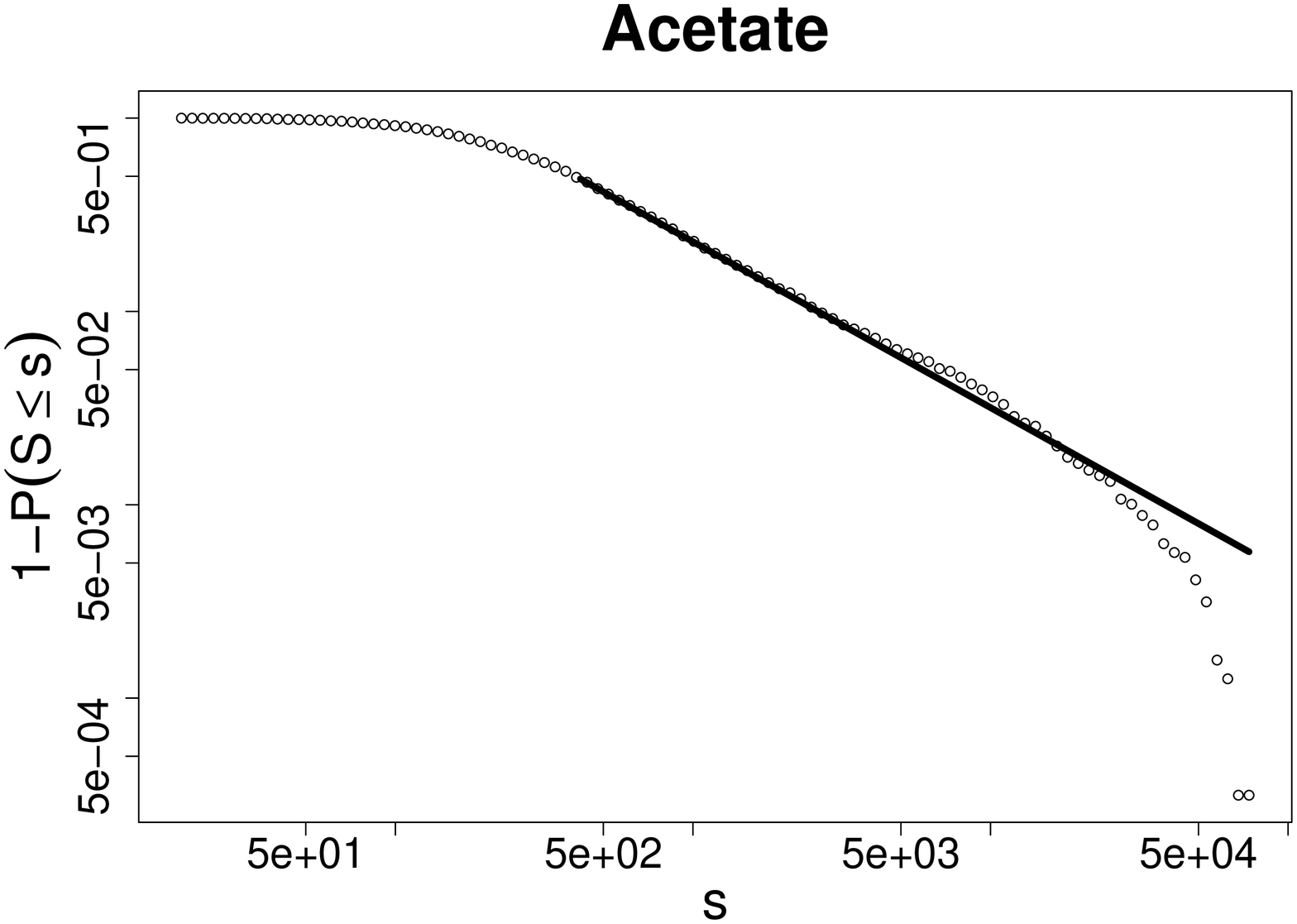}} & \\
    \end{tabular}
    \caption{\footnotesize 
      Intensity distribution of mRNA expression values in \textit{E. coli} cells from mid-log (OD ~0.2) cultures with different molecules as single carbon source: maltose, glycerol, glucose, galactose and acetate.
      Lines represent the best fitted power-law model.
      \label{midlog}}
  \end{center}
\end{figure}
%%%%%%%%%%%%%%%%%%%%%%%%%%%%%%%%%%%%%%%%%%%%%%%%%%%
\begin{figure}
  \begin{center}
    \begin{tabular}{cc}
      \multicolumn{1}{l}{\mbox{\bf a.}} & \multicolumn{1}{l}{\mbox{\bf b.}} \\
      \rotatebox{270}{\includegraphics[scale=.25]{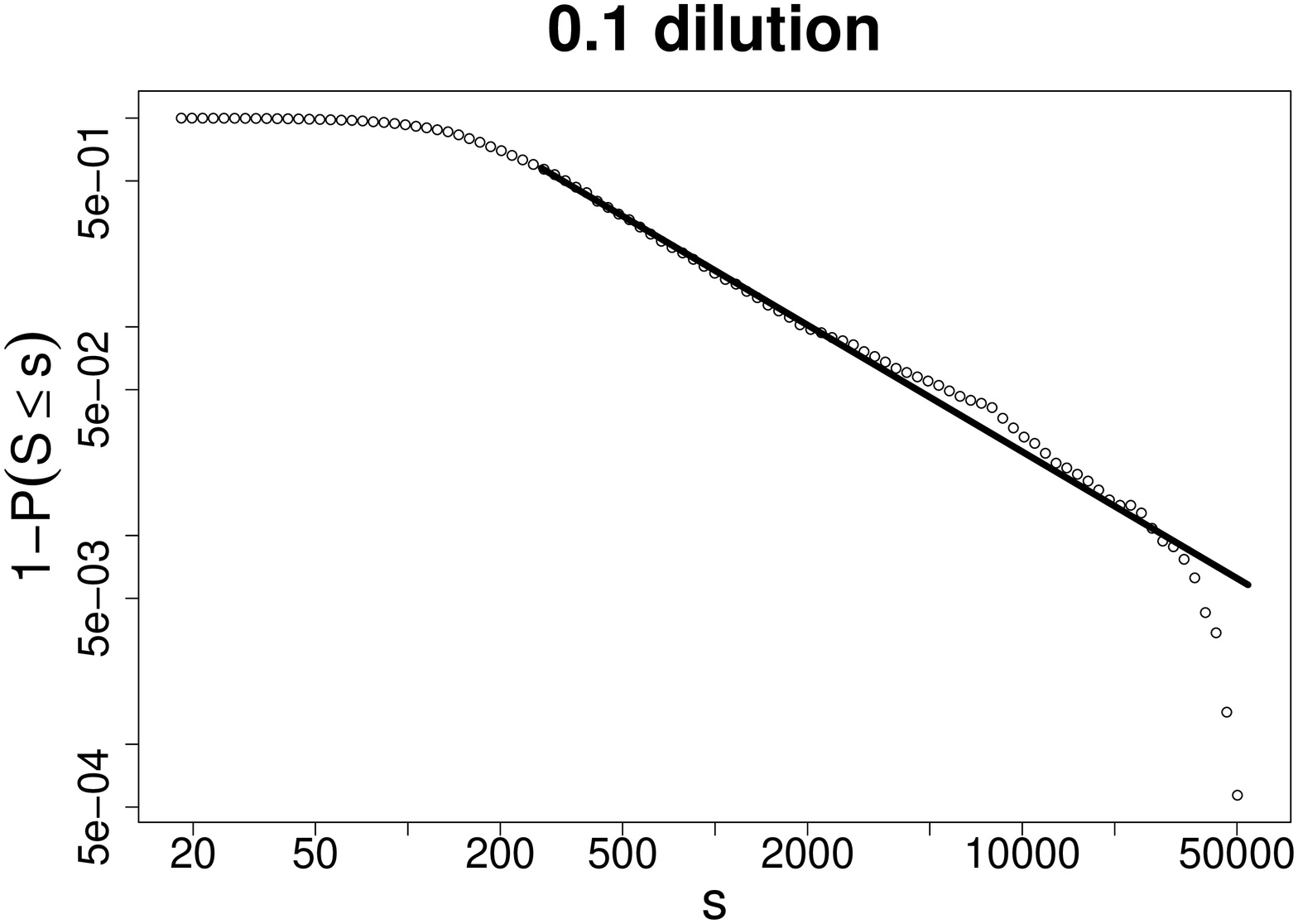}} & \rotatebox{270}{\includegraphics[scale=.25]{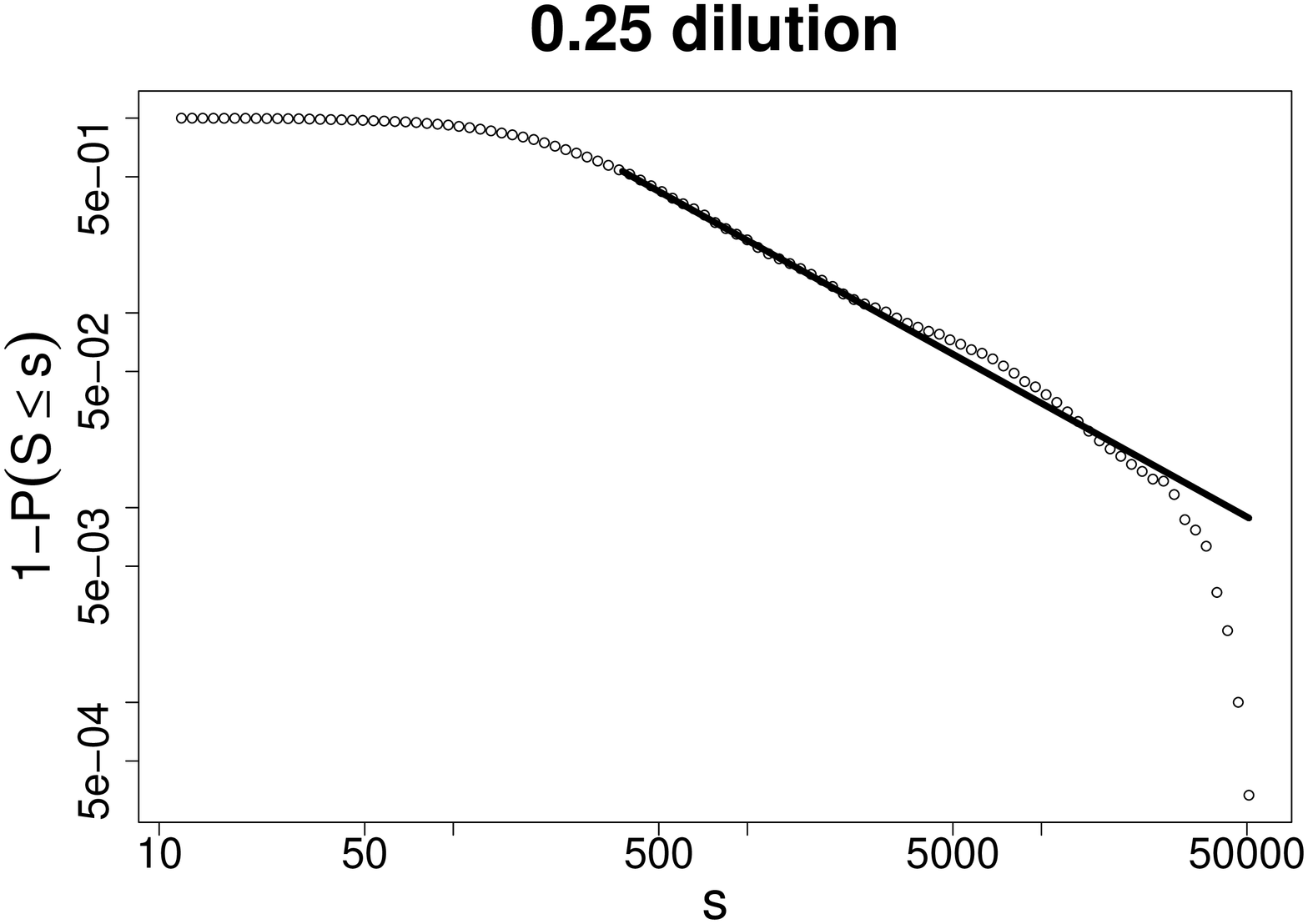}}\\
      \multicolumn{1}{l}{\mbox{\bf c.}} & \multicolumn{1}{l}{\mbox{\bf d.}} \\
      \rotatebox{270}{\includegraphics[scale=.25]{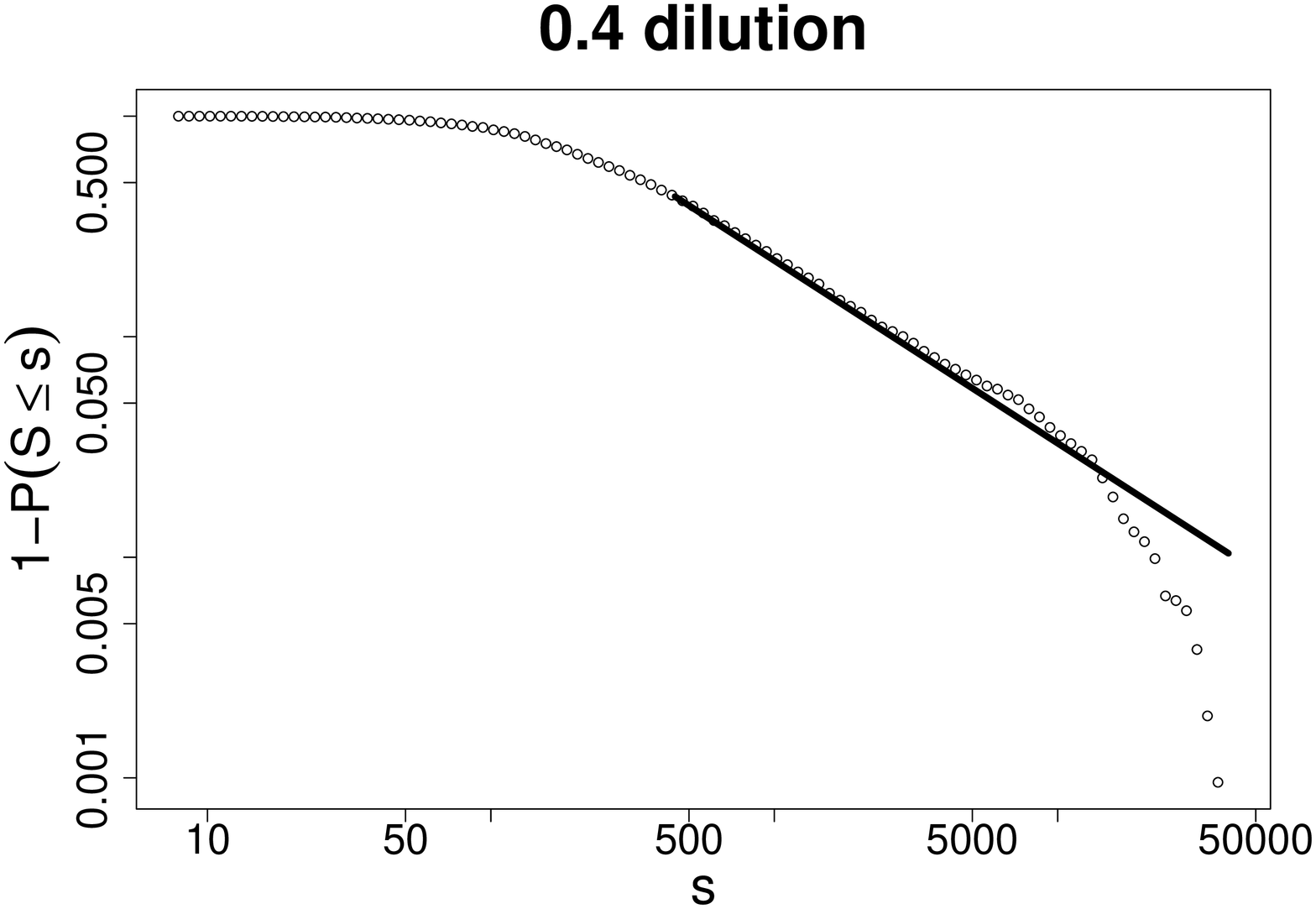}} & \rotatebox{270}{\includegraphics[scale=.25]{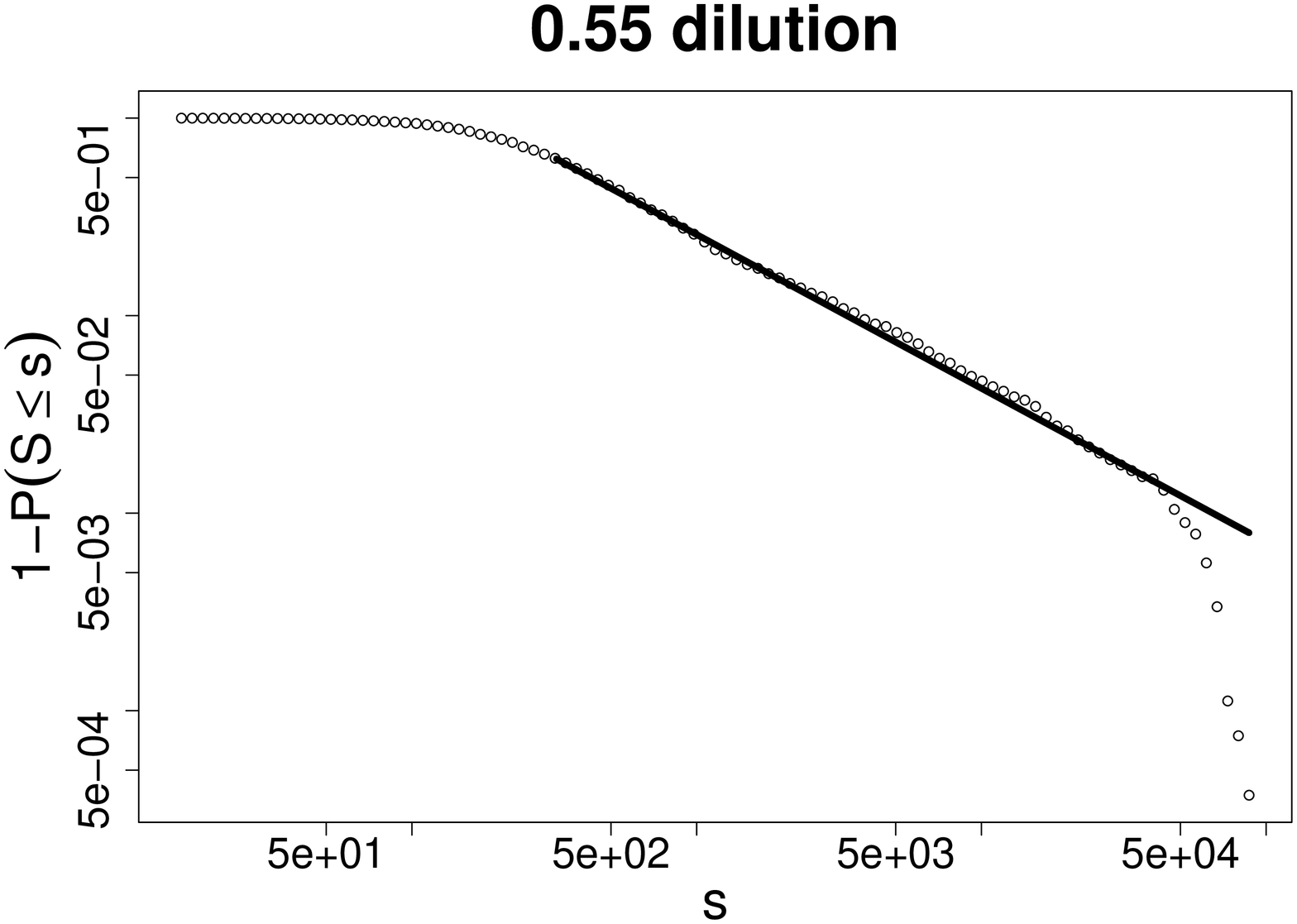}}\\
      \multicolumn{1}{l}{\mbox{\bf e.}}&\\
      \rotatebox{270}{\includegraphics[scale=.25]{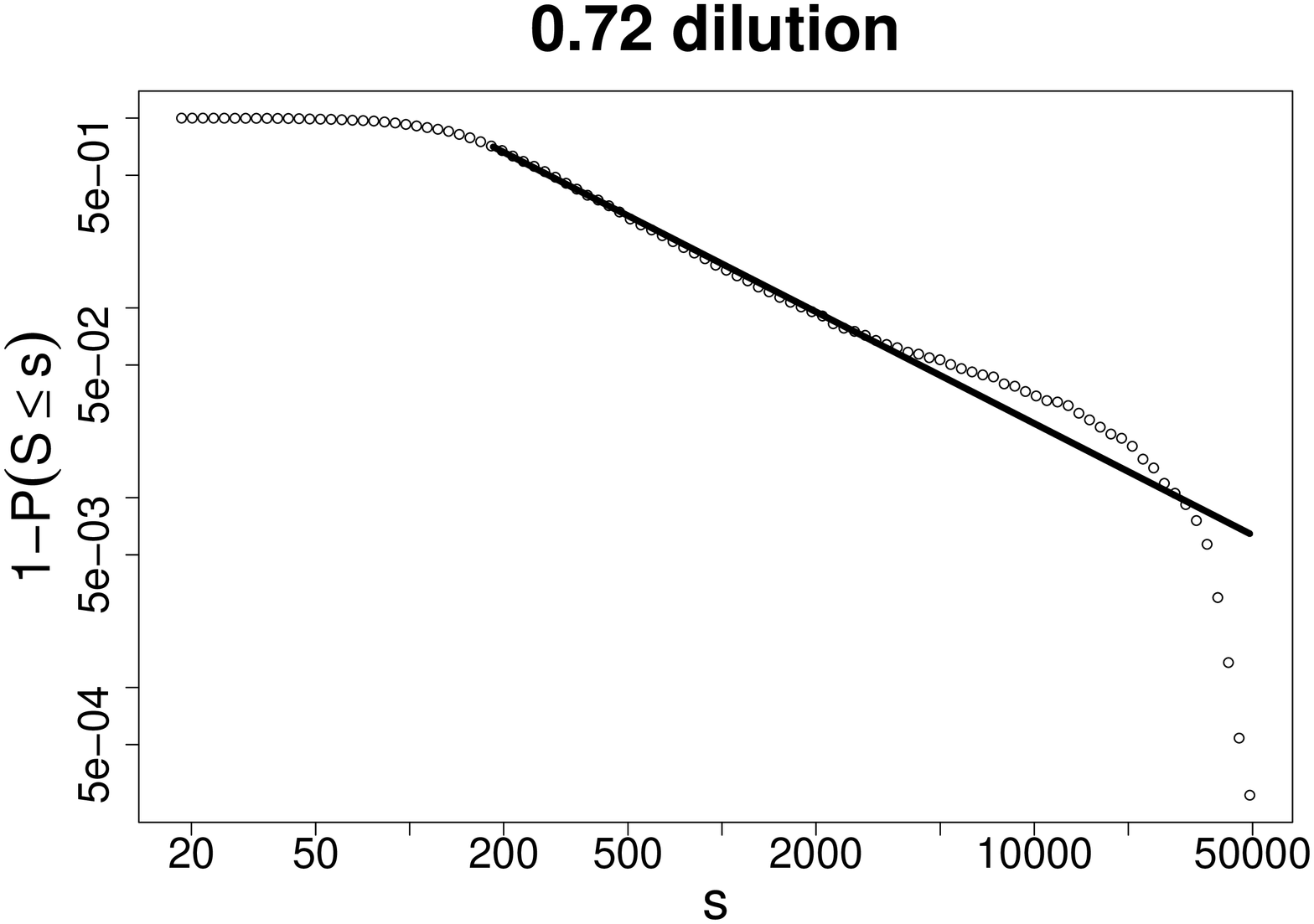}} & \\
    \end{tabular}
    \caption{\footnotesize 
      Intensity distribution of mRNA expression values in \textit{E. coli} cells from steady-state cultures with 0.2\% glucose-restricted M9 minimal medium at different dilution rates: 0.1, 0.25, 0.4, 0.55 and 0.72.
      Lines represent the best fitted power-law model.
      \label{steadyState}}
  \end{center}
\end{figure}
%%%%%%%%%%%%%%%%%%%%%%%%%%%%%%%%%%%%%%%%%%%%%%%%%%%
\begin{figure}
  \begin{center}
    \rotatebox{270}{\includegraphics[scale=.25]{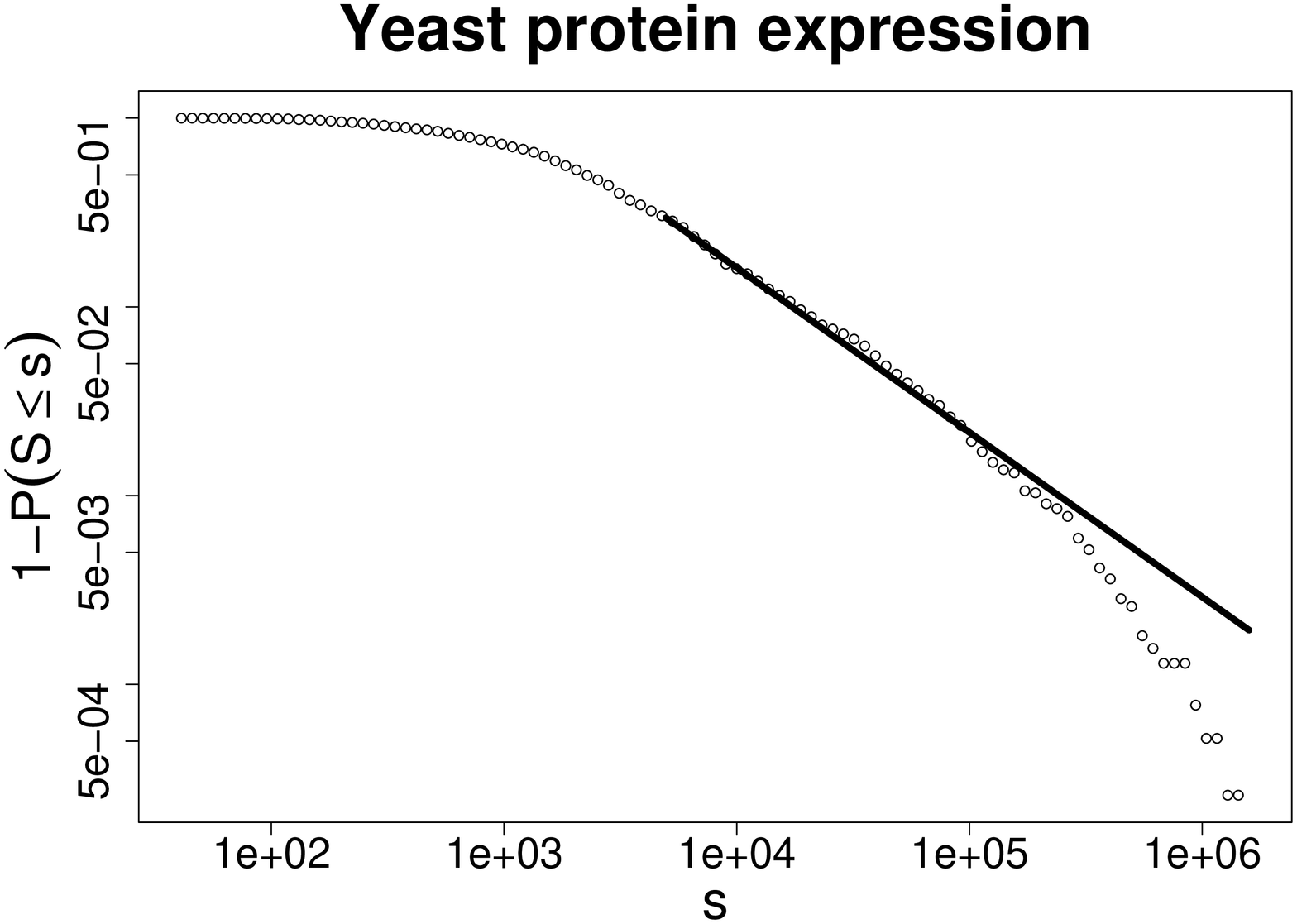}} 
    \caption{\footnotesize 
      Protein expression values distribution in \textit{S. cerevisiae} clones expressing individual GFP or TAP tagged proteins.
      Line represents the best fitted power-law model.
      \label{yeast}}
  \end{center}
\end{figure}
%%%%%%%%%%%%%%%%%%%%%%%%%%%%%%%%%%%%%%%%%%%%%%%%%%%
\begin{landscape}
  \begin{figure}
    \begin{center}
      \begin{tabular}{cccc}
        \rotatebox{270}{\includegraphics[scale=.2]{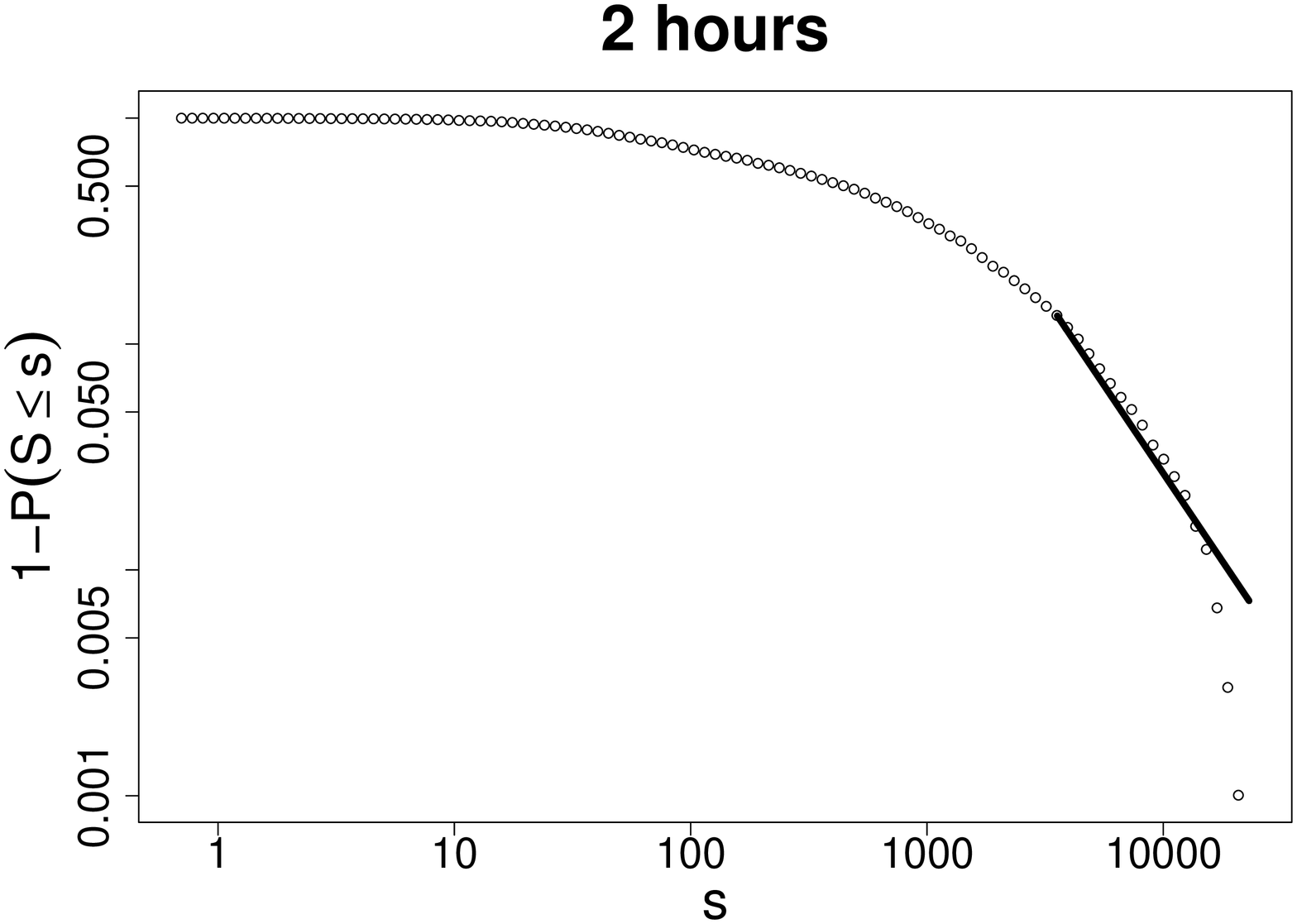}} & \rotatebox{270}{\includegraphics[scale=.2]{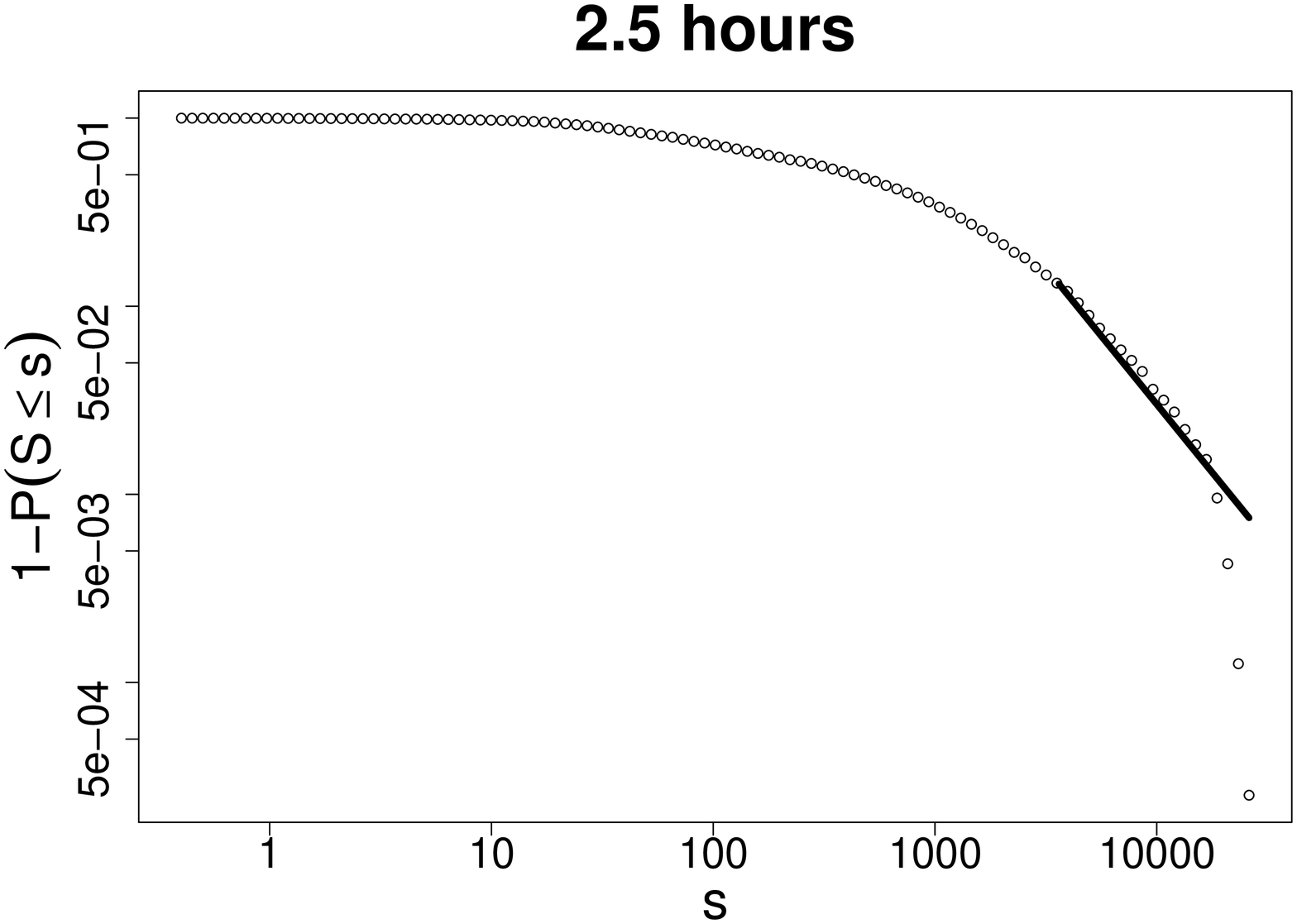}} & \rotatebox{270}{\includegraphics[scale=.2]{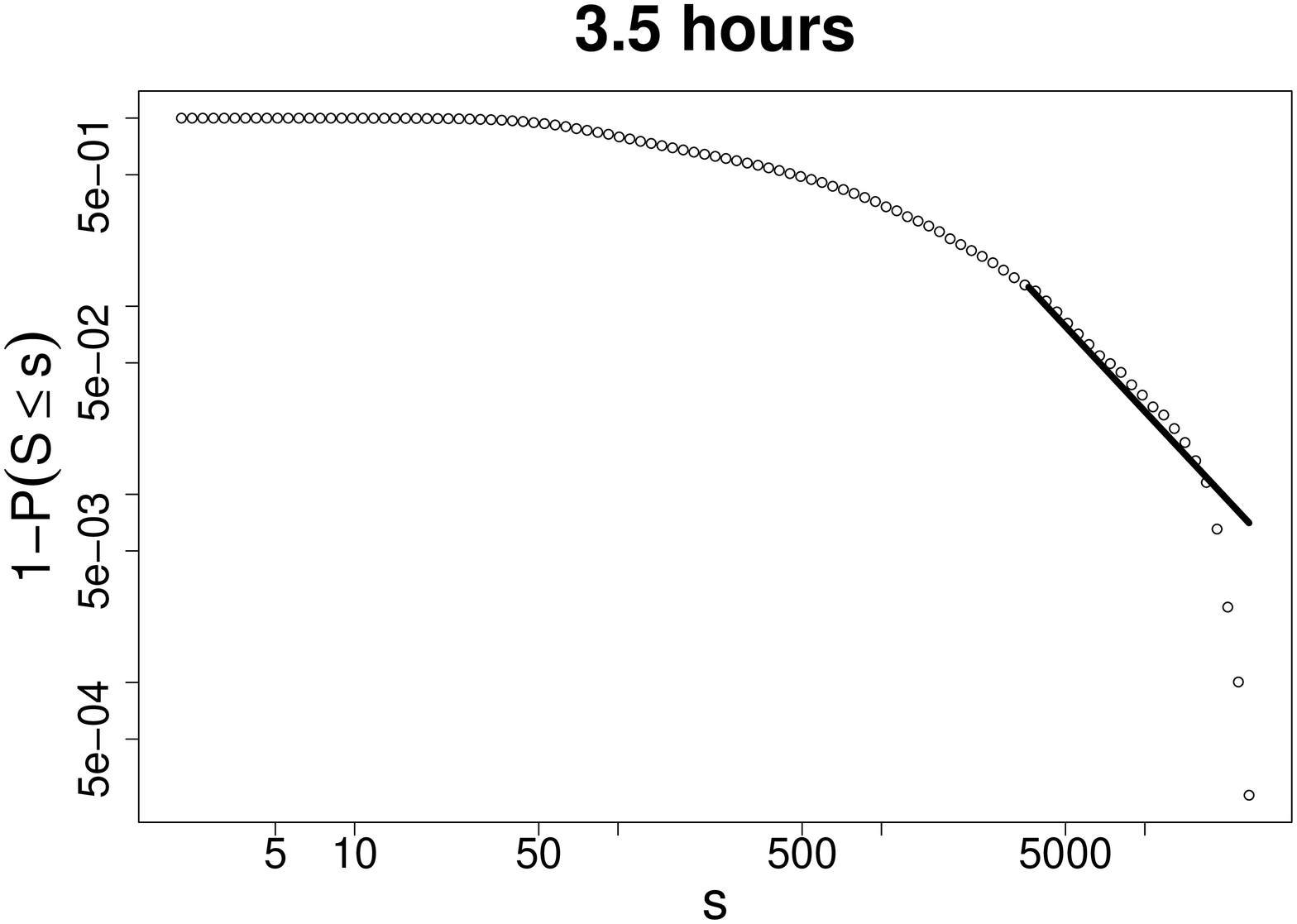}} & \rotatebox{270}{\includegraphics[scale=.2]{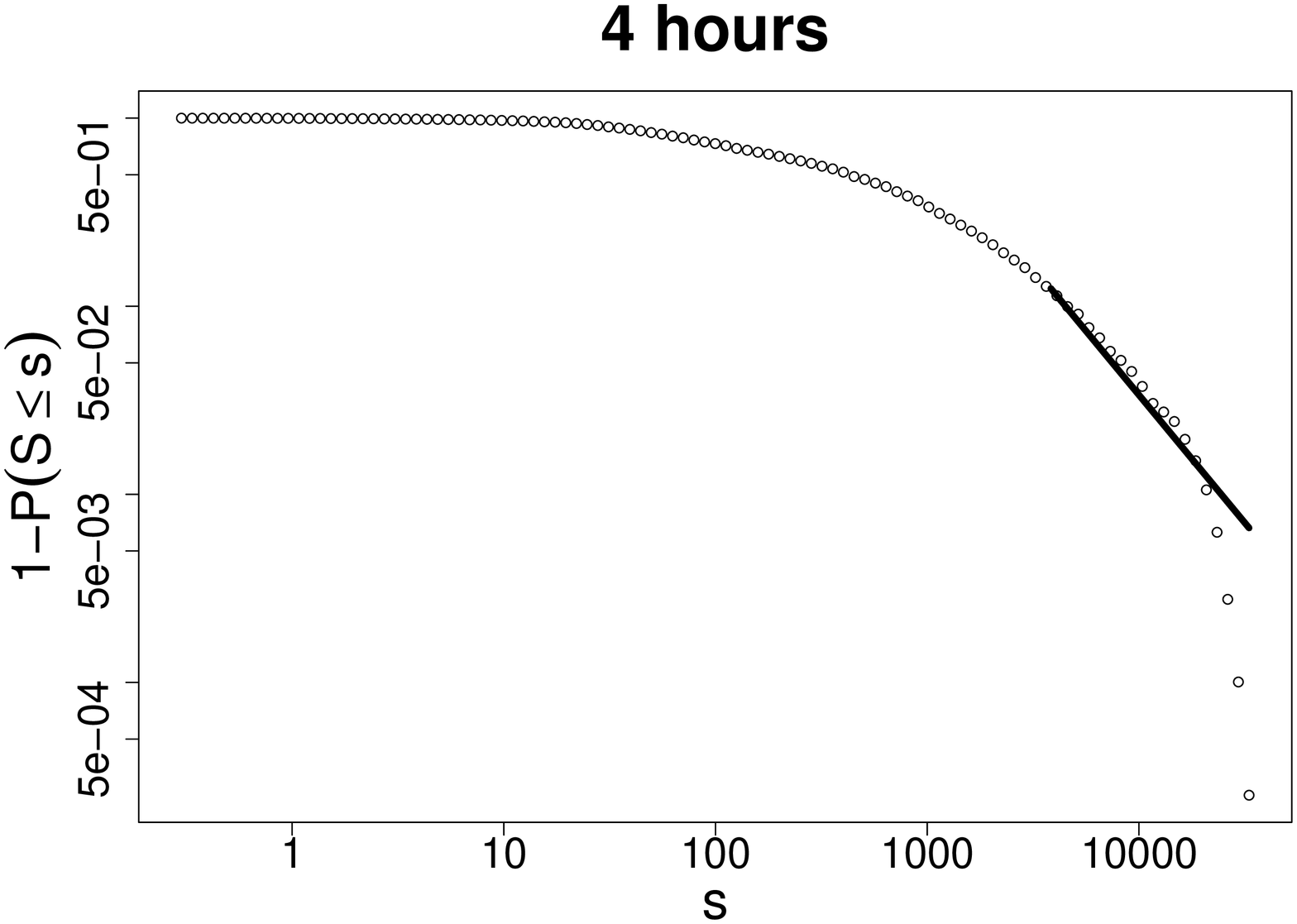}}\\
        \rotatebox{270}{\includegraphics[scale=.2]{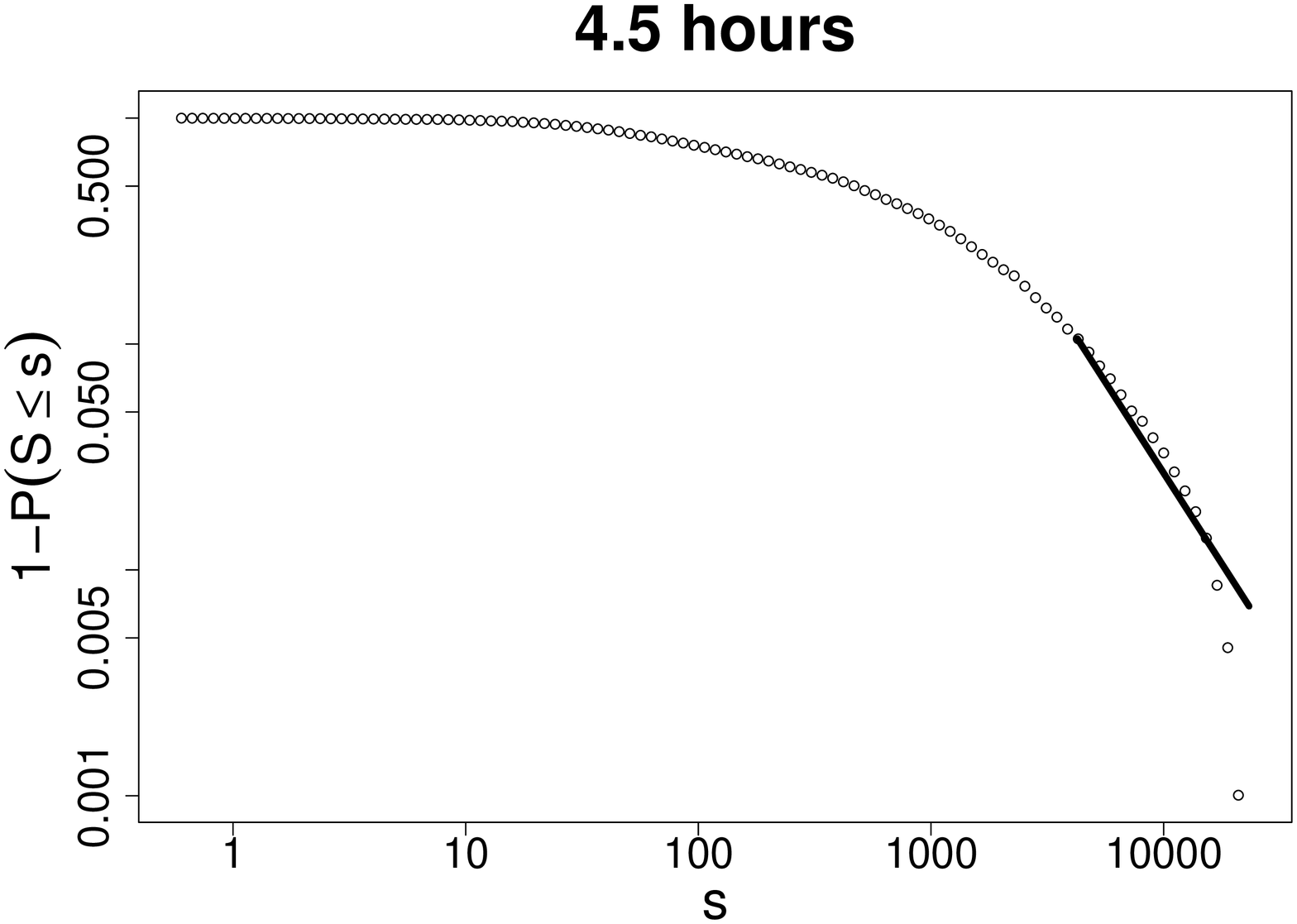}} & \rotatebox{270}{\includegraphics[scale=.2]{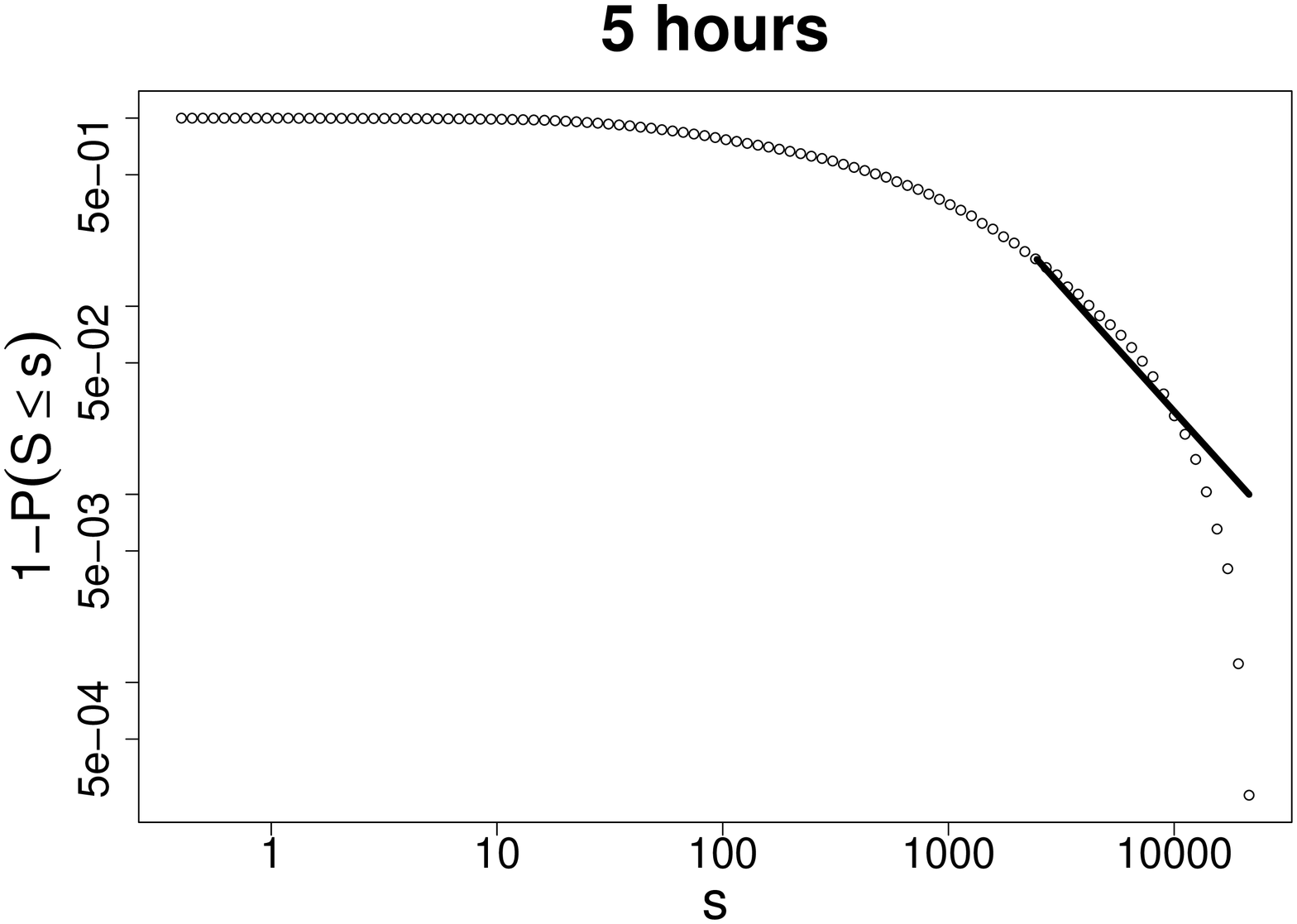}} & \rotatebox{270}{\includegraphics[scale=.2]{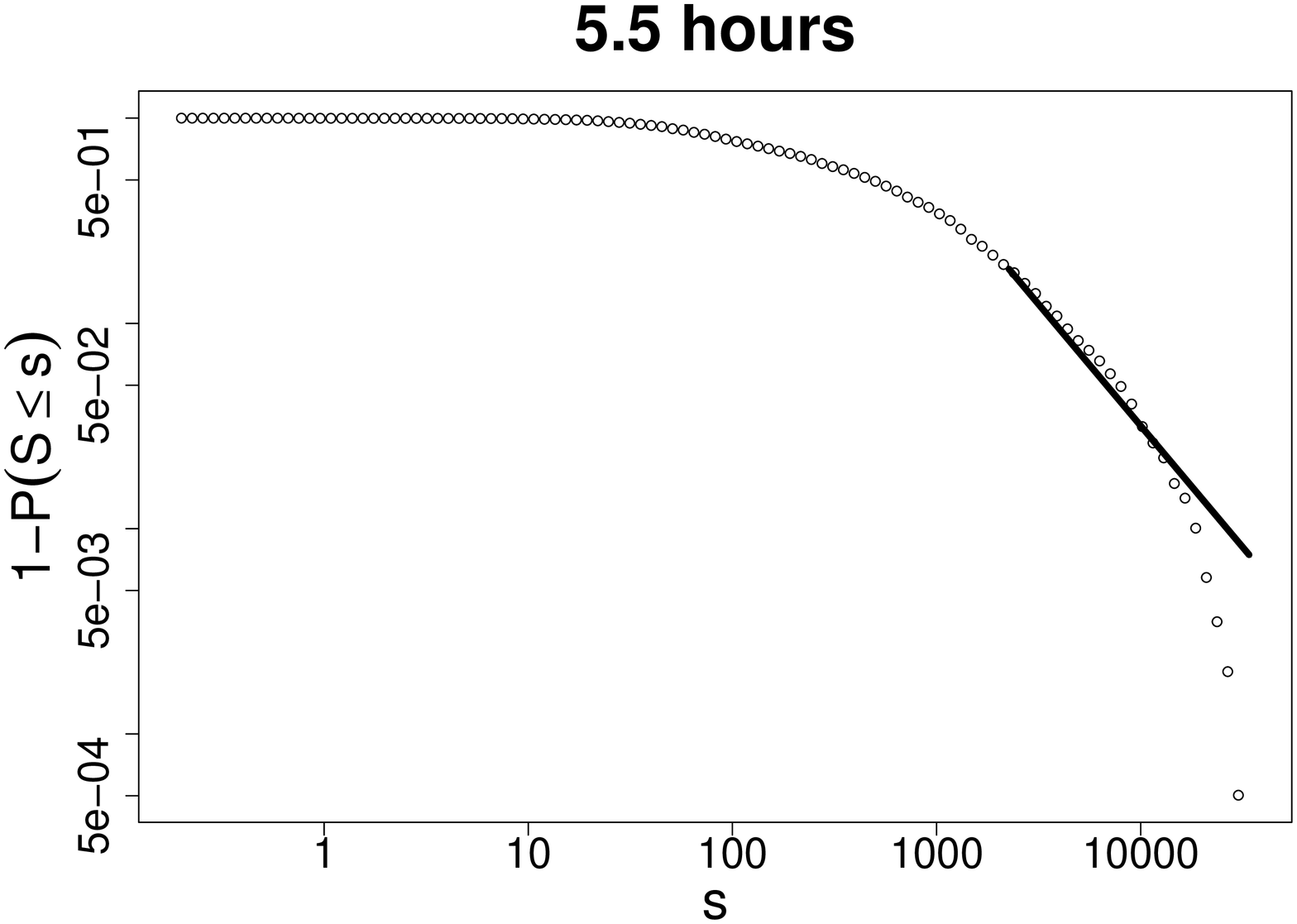}} & \rotatebox{270}{\includegraphics[scale=.2]{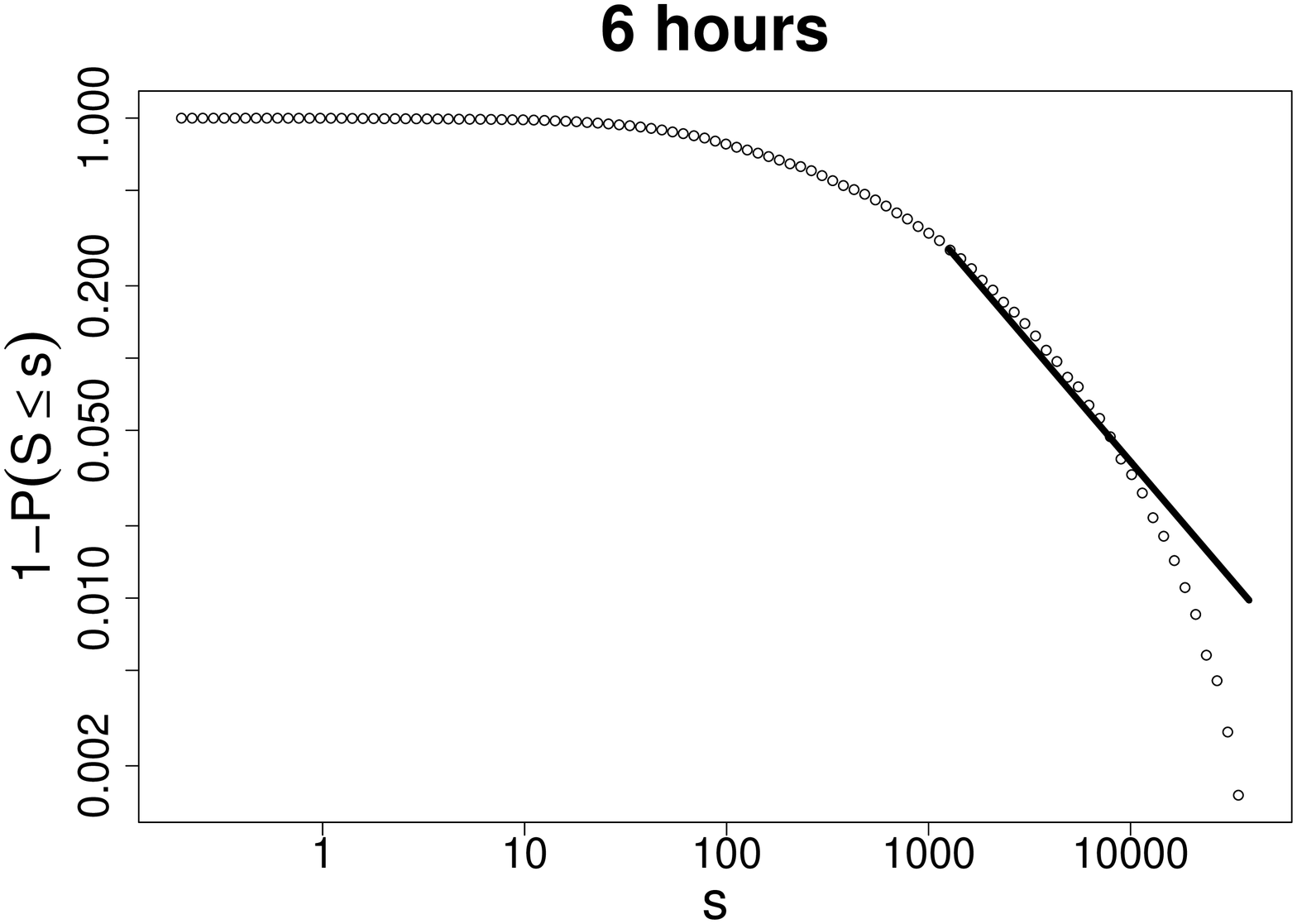}}\\
        \rotatebox{270}{\includegraphics[scale=.2]{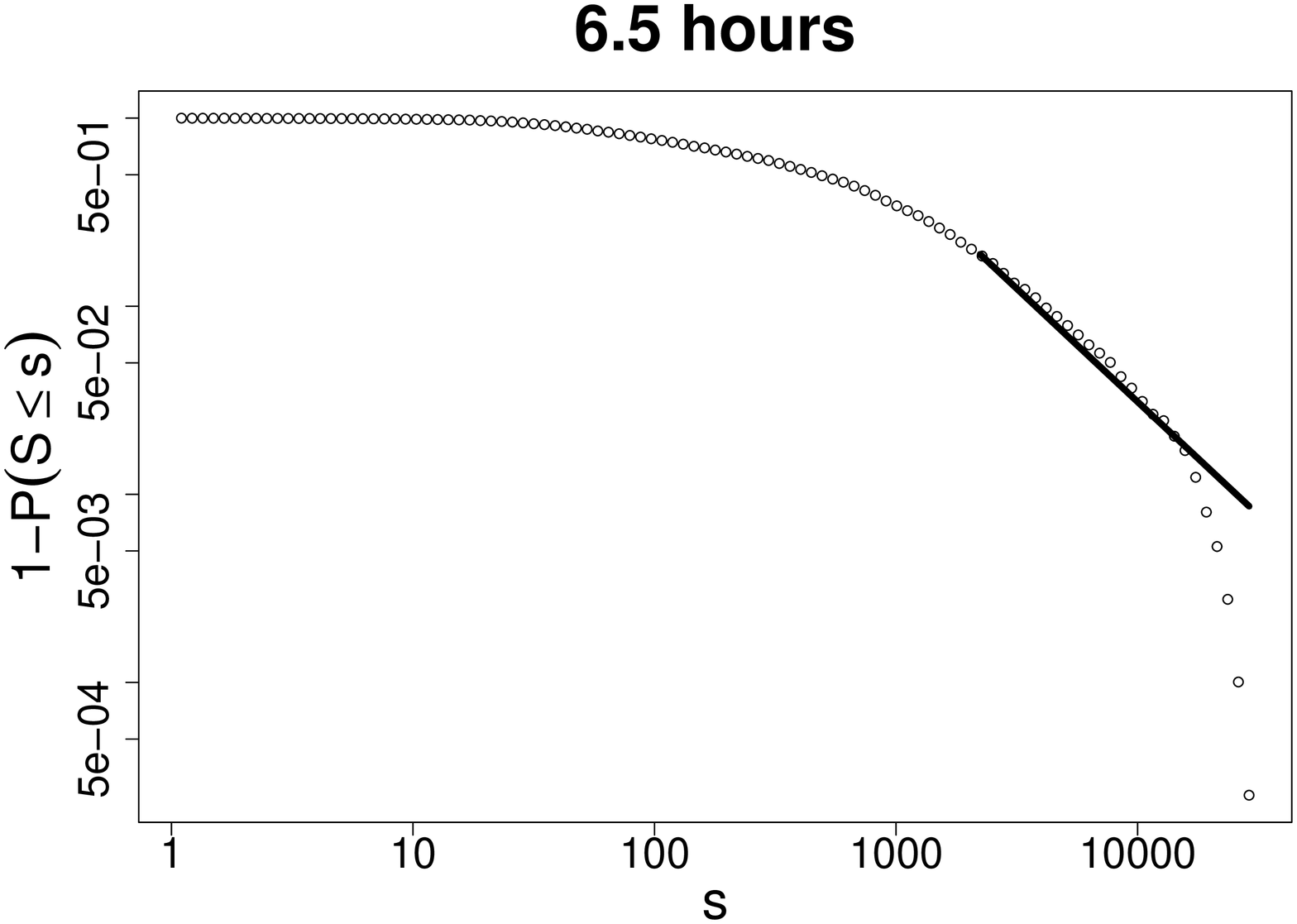}} & \rotatebox{270}{\includegraphics[scale=.2]{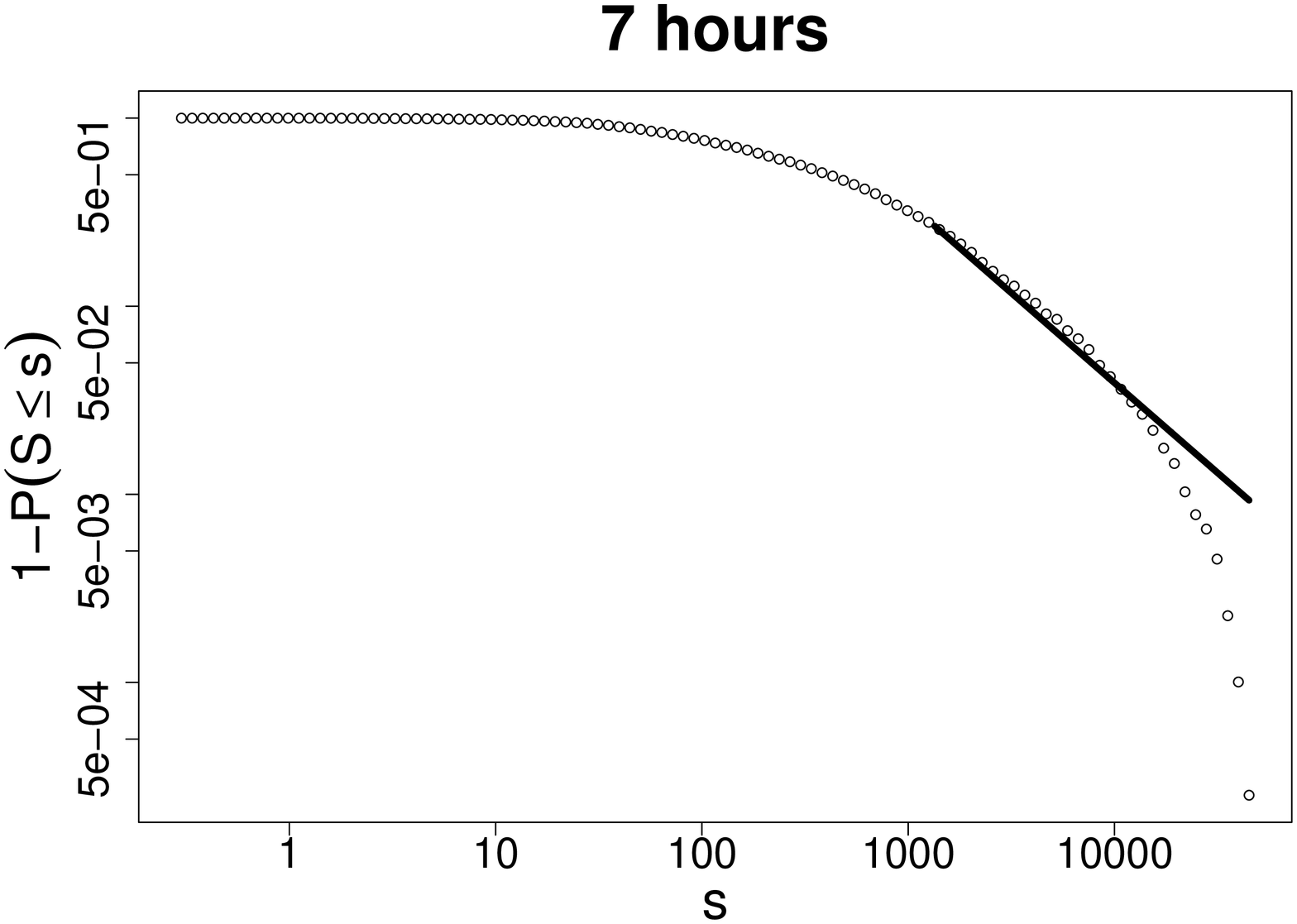}} & \rotatebox{270}{\includegraphics[scale=.2]{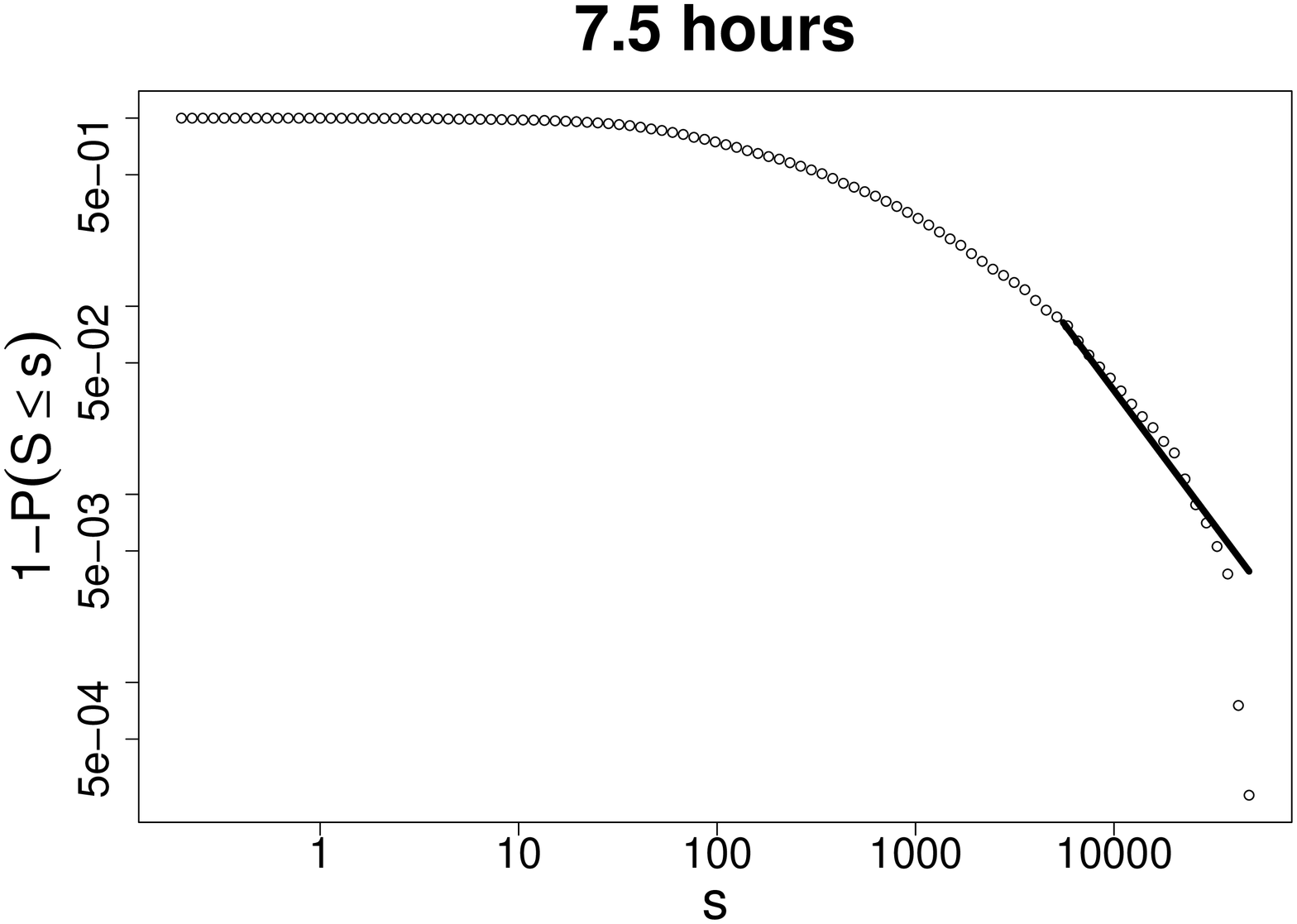}} & \rotatebox{270}{\includegraphics[scale=.2]{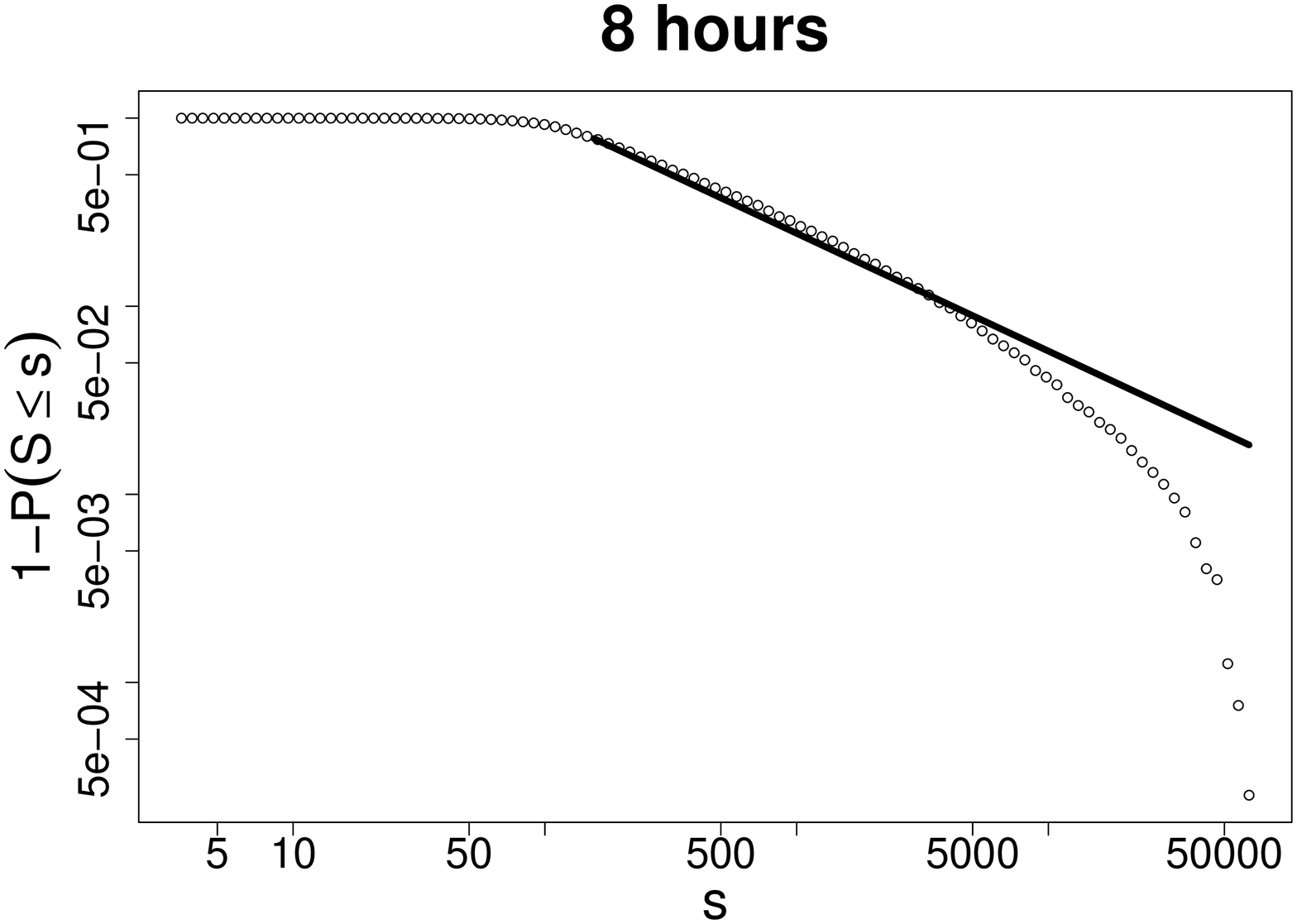}}\\
      \end{tabular}
      \caption{\footnotesize 
        Activity distributions in a dynamical transcriptome. 
        mRNA expression values distribution in \textit{E. coli} cells in a time-series growth experiment (between 2-8 hrs.) on mixed-carbon substrate medium are shown \citep{beg07}. 
        Lines represent the best fitting models for power-law models.
        \label{complex}}
    \end{center}
  \end{figure}
\end{landscape}
%%%%%%%%%%%%%%%%%%%%%%%%%%%%%%%%%%%%%%%%%%%%%%%%%%%
\begin{figure}
  \begin{center}
    \begin{tabular}{cc}
      \multicolumn{1}{l}{\mbox{\bf a.}} & \multicolumn{1}{l}{\mbox{\bf b.}} \\
      \includegraphics[scale=.3]{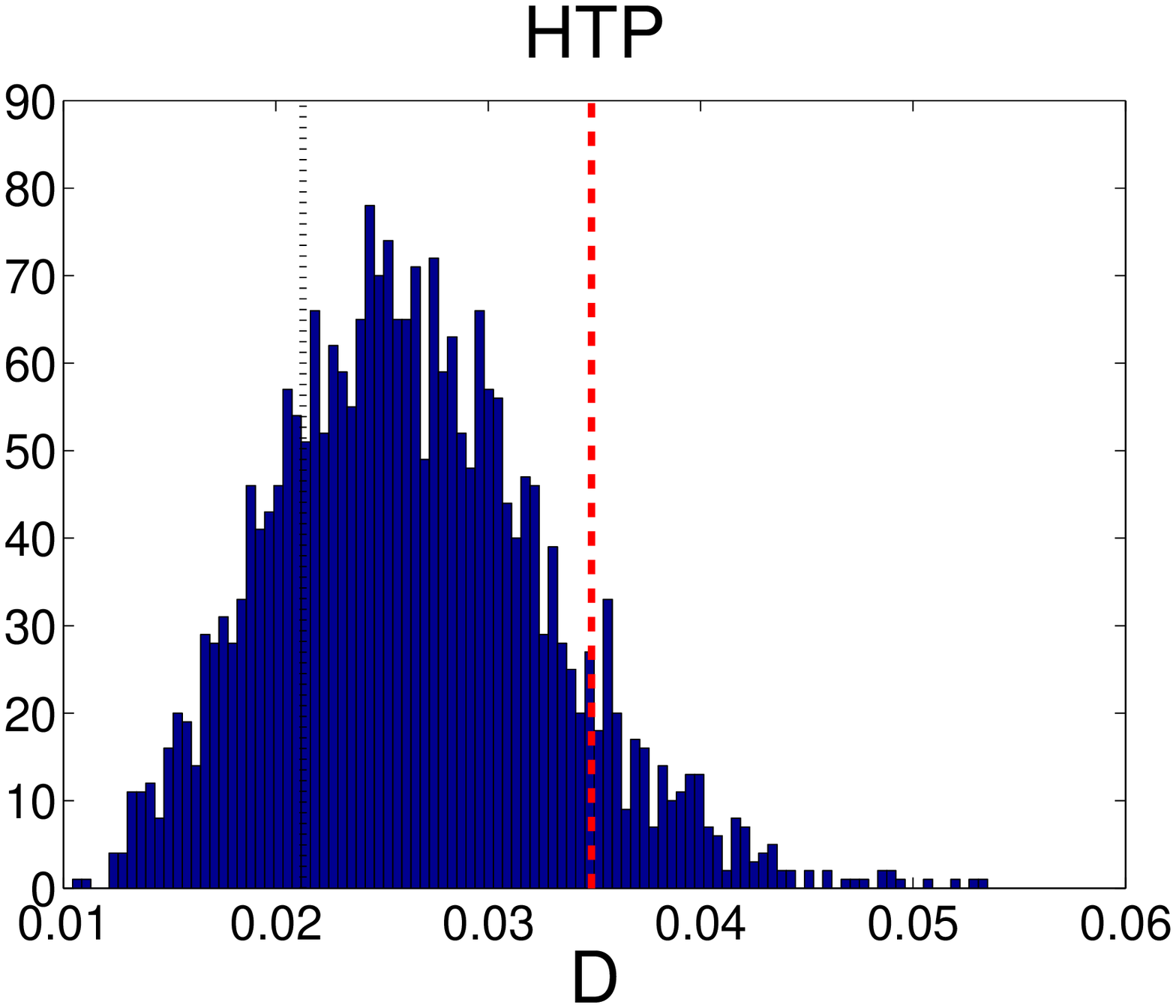}&\includegraphics[scale=.3]{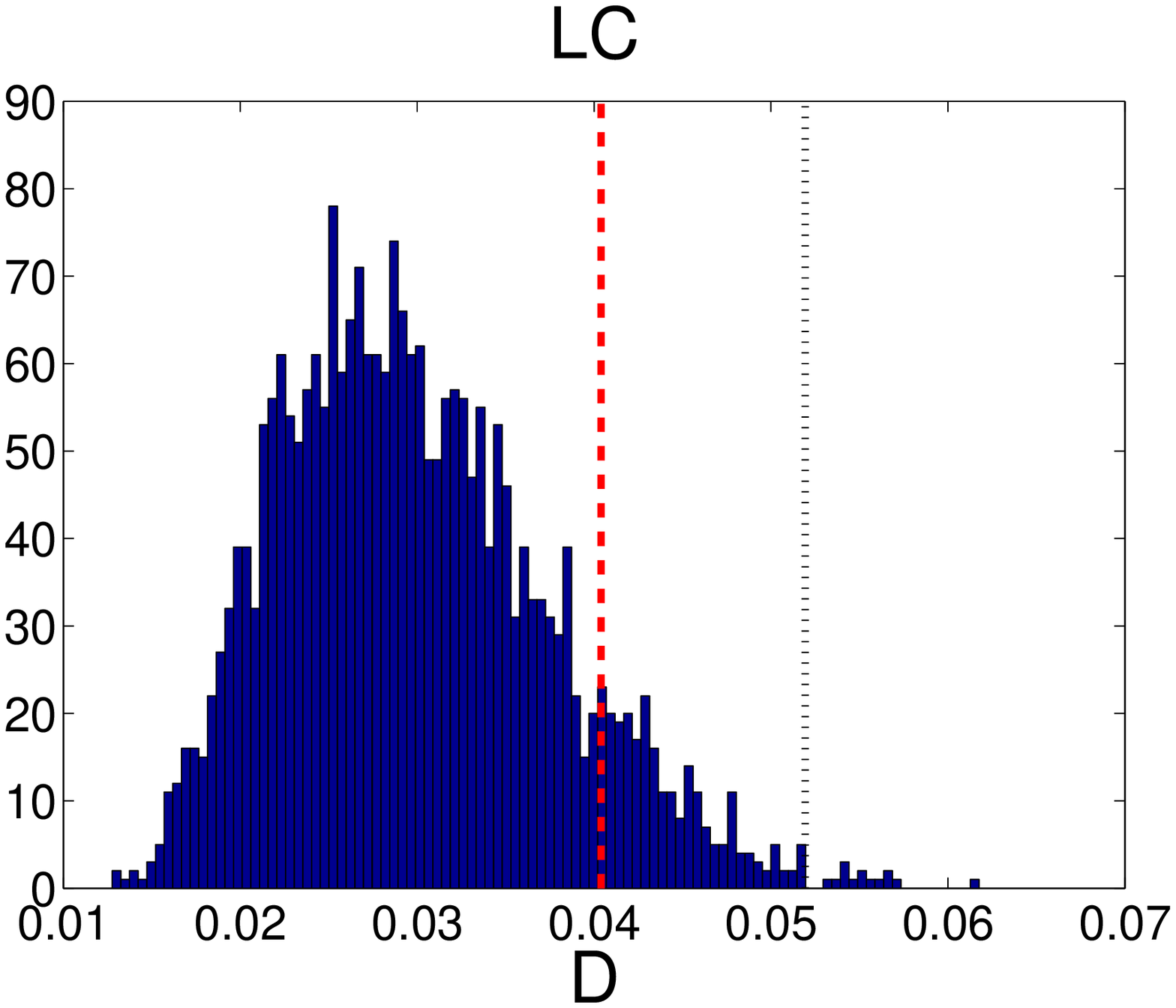}\\
      \multicolumn{1}{l}{\mbox{\bf c.}} & \multicolumn{1}{l}{\mbox{\bf d.}} \\
      \includegraphics[scale=.3]{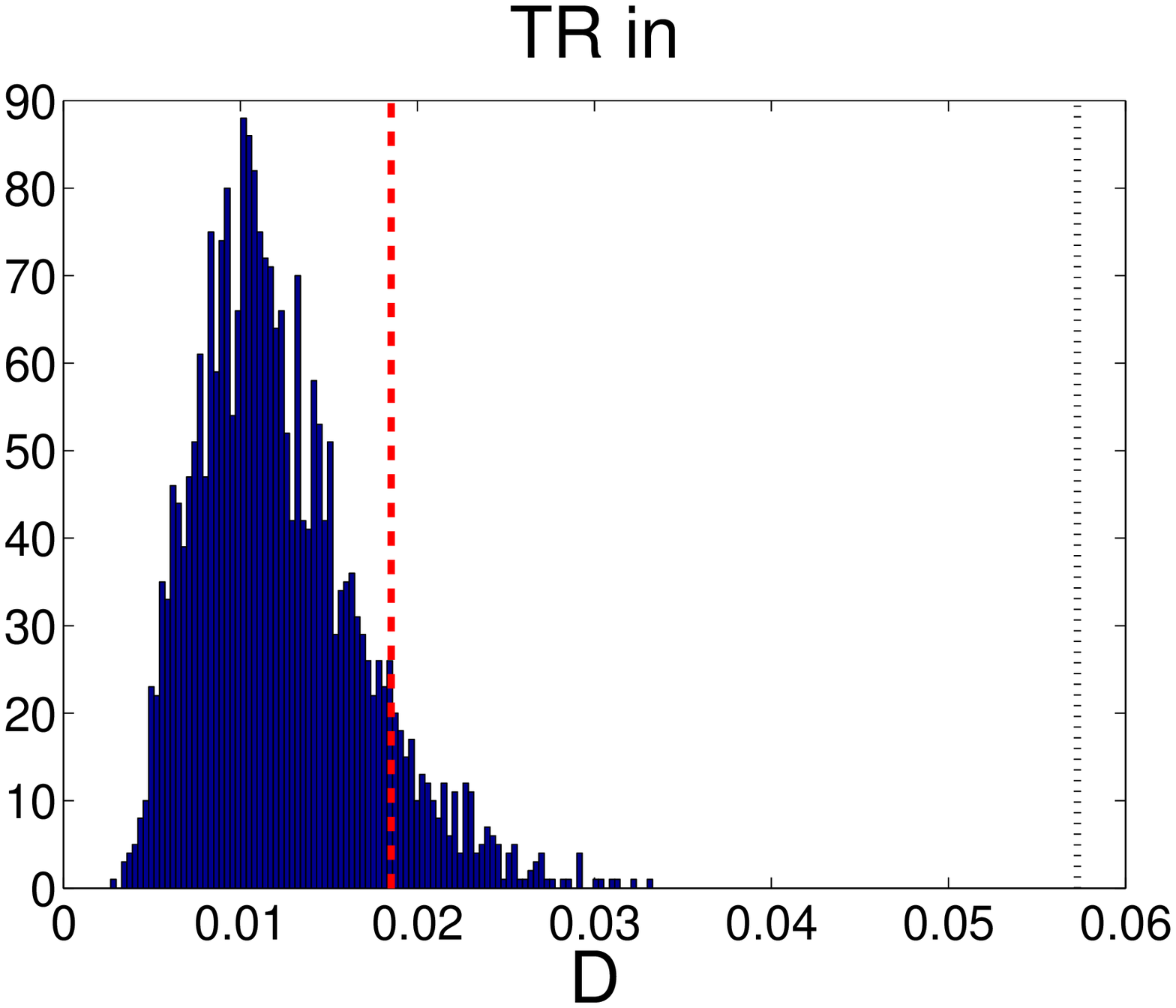}&\includegraphics[scale=.3]{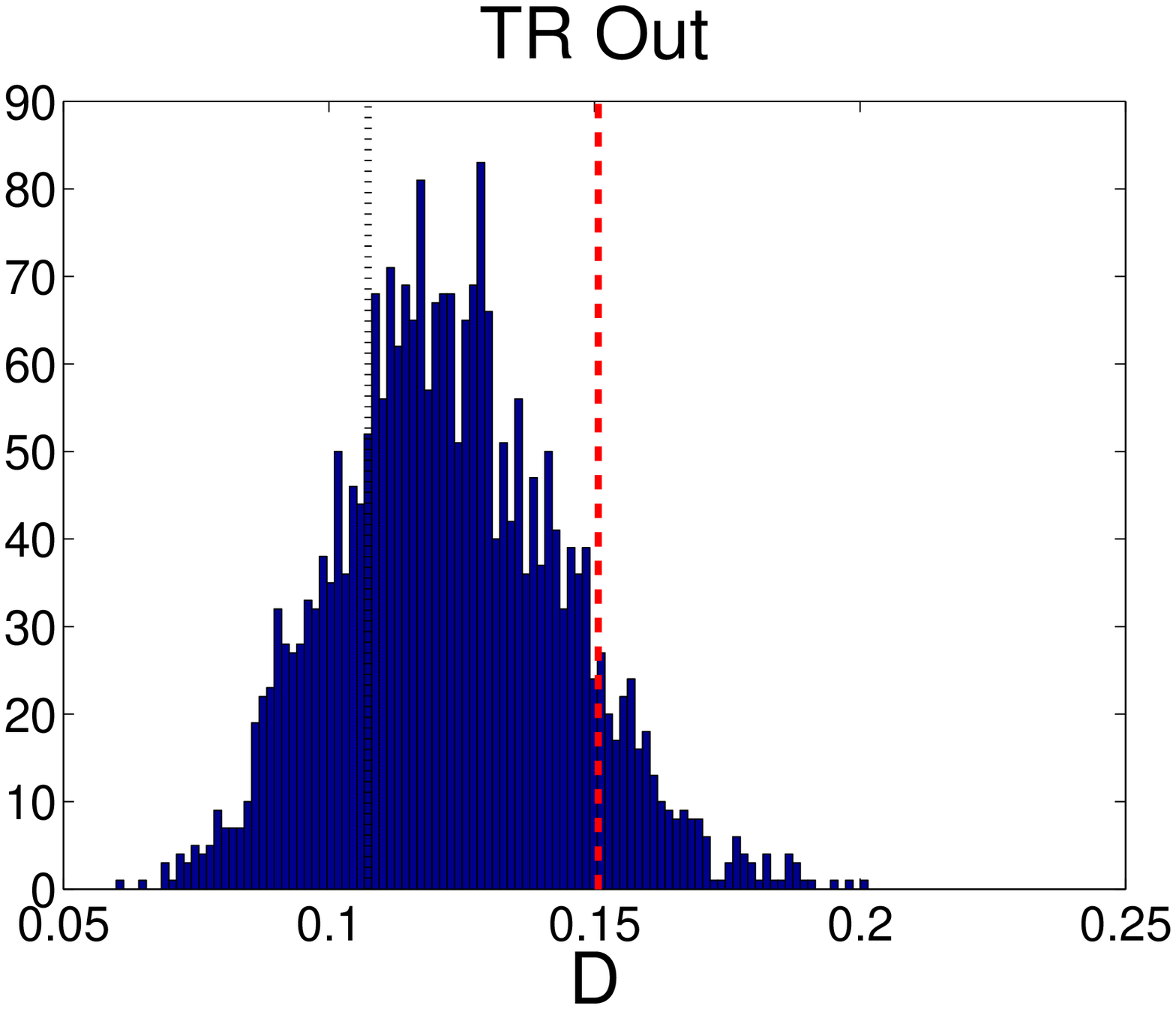}\\
      \multicolumn{1}{l}{\mbox{\bf e.}} & \multicolumn{1}{l}{\mbox{\bf f.}} \\
      \includegraphics[scale=.3]{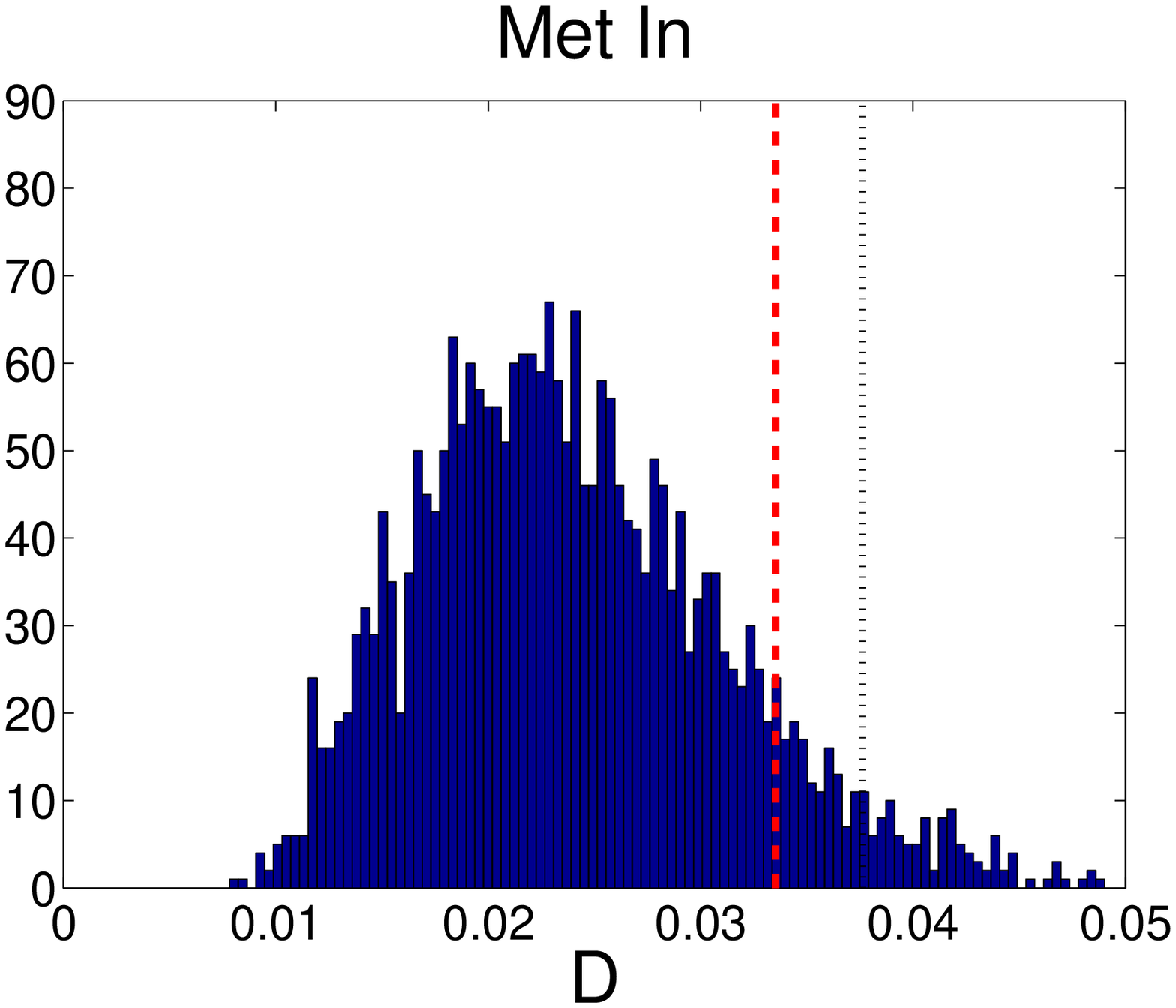}&\includegraphics[scale=.3]{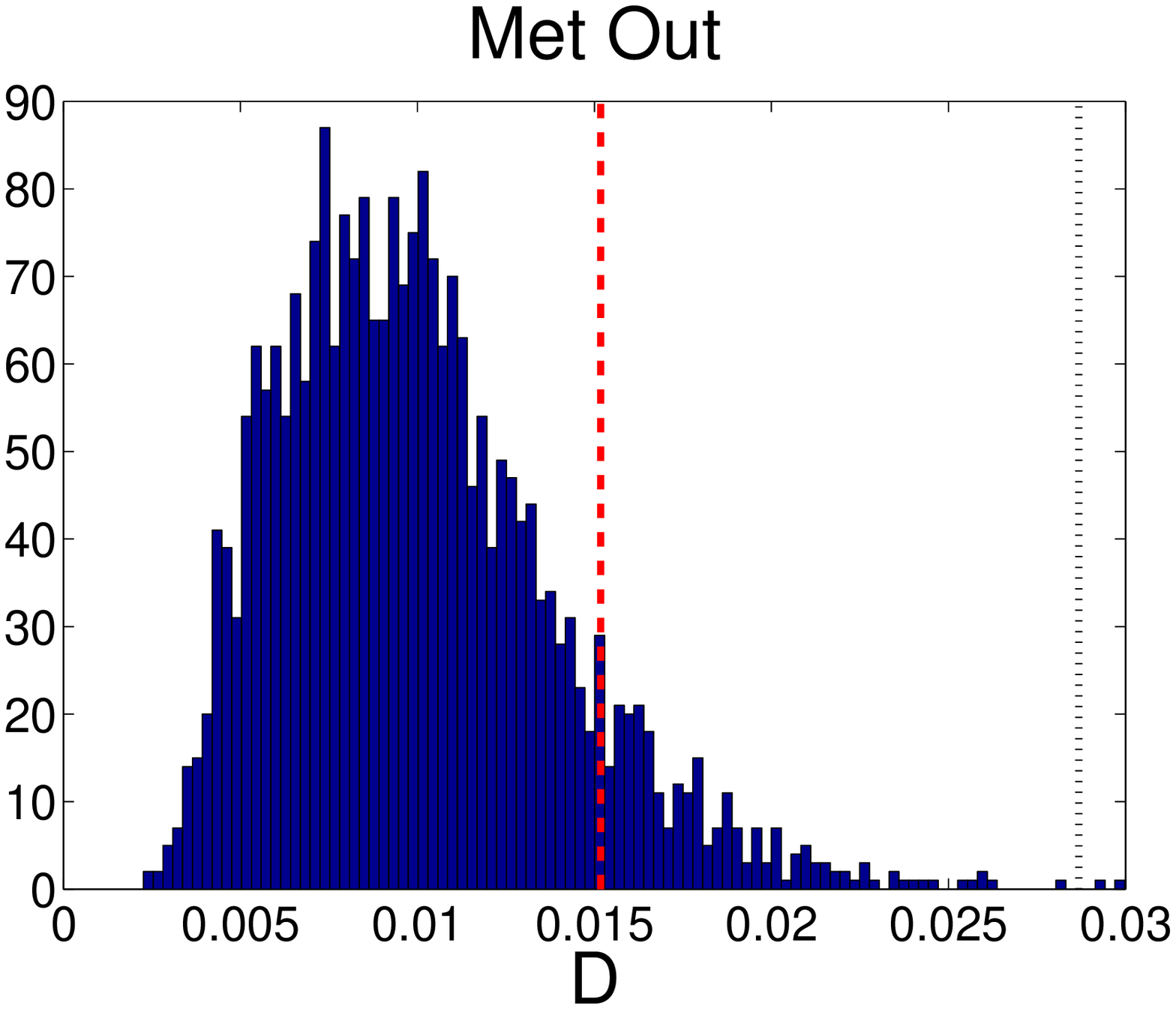}\\
    \end{tabular}
    \caption{\footnotesize 
      Distribution of the D values for the 2,500 synthetic data sets generated using Monte Carlo.
      Due to computational costs we show here the graphs for 2,500 data sets, however, 250,000 have been used to calculate the $p$-values associated to the power-law estimation.
      Red dashed line indicates the confidence level at 0.1 and black dotted line indicates the experimental value.
      \label{Ddistribution}}
  \end{center}
\end{figure}
%%%%%%%%%%%%%%%%%%%%%%%%%%%%%%%%%%%%%%%%%%%%%%%%%%%
\begin{figure}
  \begin{center}
    \includegraphics[scale=.5]{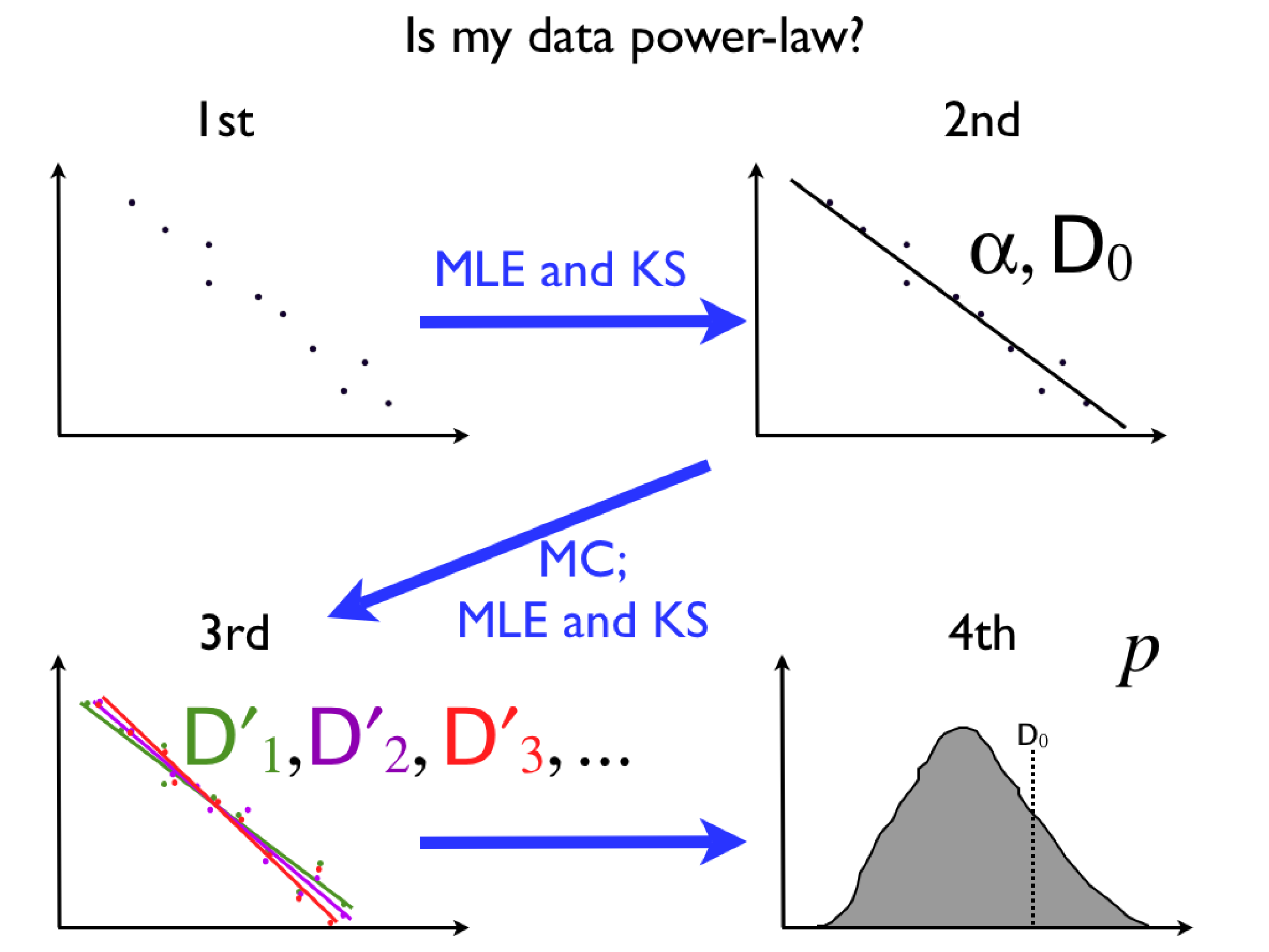}\\
    \caption{\footnotesize 
      Several steps are used to know if a data set follows a power-law distribution.
      We infer first the power-law parameters, $\alpha$ and $x_{min}$ using MLE.
      We then use the KS test to find $D_0$. 
      $D_0$ value is retrieved from two correlated distributions, the experimental distribution and the model distribution inferred from the experimental data set.
      Therefore the $p$-value associated to it is not valid.
      We generate several (250,000 in this article) \emph{synthetic} data sets from the inferred model using Monte Carlo.
      We repeat the previous step for each data set generated, MLE and KS.
      At this point we have a distribution of $D$ \emph{synthetic} values. 
      The position of $D_0$ within the distribution of \emph{synthetic} values provides the adequate $p$-value.
      The distribution of synthetic $D$ distribution and $D_0$ values for the discrete interaction data sets are shown in Figure \ref{Ddistribution}.
      \label{figureProtocol}}
  \end{center}
\end{figure}
%%%%%%%%%%%%%%%%%%%%%%%%%%%%%%%%%%%%%%%%%%%%%%%%%%% 
\begin{landscape}
  \begin{table}
    \begin{center}
      \begin{tabular}{ c | c | c  c | c  c | c  c | c  c | c}
        & Power-Law & \multicolumn{2}{|c|}{Log-Normal} & \multicolumn{2}{|c|}{Exponential} & \multicolumn{2}{|c|}{Weibull} & \multicolumn{2}{|c|}{PL + Exp.} & \\ 
        data set & $p$ & NLLR & $p$ & NLLR & $p$ &  NLLR & $p$ & LLR & $p$ & diagnostic\\ \hline
        Dil. 0.1 & 0.04 & -1.00 & 0.32 & 18.38 & \textbf{0.00} & -0.73 & 0.46 & -7.61 & \textbf{0.00} & reject\\ \hline
        Dil. 0.25 & 0.00 & -2.31 & \textbf{0.02} & 16.36 & \textbf{0.00} & -2.36 & \textbf{0.02} & -14.77 & \textbf{0.00} & reject\\ \hline
        Dil. 0.40 & 0.00 & -3.57 & \textbf{0.00} & 13.58 & \textbf{0.00} & -3.72  & \textbf{0.00} & -24.66 & \textbf{0.00} & reject\\ \hline
        Dil. 0.55 & 0.03 & -1.79 & 0.07 & 17.34 & \textbf{0.00} & -1.78 & 0.07 & -10.95 & \textbf{0.00} & reject\\ \hline
        Dil. 0.72 & 0.02 & -0.58 & 0.56 & 23.32 & \textbf{0.00} & 0.12 & 0.90 & -6.84 & \textbf{0.00} & reject\\ \hline
      \end{tabular}
      \caption{\footnotesize
        Second column shows the $p$-value associated to the differences between the data and the power-law model.
        Next four columns correspond to the log likelihood ratio tests comparing the power-law model with other plausible models.
        The normalized log likelihood ratio (NLLR) is used for non-nested functions while the raw log likelihood ratio (LLR) is used for the power-law with exponential cutoff model.
        $p$-values here are associated to the differences between the two models.
        Significant $p$-values are denoted in bold.
        \label{tableSteadyState}
      }
    \end{center}
  \end{table}
\end{landscape}
%%%%%%%%%%%%%%%%%%%%%%%%%%%%%%%%%%%%%%%%%%%%%%%%%%%
\begin{landscape}
  \begin{table}
    \begin{center}
      \begin{tabular}{ c | c | c  c | c  c | c  c | c  c | c}
        & Power-Law & \multicolumn{2}{|c|}{Log-Normal} & \multicolumn{2}{|c|}{Exponential} & \multicolumn{2}{|c|}{Weibull} & \multicolumn{2}{|c|}{PL + Exp.} & \\ 
        data set & $p$ & NLLR & $p$ & NLLR & $p$ &  NLLR & $p$ & LLR & $p$ & diagnostic\\ \hline
        Yeast & 0.00 & -1.91 & 0.06 & 8.74 & \textbf{0.00} & -1.93 & 0.05 & -8.18 & \textbf{0.00} & reject\\ \hline
      \end{tabular}
      \caption{\footnotesize
        Second column shows the $p$-value associated the differences between the data and the power-law model.
        Next four columns correspond to the log likelihood ratio tests comparing the power-law model with other plausible models.
        The normalized log likelihood ratio (NLLR) is used for non-nested functions while the raw log likelihood ratio (LLR) is used for the power-law with exponential cutoff model.
        $p$-values here are associated to the differences between the two models.
        Significant $p$-values are denoted in bold.
        \label{tableYeast}
      }
    \end{center}
  \end{table}
\end{landscape}
%%%%%%%%%%%%%%%%%%%%%%%%%%%%%%%%%%%%%%%%%%%%%%%%%%%
\begin{landscape}
  \begin{table}
    \begin{center}
      \begin{tabular}{ c | c | c  c | c  c | c  c | c  c | c}
        & Power-Law & \multicolumn{2}{|c|}{Log-Normal} & \multicolumn{2}{|c|}{Exponential} & \multicolumn{2}{|c|}{Weibull} & \multicolumn{2}{|c|}{PL + Exp.} & \\ 
        data set & $p$ & NLLR & $p$ & NLLR & $p$ &  NLLR & $p$ & LLR & $p$ & diagnostic\\ \hline
        2 hours & 0.00 & -4.76 & \textbf{0.00} & -3.67 & \textbf{0.00} & -5.02 & \textbf{0.00} & -31.57 & \textbf{0.00} & reject\\ \hline
        2.5 hours & 0.00 & -4.46 & \textbf{0.00} & -2.33 & \textbf{0.02} & -4.67  & \textbf{0.00} & -27.63 & \textbf{0.00} & reject\\ \hline
        3.5 hours & 0.00 & -4.31 & \textbf{0.00} & -2.21 & \textbf{0.03} & -4.49 & \textbf{0.00} & -25.08 & \textbf{0.00} & reject\\ \hline
        4 hours & 0.00 & -4.08 & \textbf{0.00} & -1.60 & 0.11 & -4.31 & \textbf{0.00} & -25.82 & \textbf{0.00} & reject\\ \hline
        4.5 hours & 0.00 & -4.56 & \textbf{0.00} & -4.37 & \textbf{0.00} & -4.79 & \textbf{0.00} & -29.10 & \textbf{0.00} & reject\\ \hline
        5 hours & 0.00 & -5.91 & \textbf{0.00} & -4.37 & \textbf{0.00} & -6.26 & \textbf{0.00} & -52.34 & \textbf{0.00} & reject\\ \hline
        5.5 hours & 0.00 & -5.22 & \textbf{0.00} & -0.48 & 0.63 & -5.48 & \textbf{0.00} & -38.72 & \textbf{0.00} & reject\\ \hline
        6 hours & 0.00 & -6.38 & \textbf{0.00} & 2.74 & \textbf{0.00} & -6.70 & \textbf{0.00} & -54.69 & \textbf{0.00} & reject\\ \hline
        6.5 hours & 0.00 & -4.73 & \textbf{0.00} & 0.50 & 0.61 & -4.99 & \textbf{0.00} & -33.91 & \textbf{0.00} & reject\\ \hline
        7 hours & 0.00 & -5.42 & \textbf{0.00} & 4.48 & \textbf{0.00} & -5.69 & \textbf{0.00} & -43.84 & \textbf{0.00} & reject\\ \hline
        7.5 hours & 0.00 & -2.88 & \textbf{0.00} & -0.02 & 0.99 & -3.01 & \textbf{0.00} & -12.29 & \textbf{0.00} & reject\\ \hline
        8 hours & 0.00 & -8.20 & \textbf{0.00} & 17.77 & \textbf{0.00} & -8.60 & \textbf{0.00} & -98.07 & \textbf{0.00} & reject\\ \hline
      \end{tabular}
      \caption{\footnotesize
        Second column shows the $p$-value associated the differences between the data and the power-law model.
        Next four columns correspond to the log likelihood ratio tests comparing the power-law model with other plausible models.
        The normalized log likelihood ratio (NLLR) is used for non-nested functions while the raw log likelihood ratio (LLR) is used for the power-law with exponential cutoff model.
        $p$-values here are associated to the differences between the two models.
        Significant $p$-values are denoted in bold.
        \label{tableComplex}
      }
    \end{center}
  \end{table}
\end{landscape}
%%%%%%%%%%%%%%%%%%%%%%%%%%%%%%%%%%%%%%%%%%%%%%%%%%%